\RequirePackage{ifpdf}
\documentclass[hyper,letterpaper]{JHEP3}
\usepackage{amssymb,amsfonts,bm,amsmath,empheq}
\usepackage{cite}
\usepackage{graphicx}
\usepackage{multirow}
\usepackage{verbatim}
\usepackage{appendix}

\usepackage{url}
\usepackage{float}

\newcommand{\bea}{\begin{eqnarray}}
\newcommand{\eea}{\end{eqnarray}}
\newcommand{\be}{\begin{equation}}
\newcommand{\ee}{\end{equation}}

\def\bra#1{\left\langle\,#1\,\right|}
\def\ket#1{\left|\,#1\,\right\rangle}

\def\IZ{\mathbb {Z}}

\def\IC{\mathbb {C}}

\renewcommand{\bar}{\overline}
\renewcommand{\hat}{\widehat}

\def\bra#1{\left\langle\,#1\,\right|}
\def\ket#1{\left|\,#1\,\right\rangle}

\def\vect#1{\hbox{\boldmath{$#1$}}}
\def\sr#1{\scriptscriptstyle\rm #1}

\makeatletter
\def\Left#1#2\Right{\begingroup%
   \def\ts@r{\nulldelimiterspace=0pt \mathsurround=0pt}%
   \let\@hat=#1%
   \def\sht@im{#2}%
   \def\@t{{\mathchoice{\def\@fen{\displaystyle}\k@fel}%
          {\def\@fen{\textstyle}\k@fel}%
          {\def\@fen{\scriptstyle}\k@fel}%
          {\def\@fen{\scriptscriptstyle}\k@fel}}}%
   \def\g@rin{\ts@r\left\@hat\vphantom{\sht@im}\right.}%
   \def\k@fel{\setbox0=\hbox{$\@fen\g@rin$}\hbox{%
      $\@fen \kern.3875\wd0 \copy0 \kern-.3875\wd0%
      \llap{\copy0}\kern.3875\wd0$}}%
      \def\pt@h{\mathopen\@t}\pt@h\sht@im%
      \Right}%
\def\Right#1{\let\@hat=#1%
   \def\st@m{\mathclose\@t}%
   \st@m\endgroup}
\makeatother

\title{From CFT to Ramond super-quantum curves}

\author{Pawe{\l} Ciosmak$^{1}$, Leszek Hadasz$^{2}$, Zbigniew Jask\'olski$^{3}$, Masahide Manabe$^{4}$ and Piotr Su{\l}kowski$^{5,6}$
\\
$^1$ Faculty of Mathematics, Informatics and Mechanics, University of Warsaw, ul. Banacha 2, 02-097 Warsaw, Poland  \\
$^2$ M.\ Smoluchowski Institute of Physics, Jagiellonian University, ul. {\L}ojasiewicza 11, 30-348 Krak{\'o}w, Poland  \\
$^3$ Institute of Theoretical Physics, University of Wroc{\l}aw, pl. M. Borna 1, 95-204~Wroc{\l}aw, Poland \\
$^4$ Max-Planck-Institut f\"ur Mathematik, Vivatsgasse 7, 53111 Bonn, Germany \\
$^5$ Faculty of Physics, University of Warsaw, ul. Pasteura 5, 02-093 Warsaw, Poland  \\
$^6$ Walter Burke Institute for Theoretical Physics, California Institute of Technology, Pasadena, CA 91125, USA}

\abstract{As we have shown in the previous work, using the formalism of matrix and eigenvalue models, to a given classical algebraic curve one can associate an infinite family of quantum curves, which are in one-to-one correspondence with singular vectors of a certain (e.g. Virasoro or super-Virasoro) underlying algebra. In this paper we reformulate this problem in the language of conformal field theory. Such a reformulation has several advantages: it leads to the identification of quantum curves more efficiently, it proves in full generality that they indeed have the structure of singular vectors, it enables identification of corresponding eigenvalue models. Moreover, this approach can be easily generalized to other underlying algebras. To illustrate these statements we apply the conformal field theory formalism to the case of the Ramond version of the super-Virasoro algebra. We derive two classes of corresponding Ramond super-eigenvalue models, construct Ramond super-quantum curves that have the structure of relevant singular vectors, and identify underlying Ramond super-spectral curves.
We also analyze Ramond multi-Penner models and show that they lead to supersymmetric generalizations of BPZ equations.
\\
\\
\\
\\
\\
\\
\\
\\
\\
{\tt CALT-2017-070}}

\begin{document}

%******************************************************************************************
%******************************************************************************************
%******************************************************************************************

\newpage

\section{Introduction}  \label{sec-intro}

{\em Quantum curves} are intriguing objects, identified originally in various problems related to string theory and supersymmetric gauge theories \cite{ADKMV,DHSV,DHS,ACDKV,Kozcaz:2010af}, and analyzed from various mathematical perspectives e.g. in \cite{DijkgraafFuji-2,Borot:2012cw,superA,Dunin-Barkowski:2013wca,Schwarz:2014hfa,Norbury-quantum,Marino:2015nla,Dumitrescu:2015mpa,Bouchard:2016obz,Bouchard:2016uud,Belliard:2016cyc,Fuji:2017vzs}. In general, quantum curves take form of differential operators $\widehat{A}(\hat x, \hat y)$ imposing Schroedinger-like equations on appropriately defined wave-functions $\Psi(x)$
\be
\widehat{A}(\hat x, \hat y)\Psi(x) = 0.      \label{Ahatxy}
\ee
The operators $\hat x$ and $\hat y$ satisfy the commutation relation
\be
[\hat  y, \hat x] = \hbar,
\ee
so that $\hat y$ can be identified as $\hbar\partial_x$. In the limit $\hbar\to 0$ the operators $\hat x$ and $\hat y$ reduce to complex commuting variables $x$ and $y$, and the quantum curve equation (\ref{Ahatxy}) reduces to a ``classical'' algebraic curve
\be
A(x,y) = 0.     \label{Axy}
\ee

%Quantum curves are intimately related to matrix models and the formalism of topological recursion \cite{eyn-or}.
Conjecturally, in all situations where quantum curves arise, their form can be determined by means of the {\em topological recursion} \cite{abmodel}, which can be regarded as a reformulation and generalization of {\em loop equations} in matrix models \cite{eyn-or}. From this perspective the classical curve (\ref{Axy}) is identified as an algebraic curve that provides the initial condition for the topological recursion. In case the corresponding matrix model is known, the curve (\ref{Axy}) is identified as its {\em spectral curve}, and the wave-function $\Psi(x)$ is identified as a determinant expectation value $\langle\textrm{det}(x-M)\rangle$, where $\langle\, \cdot\, \rangle$ denotes an expectation value computed by integrating over matrices $M$ from an appropriate ensemble. Therefore, using the topological recursion or matrix model formalism, to a given algebraic curve one can associate the corresponding quantum curve.

In fact, it turns out that to a given algebraic curve one can assign not only one, but the whole family of quantum curves, which have the structure of {\em singular vectors} of the underlying symmetry algebra \cite{Manabe:2015kbj}. For curves related to hermitian matrix models, or the original topological recursion formulation \cite{eyn-or}, this symmetry algebra is the Virasoro algebra, and corresponding quantum curves have the structure of Virasoro singular vectors, as found in \cite{Manabe:2015kbj}. In the present paper we also refer to these curves as {\em Virasoro quantum curves}. In this case the determinant form of the wave-function is generalized so that it depends on an additional parameter $\alpha$; however consistent quantum curve equations arise only for certain discrete values of this parameter, which coincide with Virasoro degenerate momenta. One can also consider $\beta$-deformed version of these results -- this also leads to a discrete family of quantum curves with the structure of Virasoro singular vectors, however in this case the Virasoro algebra has an arbitrary central charge, parametrized by the parameter $\beta$. The matrix (or eigenvalue) model form of the wave-function, depending on both parameters $\alpha$ and $\beta$, is referred to as {\em $\alpha/\beta$-deformed matrix integral} in \cite{Manabe:2015kbj}. These results can be regarded as a manifestation of general, intimate links between matrix models and the Virasoro algebra; for example it has been known for a long time, that matrix model loop equations -- and the topological recursion itself -- can be rewritten in the form of {\em Virasoro constraints} \cite{Fukuma:1990jw,Dijkgraaf:1990rs,Awata:1994xd}.

The above results have been generalized to a supersymmetric case in \cite{Ciosmak:2016wpx}, by considering ($\beta$-deformed) super-eigenvalue models for the Neveu-Schwarz sector \cite{AlvarezGaume:1991jd,Becker:1992rk,McArthur:1993hw,Plefka:1996tt,Semenoff:1996vm,Itoyama:2003mv}. These models generalize eigenvalue representation of hermitian matrix models in such a way, that the underlying algebra takes form of the Neveu-Schwarz version of the super-Virasoro algebra; in particular corresponding loop equations can be rewritten as {\em super-Virasoro constraints}. Consequently, to a super-eigenvalue model one can associate an infinite family of {\em super-quantum curves}, which have the structure of Neveu-Schwarz singular vectors of the super-Virasoro algebra. In the classical limit, such super-quantum curves reduce to supersymmetric algebraic curves, which are interesting in their own right \cite{Rabin:1987rg,Rabin:1993bw}.

To sum up, to a given classical (possibly supersymmetric) curve one can associate an infinite family of quantum curves, which have the structure of singular vectors of the underlying algebra. This result was found in \cite{Manabe:2015kbj,Ciosmak:2016wpx} upon the analysis of eigenvalue models, which provide a representation (or generalization) of matrix models; for a summary see also \cite{Ciosmak:2017omd}.

The aim of the present paper is twofold. First, we clarify the role of conformal field theory in the description of quantum curves. In particular, we rederive (in Virasoro and Neveu-Schwarz case) quantum curves using only conformal field theory techniques (instead of eigenvalue models). The main feature of this approach is the fact, that the singular vector structure of quantum curves follows automatically; being a consequence of the conformal field theory construction, the singular vector structure of quantum curves postulated in \cite{Manabe:2015kbj,Ciosmak:2016wpx} is therefore proven. Moreover, this approach has certain calculational advantages, and can be rather easily extended to more general algebras (possibly, although not necessarily, corresponding to more general matrix models, for example multi-matrix models, or their deformations e.g. see \cite{Chiantese:2003qb,Awata:2010yy}). Second, using the conformal field theory approach -- and to illustrate its power -- we find an infinite family of super-quantum curves corresponding to the Ramond sector of the super-Virasoro algebra. From this analysis we also derive the super-eigenvalue models representing the Ramond sector, whose form is not obvious to postulate a priori.

More precisely, we find two types of super-eigenvalue models in the Ramond sector, and corresponding two types of quantum curves, which have respectively the structure of Neveu-Schwarz or Ramond singular vectors, and which we call respectively {\em Ramond-NS} and {\em Ramond-R super-quantum curves}. Furthermore, we illustrate equivalence of conformal field theory calculations and eigenvalue models by showing that the same super-quantum curves arise in both approaches. We also find corresponding classical super-spectral curves that encode eigenvalue distribution in the super-eigenvalue model. Finally, we consider the special case of Penner-like potentials, and show that Ramond super-quantum curves in this case take form of supersymmetric versions of BPZ equations \cite{Belavin:1984vu}. The identification of the Ramond super-quantum and super-spectral curves generalizes the analysis in \cite{Ciosmak:2016wpx}, which was restricted to the Neveu-Schwarz sector.

We stress, that various observations and properties of quantum curves discussed in \cite{Manabe:2015kbj,Ciosmak:2016wpx,Ciosmak:2017omd} also hold (or are expected to hold) for Ramond super-quantum curves found in this paper. In particular, Ramond super-quantum curves are ``quantum'' in a double sense, and reduce to ``classical'' objects in two different limits: the {\em 't Hooft limit} corresponding to an infinite number of eigenvalues $N$, and the classical conformal field theory limit corresponding to infinite value of the parameter $\beta$. These two quantum structures have an analogous role, which is manifest after replacing parameters $1/N$ and $\beta$ by familiar parameters  $\epsilon_1$ and $\epsilon_2$, encoding the {\em Omega-background} in gauge theory interpretation. Analogous two quantum structures have been also discussed e.g. in \cite{Teschner:2010je} in the context of Langlands duality. Another important feature of quantum curves at higher levels is that, in the classical 't Hooft limit, they factorize into a product of several classical (spectral) curves.

\bigskip

Let us summarize the most important results and formulae of this work, by presenting conformal field theory formulation of various eigenvalue models that we derive and analyze, and corresponding wave-functions and quantum curve equations. We hope that this short summary could be helpful for a reader; details of the notation are explained in the main text of the manuscript.

First, in section \ref{sec-Virasoro} we recall that the usual $\beta$-deformed hermitian matrix model has the eigenvalue representation, which can be realized as the following integrated expectation value in the free boson theory, see (\ref{ZV-intro})-(\ref{ZV-CFT-intro})
\be
Z  = \int\!d^N\!z\ \bra{V_{N\sqrt{\beta},\vect{\scriptstyle t}}}\prod\limits_{a=1}^N\left.\left.\mathsf{E}^{-\sqrt{\beta}}(z_a)\right|0\right\rangle
= \int\!d^N\!z\ \Delta(\vect{z})^{2\beta}\,{\rm e}^{-\frac{\sqrt{\beta}}{\hbar}\sum\limits_{a=1}^NV(z_a)},
\ee
where vertex operators $\mathsf{E}^{\alpha}(x)$ are defined in (\ref{E-vertexop}), and $\big\langle V_{N\sqrt{\beta},\vect{\scriptstyle t}}\big|$ is the coherent state (\ref{bra-Virasoro}). The eigenvalue model in the right side of the equation involves the usual Vandermonde determinant $\Delta(\vect{z}) = \prod_{a<b} (z_a-z_b)$. The wave-function (\ref{Virasoro}) is then defined as a correlator which involves an additional insertion of $x$- and $\alpha$-dependent vertex operator $\mathsf{E}^{\frac{\alpha}{\hbar}}(x)$
\be
\begin{split}
\widehat{\chi}_\alpha(x) & =
\int\!d^N\!z\ \bra{V_{N\sqrt{\beta}-\alpha/\hbar,\vect{\scriptstyle t}}}\mathsf{E}^{\frac{\alpha}{\hbar}}(x)\prod\limits_{a=1}^N\left.\left.\mathsf{E}^{-\sqrt{\beta}}(z_a)\right|0\right\rangle = \\
& = {\rm e}^{\frac{\alpha}{\hbar^2}V(x)}\int\!d^N\!z\ \prod\limits_{a=1}^N(x-z_a)^{-\frac{2\alpha\sqrt\beta}{\hbar}}\,\Delta(\vect{z})^{2\beta}\,{\rm e}^{-\frac{\sqrt{\beta}}{\hbar}\sum\limits_{a=1}^NV(z_a)}.
\end{split}
\ee
This wave-function, for special discrete values of $\alpha$, satisfies quantum curve equations that take form of Virasoro singular vectors, as shown in section \ref{ssec-Virasoro-quantum-curves}.

In section \ref{sec-NS} we construct the eigenvalue model and quantum curves in the Neveu-Schwarz (NS) sector of the super-Virasoro algebra. The eigenvalue model has the following form and conformal field theory representation, see (\ref{Z-NS-1})-(\ref{Z-NS-2})
\begin{equation}
Z =  \left\langle\left.\left. V_{N\sqrt{\beta},\hbox{{\boldmath $\scriptstyle t$},{\boldmath $\scriptstyle \xi$}}}\right|\mathsf{Q}_{{\sr NS}}^N\right|0\right\rangle
= \int\! d^N\!z\, d^N\!\theta\ \Delta_{{\sr NS}}(\vect{z},\vect{\theta})^{\beta}\,
{\rm e}^{-\frac{\sqrt\beta}{\hbar}\sum_{a=1}^NV(z_a,\theta_a)},
\end{equation}
where the potential involves both commuting times $t_m$ and anti-commuting times $\xi_{m+1/2}$
\be
V(z,\theta) = V_{\sr B}(z) + V_{\sr F}(z)\theta, \qquad
V_{\sr B}(z) \; = \; \sum\limits_{m=0}^\infty t_m z^m, \qquad
V_{\sr F}(z) \; = \;  \sum\limits_{m=0}^\infty \xi_{m+1/2} z^m.
\ee
The NS screening charge $\mathsf{Q}_{{\sr NS}}$ and the coherent state $\big\langle V_{N\sqrt{\beta},\hbox{{\boldmath $\scriptstyle t$},{\boldmath $\scriptstyle \xi$}}} \big|$ are defined respectively in (\ref{screening:charge:NS}) and (\ref{coherent:bra:NS}), and the NS version of the Vandermonde determinant takes form (\ref{Vandermonde-NS})
\begin{align}
\Delta_{{\sr NS}}(\vect{z},\vect{\theta}) = \prod\limits_{1\le a < b \le N}\left(z_a-z_b-\theta_a\theta_b\right).
\end{align}
The NS wave-function in this model (\ref{wave:function:NS})-(\ref{wave:function:NS-2}) is defined as the expectation value with an additional insertion of the vertex operator superfield $\Phi^{\frac{\alpha}{\hbar}}(x,\theta)$ defined in (\ref{NS:superfield})
\begin{align}
\begin{split}
\widehat\chi_\alpha(x,\theta) &=
\left\langle\left.\left. V_{N\sqrt{\beta}-\alpha/\hbar,\hbox{{\boldmath $\scriptstyle t$},{\boldmath $\scriptstyle\xi$}}}\right|\Phi^{\frac{\alpha}{\hbar}}(x,\theta)\mathsf{Q}_{{\sr NS}}^N\right|0\right\rangle = \\
& = {\rm e}^{\frac{\alpha}{\hbar^2}V(x,\theta)}
\int\! d^N\!z\, d^N\!\theta\
\prod\limits_{a=1}^N
(x-z_a-\theta\theta_a)^{-\frac{\alpha\sqrt{\beta}}{\hbar}}
\Delta_{{\sr NS}}(\vect{z},\vect{\theta})^{\beta}\,
{\rm e}^{-\frac{\sqrt\beta}{\hbar}\sum_{b=1}^NV(z_b,\theta_b)},
\end{split}
\end{align}
and it satisfies equations which take form of super-Virasoro Neveu-Schwarz singular vectors, derived in section \ref{ssec-NS-quantum-curves}.

In section \ref{sec-Ramond} we start considering the Ramond sector of the super-Virasoro algebra. We explain that there are two natural eigenvalue models and wave-functions that can be considered, having the schematic structure given in (\ref{chi-R-NS-R}) and (\ref{chi-NS-R-R}), which satisfy quantum curve equations that take form of either Neveu-Schwarz or Ramond singular vectors. We call these models Ramond-NS and Ramond-R respectively. The Ramond-NS eigenvalue model is introduced in (\ref{Z-Ramond})
\be
Z  = \left.\left.\left\langle V^+_{N\sqrt{\beta},\hbox{{\boldmath $t$},{\boldmath $\xi$}}}\right|\mathsf{Q}_{{\sr R}}^N\right|0,+\right\rangle
= \int\! d^N\!z\, d^N\!\theta\
\Delta_{\sr R}(\vect{z},\vect{\theta})^\beta\,
{\rm e}^{-\frac{\sqrt{\beta}}{\hbar}\sum_{a=1}^N\! V_{\sr R}(z_a,\theta_a)},
\ee
with the coherent state $\big\langle V^+_{N\sqrt{\beta},\hbox{{\boldmath $t$},{\boldmath $\xi$}}}\big|$ and the screening charge $ \mathsf{Q}_{{\sr R}}$ defined respectively in (\ref{coherent:bra:R}) and (\ref{screening:charge:R}). The Ramond-NS version of the Vandermonde determinant takes form (\ref{Vandermonde-Ramond})
\begin{equation}
\Delta_{\sr R}(\vect{z},\vect{\theta}) = \prod_{1\le a < b \le N}  \left(z_a-z_b-\frac{z_a+z_b}{2\sqrt{z_az_b}}\,\theta_a\theta_b\right).
\end{equation}
The Ramond-NS wave-function, of the schematic form (\ref{chi-R-NS-R}), is introduced in (\ref{chi-alpha-Ramond}) in the conformal field theory language as the expectation value involving the vertex operator superfield $\Phi^{\frac{\alpha}{\hbar}}(x,\theta)$ defined in (\ref{Phi-Ramond})
\be
\widehat\chi_\alpha(x,\theta) = \left.\left.\left\langle V^+_{N\sqrt{\beta}-\frac{\alpha}{\hbar},\hbox{{\boldmath $t$},{\boldmath $\xi$}}}\right|\Phi^{\frac{\alpha}{\hbar}}(x,\theta)
\mathsf{Q}_{{\sr R}}^N
\right|0,+\right\rangle
\ee
and it is written down more explicitly, in the eigenvalue representation, in (\ref{chi_hat_w}). Quantum curves that annihilate such a wave-function are derived in section \ref{ssec-wavefunction-Ramond} and they take form of Neveu-Schwarz singular vectors (however with a specific representation of NS algebra, relevant for the Ramond sector that we are considering). In section \ref{sec-matrix} we rederive these quantum curves using techniques of matrix and eigenvalue models; in addition, in section \ref{ssec-spectralcurve} we derive super-spectral curves (i.e. spectral curves of the Ramond-NS eigenvalue model), and in section \ref{ssec-Ramond-NS-Penner} we analyze the Ramond-NS eigenvalue model with the specific multi-Penner potential.

Furthermore, in section \ref{sec-Ramond-R} we analyze the Ramond-R model, with the wave-function of the schematic form (\ref{chi-NS-R-R}). In this case we consider directly the wave-function for the model with the Penner-like potential, which we find to take form (\ref{psi-Ramond-R-1})-(\ref{psi-Ramond-R-2})
\be
\chi^{\sr R}_{\alpha}(x,\xi) = x^{1/8} (x-w)^{-\frac{\alpha\gamma}{\hbar^2}}
{\rm e}^{-\frac{\gamma\xi\eta}{\hbar^2}\frac{\sqrt{x}}{\sqrt{w(x-w)}}}\,\int\! \Big( {\Psi}_{+}(x,\vect{z},\vect{\theta}) + \frac{\sqrt{2}}{\hbar}{\rm e}^{\frac{i\pi}{4}}\,\xi\,{\Psi}_{-}(x,\vect{z},\vect{\theta}) \Big)\,d^N\!z\, d^N\!\theta,
\ee
where the $x$-dependence arises from the insertion of the Ramond chiral primary fields ${\sf R}_{\pm}^{\frac{\alpha}{\hbar}}(x)$ defined in (\ref{Ramond:chiral:fields})
\begin{align}
\begin{split}
{\Psi}_{\pm}(x,\vect{z},\vect{\theta}) &= \bra{\alpha_0}{\sf R}_{\pm}^{\frac{\alpha}{\hbar}}(x)\Phi^{\frac{\gamma}{\hbar}}(w,\eta)\prod\limits_{a=1}^N\Phi^{-\sqrt{\beta}}(z_a,\theta_a)\ket{\sigma_+} = \\
&= \Theta_{\pm}(x)\, (x-w)^{\frac{\alpha\gamma}{\hbar^2}}\,
%\prod\limits_{a=1}^N(x-z_a)^{-\frac{\alpha\sqrt{\beta}}{\hbar}}\,
\Delta_{{\sr R},x}(\vect{z},\vect{\theta})^\beta\,
{\rm e}^{-\frac{\sqrt{\beta}}{\hbar}\sum_{a=1}^N
\left(V_{{\sr B},x}(z_a) +V_{{\sr F},x}(z_a)\theta_a\right)},
\end{split}
\end{align}
see (\ref{PsiR:definition}) and (\ref{Ramond-R-eigenvalue}). The functions $\Theta_{\pm}(x)$ are given in (\ref{ramond_theta_def}), and the Ramond-R version of the Vandermonde determinant takes form (\ref{Vandermonde-Ramond-R})
\begin{align}
\Delta_{{\sr R},x}(\vect{z},\vect{\theta}) =  \prod\limits_{a<b}
\left(z_a-z_b -\left(\sqrt{\frac{z_a(x-z_b)}{z_b(x-z_a)}}+\sqrt{\frac{z_b(x-z_a)}{z_a(x-z_b)}}\right)\frac{\theta_a\theta_b}{2}\right).
\end{align}
Quantum curves for the Ramond-R model take form of Ramond singular vectors and we derive them in section \ref{subsec_rr_sq}. Finally, in section \ref{sec-matrix-Ramond-R} we reconsider the Ramond-R eigenvalue model from the matrix model perspective, and using matrix (or eigenvalue) model techniques we rederive Ramond-R super-quantum curves.

\bigskip

Once more, and more succinctly, the plan of this paper is as follows. In section \ref{sec-Virasoro} we derive, from the viewpoint of conformal field theory, quantum curves corresponding to the underlying Virasoro algebra. In section \ref{sec-NS} we similarly derive quantum curves corresponding to the Neveu-Schwarz sector of the super-Virasoro algebra. In section \ref{sec-Ramond}, starting from the conformal field theory formalism we analyze the Ramond sector of super-Virasoro algebra, and derive an eigenvalue model for the Ramond-NS sector, as well as Ramond-NS super-quantum curves. In section \ref{sec-Ramond-R} we derive eigenvalue model and super-quantum curves in the Ramond-R sector using conformal field theory approach. In section \ref{sec-matrix} we rederive super-quantum curves in the Ramond-NS sector using techniques of matrix or eigenvalue models; among others, in this section we also find the Ramond super-spectral curve, and confirm that it agrees with the classical limit of the Ramond super-quantum curve. Similarly, in section \ref{sec-matrix-Ramond-R} we derive super-quantum curves of Ramond-R type using techniques of eigenvalue models. In the appendix we collect various proofs and computations.

%******************************************************************************************
%******************************************************************************************
%******************************************************************************************
%******************************************************************************************

\section{From conformal field theory to Virasoro quantum curves}   \label{sec-Virasoro}

We start our analysis from a discussion of the Virasoro algebra, which is the underlying algebra of a ($\beta$-deformed) hermitian matrix (or eigenvalue) model
\be
Z  = \int\!d^N\!z\ \Delta(\vect{z})^{2\beta}\,{\rm e}^{-\frac{\sqrt{\beta}}{\hbar}\sum\limits_{a=1}^NV(z_a)}, \label{ZV-intro}
\ee
where $d^N\!z=\prod_{a=1}^N dz_a$, $\Delta(z)=\prod_{a<b}(z_a-z_b)$ is  the Vandermonde determinant, and we consider a generic potential $V(z) = \sum_{m=0}^\infty t_m z^m$. We recall first that such a model can be defined by the following expectation value, which is written completely in terms of conformal field theory quantities
\be
Z  = \int\!d^N\!z\ \bra{V_{N\sqrt{\beta},\vect{\scriptstyle t}}}\prod\limits_{a=1}^N\left.\left.\mathsf{E}^{-\sqrt{\beta}}(z_a)\right|0\right\rangle, \label{ZV-CFT-intro}
\ee
for an appropriately defined state $\langle V_{N\sqrt{\beta},\vect{\scriptstyle t}}|$ and vertex operators $\mathsf{E}^{-\sqrt{\beta}}(z_a)$. Similarly wave-functions, defined as determinant-like expectation values from matrix model viewpoint, can be expressed in terms of conformal field theory quantities as follows\footnote{The parameter $\alpha=\alpha_{\textrm{here}}$ in this section is related with the parameter $\alpha=\alpha_{\textrm{ms}}$ in \cite{Manabe:2015kbj} by $\alpha_{\textrm{ms}}=2\alpha_{\textrm{here}}$ (in $\hbar=1$ unit).}
\begin{equation}
\begin{split}
\widehat{\chi}_\alpha(x)
& =  \int\!d^N\!z\ \bra{V_{N\sqrt{\beta}-\alpha/\hbar,\vect{\scriptstyle t}}}\mathsf{E}^{\frac{\alpha}{\hbar}}(x)\prod\limits_{a=1}^N\left.\left.\mathsf{E}^{-\sqrt{\beta}}(z_a)\right|0\right\rangle = \\
& =  {\rm e}^{\frac{\alpha}{\hbar^2}V(x)}\int\!d^N\!z\ \prod\limits_{a=1}^N(x-z_a)^{-\frac{2\alpha\sqrt\beta}{\hbar}}\,\Delta(\vect{z})^{2\beta}\,{\rm e}^{-\frac{\sqrt{\beta}}{\hbar}\sum\limits_{a=1}^NV(z_a)}.  \label{psi-intro}
\end{split}
\end{equation}
These wave-functions are annihilated by quantum curve operators that we are after only for special values of the parameter $\alpha$, which correspond to the degenerate momenta.

We recall now a general form of quantum curves associated to the underlying Virasoro algebra. As found in \cite{Manabe:2015kbj} by generalizing a discussion in \cite{ACDKV}, such curves can be determined simply by writing expressions for Virasoro singular vectors (at arbitrary level) in terms of the following representation of Virasoro generators
\begin{equation}
{\widehat L}_{-1} = \partial_x, \qquad {\widehat L}_{-n} = \frac{1}{\hbar^2(n-2)!} \Big(\frac14\partial^{n-2}_x\left(V'(x)\right)^2 + \frac{Q\hbar}{2}\partial_x^nV(x) + \partial_x^{n-2}\widehat{f}_t(x) \Big),   \label{hat:L:minus:n-intro}
\end{equation}
for $n \ge 2$, where
\begin{equation}
Q\equiv \beta^{-\frac12}-\beta^{\frac12},
\end{equation}
and $\widehat{f}_t(x)$ is a partial differential operator defined by (\ref{operator:f:definition}).
Moreover, it is useful to take advantage of universal expressions for singular vectors (up to a given level) that have been found in \cite{Manabe:2015kbj}. These expressions depend on the parameter $\alpha$ and reduce to  the expression for a singular vector labeled by integers $r,s$ upon specialization of $\alpha$ to
\be
\alpha_{r,s} = \frac{r-1}{2}\beta^{-\frac12} - \frac{s-1}{2}\beta^{\frac12},\qquad r,s\in\mathbb{Z},\quad r,s\geq1. \label{alpha-rs-intro}
\ee
For example, singular vectors at level 2 can be obtained from the expression
\be
%\widehat{A}^{2}(\alpha) =
L_{-1}^2 - 4\alpha^2L_{-2}   \label{A2-vector-intro}
\ee
upon specialization $\alpha=\alpha_{2,1}$ or $\alpha_{1,2}$. Moreover, for $\alpha=\alpha_{1,1}=0$ this expression reduces to the singular vector $L_{-1}$ at level 1 (up to an additional $L_{-1}$). Substituting the representation (\ref{hat:L:minus:n-intro}) in the formula (\ref{A2-vector-intro}),
%$\widehat{A}^{2}(\alpha)$,
it follows that quantum curves at level 2 (and 1) arise from the expression
\be
\widehat{A}^{2} = \partial^2_x     -\frac{\alpha^2}{\hbar^4}
\left(\left(V'(x)\right)^2 + 2Q\hbar V''(x) + 4\widehat{f}_t(x) \right)    \label{A2-intro}
\ee
(where the superscript of $\widehat{A}^{2}$ denotes level 2) upon the specialization $\alpha = 0$, $\frac{\hbar\sqrt{\beta}}{2}$, or $-\frac{\hbar}{2\sqrt{\beta}}$; note that in expressions for quantum curves we combine values of $\alpha_{r,s}$ with an additional factor of $\hbar$ when compared to (\ref{alpha-rs-intro}) -- this follows from a factor $\hbar$ in the vertex operator $\mathsf{E}^{\frac{\alpha}{\hbar}}$ in (\ref{psi-intro}), which is natural to include from the matrix model perspective. Similarly, quantum curves at level 3 (or lower levels) arise from the expression
\be
\widehat{A}^{3} = \partial^3_x  -4\frac{\alpha^2}{\hbar^2}\partial_x{\widehat L}_{-2} + \frac{2\alpha^2(2\alpha(2\alpha+Q\hbar)-\hbar^2)}{\hbar^4} {\widehat L}_{-3}    \label{A3-intro}
\ee
upon the specialization
$\alpha = 0$, $\frac{\hbar\sqrt{\beta}}{2}$, $-\frac{\hbar}{2\sqrt{\beta}}$, $\hbar\sqrt{\beta}$, or $-\frac{\hbar}{\sqrt{\beta}}$, and with $L_{-2}$ and $L_{-3}$ given in (\ref{hat:L:minus:n-intro}). Such quantum curves, at level $n=rs$, annihilate the wave-function (\ref{psi-intro}), $\widehat{A}^n\psi_{\alpha}(x) = 0$, for relevant specialization $\alpha=\alpha_{r,s}$.

While in \cite{Manabe:2015kbj} the representation of Virasoro generators (\ref{hat:L:minus:n-intro}) and the form of quantum curves, such as (\ref{A2-intro}) and (\ref{A3-intro}), was derived using matrix model formalism, in this section we rederive these results from purely conformal field theory viewpoint. We also explain, from purely conformal field theory perspective, where expressions for singular vectors, such as (\ref{A2-vector-intro}) and its higher level generalizations, come from. To this aim it is of advantage to consider the background charge representation of the Virasoro algebra, which is therefore the starting point of our analysis. The approach presented in this section will be generalized to the supersymmetric case in the following sections, which ultimately will enable us to derive super-eigenvalue models for the Ramond sector and corresponding Ramond super-quantum curves.

%******************************************************************************************
%******************************************************************************************

\subsection{Background charge representation of the Virasoro algebra}

Consider the Heisenberg algebra
\begin{equation}
\label{Heisenberg}
[{\sf a}_m,{\sf a}_n] = \frac12 m \delta_{m+n,0}, \qquad m,n \in {\mathbb Z},
\end{equation}
and its highest weight states $\ket{\alpha}$ with $\alpha \in {\mathbb R}$ defined by
\begin{align}
\label{charged:vacuumI}
{\sf a}_m\ket{\alpha}  =  0, \hskip 10pt m > 0, \hskip 30pt {\sf a}_0\ket{\alpha} \; = \; \alpha\ket{\alpha}.
\end{align}
In what follows, to avoid problems with the Dirac notation, we also use the notation $\mu_\alpha \equiv \ket{\alpha}$. Denote by ${\cal H}_\alpha$ the free vector space spanned by vectors of the form
\begin{align}
\label{Fock:space:basis}
{\sf a}_{-J}\ket{\alpha} \equiv {\sf a}_{-j_1}{\sf a}_{-j_2}\cdots {\sf a}_{-j_l}\ket{\alpha},
\qquad 0 < j_1 \le j_2 \le \ldots \le j_l.
\end{align}
These vectors form the {canonical basis} in ${\cal H}_\alpha$. The space ${\cal H}_\alpha$ has a natural $\mathbb{Z}$-grading
\begin{equation}
\label{Zgrading}
{\cal H}_\alpha = \bigoplus\limits_{n\ge 0} {\cal H}_\alpha^n,
\qquad {\cal H}_\alpha^n\equiv {\rm span} \Big\{ {\sf a}_{-J}\ket{\alpha}: |J|\equiv \sum\limits_{k=1}^l j_k =n
\Big\}.
\end{equation}
Fixing a real parameter $Q$, one can define a hermitian pairing
\[
( \;.\;,\;.\;)_{\alpha,Q} :\ {\cal H}_{Q-\alpha} \times {\cal H}_\alpha \to \mathbb{C},
\]
by requiring
\begin{align}
\label{bilinear:form:Virasoro}
{\sf a}_n^\dag = -{\sf a}_n, \;\; n \neq 0,
\;\;\;
{\sf a}_0^\dag = Q-{\sf a}_0,
\;\;\;
(\,\mu_{Q-\alpha}\,,\,\mu_{\alpha})_{\alpha,Q} =1.
\end{align}
The {Heisenberg module} ${\cal H}_{\alpha,Q}$ is the pair of representations $\langle {\cal H}_{Q-\alpha},{\cal H}_\alpha\rangle $ endowed with the pairing $( \;.\;,\;.\;)_{\alpha,Q}$.

Furthermore, consider the Virasoro algebra with the central charge $c$
\begin{equation}
\label{virasoro}
[{\sf L}_m, {\sf L}_n ] = (m-n) {\sf L}_{m+n} +\frac{c}{12}(m^3-m)\delta_{m+n,0},\qquad m,n\in \mathbb{Z}.
\end{equation}
The Verma module ${\cal V}_{\Delta,c}$ with the highest weight $\Delta$ is defined as the representation of the Virasoro algebra generated by the vectors of the form
\begin{align}
\label{Verma:space:basis}
{\sf L}_{-I}\ket{\Delta} \equiv {\sf L}_{-i_1}{\sf L}_{-i_2}\cdots {\sf L}_{-i_l}\ket{\Delta},
\qquad   0 < i_1 \le i_2 \le \ldots \ i_l,
\end{align}
where the highest weight vector $\ket{\Delta}$ satisfies
\begin{align}
\label{highestweight}
{\sf L}_m\ket{\Delta} = 0, \qquad m > 0,\qquad {\sf L}_0 \ket{\Delta}= \Delta\ket{\Delta}.
\end{align}
The vectors (\ref{Verma:space:basis}) form the {canonical basis} in ${\cal V}_{\Delta,c}$. The Verma module ${\cal V}_{\Delta,c}$ has a natural $\mathbb{Z}$-grading
\begin{equation}
\label{ZgradingV}
{\cal V}_{\Delta,c}= \bigoplus\limits_{n\geq 0} {\cal V}_{\Delta,c}^n,
\qquad   {\cal V}_{\Delta,c}^n\equiv {\rm span} \left\{ L_{-I}\ket{\alpha}: |I| =n
\right\}.
\end{equation}
The Virasoro Verma module ${\cal V}_{\Delta,c}$ is endowed with the hermitian { Schapovalov form} $( \;.\;,\;.\;)_{\Delta,c}$, defined by the conditions
\begin{align}
 {\sf L}_m^\dag = {\sf L}_{-m}, \;\; m \in \mathbb{Z},
\qquad (\,\nu_\Delta\,,\nu_{\Delta})_{\Delta,c} =1,\qquad \nu_\Delta \equiv \ket{\Delta}.
\nonumber
\end{align}
$\mathbb{Z}$-grading (\ref{ZgradingV}) is orthogonal with respect to this form. We say that $\xi \in {\cal V}_{\Delta, c}$ is a {null vector} if it is orthogonal with respect to the Schapovalov form to all vectors in ${\cal V}_{\Delta, c}$. A null vector $\xi \in {\cal V}_{\Delta, c}$ is called a {singular vector} if it satisfies the highest weight state condition
\[
{\sf L}_n \xi=0,\qquad  n>0.
\]

Let
\begin{equation}
\label{cQrel}
c= 1- 6Q^2,\qquad  \Delta=\alpha(\alpha-Q)\equiv \Delta_\alpha.
\end{equation}
In the case under consideration $Q$ is real, so $c\leq 1$. The {background charge representation} on ${\cal H}_{\alpha}$ of the Virasoro algebra with the central charge $c$ and the highest weight $\Delta_\alpha$ is defined by the map $\sigma_{\alpha, Q}$ between the universal enveloping algebras of the Virasoro algebra and the Heisenberg algebra as
\begin{align}
\nonumber
\sigma_{\alpha, Q} &\ :\ {\rm End}\left({\cal V}_{\Delta, c}\right) \ni {\sf L}_m \ \to \ L(\alpha)_m\in {\rm End}({\cal H}_{\alpha}),
\\
\label{Virasoro:alebra:rep}
L(\alpha)_0 & =
2\sum\limits_{n=1}^\infty{\sf a}_{-n}{\sf a}_n +
\alpha\left(\alpha- Q\right),
\\
\nonumber
L(\alpha)_m& =
\sum\limits_{n\neq 0,m}\!{\sf a}_{m-n}{\sf a}_n +\left(2\alpha - (m+1)Q\right){\sf a}_m, \hskip 1cm m \neq 0.
\end{align}
This map satisfies
\begin{equation}
\label{homprop}
\sigma_{\alpha, Q}({\sf L}_m {\sf L}_n) = \sigma_{\alpha, Q}({\sf L}_m)\sigma_{\alpha, Q}({\sf L}_n).
\end{equation}
Using $\sigma_{\alpha,Q}$ we define the transition map
\[
S_{\alpha,Q}: {\cal V}_{\Delta_\alpha, c} \to {\cal H}_{\alpha}
\]
by its action on the canonical basis in ${\cal V}_{\Delta_\alpha,c}$
\begin{equation}
\label{S:map:definition}
S_{\alpha,Q} {\sf L}_{-I}\ket{\Delta} = L(\alpha)_{-i_1}L(\alpha)_{-i_2}\cdots L(\alpha)_{-i_l}\ket{\alpha}.
\end{equation}
The transition map  $S_{\alpha,Q}$ has the following properties, easily inferred from its definition:
\begin{enumerate}
\item
$S_{\alpha,Q}$ is a homomorphism of the Virasoro algebra representations.
\item
$S_{\alpha,Q}$ preserves the $\mathbb{Z}$ grading
$
S_{\alpha,Q}({\cal V}_{\Delta_\alpha, c}^n)\subset {\cal H}^n_{\alpha}
$
hence
\[
S_{\alpha,Q} = \bigoplus\limits_{n\ge 0} S_{\alpha,Q}^n,\;\;\;\;
S^n_{\alpha,Q}:{\cal V}_{\Delta_\alpha, c}^n\to  {\cal H}^n_{\alpha}.
\]
\item
$S_{\alpha,Q}$ is compatible with the hermitian form in ${\cal V}_{\Delta_\alpha, c}$ and the hermitian pairing in ${\cal H}_{\alpha,Q},$ i.e. for any  $ \xi,\xi'\in {\cal V}_{\Delta_\alpha, c}$
\[
(\xi,\xi')_{\Delta_\alpha,c} = (S_{Q-\alpha,Q}\xi,S_{\alpha,Q}\xi')_{\alpha,Q}.
\]
\item
The kernel of $S_{\alpha,Q}$ is the subspace of all null vectors in ${\cal V}_{\Delta_\alpha,c}$.
\end{enumerate}

Consider the matrix of $S_{\alpha,Q}$ with respect to the canonical bases in ${\cal V}_{\Delta_\alpha,c}$
and ${\cal H}_{\alpha}$
\begin{align}
\label{transition_matrix}
 L(\alpha)_{-I} \mu_{\alpha}=
\sum\limits_{|J|=n}  [S^{\,n}_{\alpha,Q}]_{IJ}\, {\sf a}_{-J}  \mu_{\alpha}.
\end{align}
Matrices $S^{\,n}_{\alpha,Q}$ were studied (in different parameterization) in \cite{Kato:1987qda}, where the formula for their determinant was found
\begin{equation}
\label{kac_for_s}
\det S^{\,n}_{\alpha,Q} = {\rm const} \begin{array}[t]{c}
{\displaystyle\prod} \\[-6pt]
{\scriptscriptstyle
1 \le\, rs \,\le\, n}
\\[-8pt]
\scriptscriptstyle 0< r,s
\end{array}\!\!\!\!
\left( \alpha-\alpha_{r,s}\right)^{p(n-{rs})}.
\end{equation}
Here $\beta$ is related to $Q$ by $Q = \beta^{-\frac12} - \beta^{\frac12}$, values of $\alpha_{r,s}$ agree with those in (\ref{alpha-rs-intro})
\begin{align}
\label{alpha:rs}
\alpha_{r,s} = \frac{r-1}{2}\beta^{-\frac12} - \frac{s-1}{2}\beta^{\frac12},\qquad r,s\in\mathbb{Z},\quad r,s\geq1,
\end{align}
and $p(n)$ is the number of partitions of $n$, which can be read off from the generating function
\[
\sum\limits_{n=0}^{\infty} p(n)x^n= \prod\limits_{m=1}^{\infty} \frac{1}{1-x^m}.
\]
As proven in \cite{Kato:1987qda}, for generic values of $Q$ matrix elements of the matrix inverse to $S^{\,n}_{\alpha,Q}$ have at most simple poles at $\alpha=\alpha_{r,s}.$ Arguments from linear algebra then show that null vectors $\xi_{\textrm{null}}^n \in {\cal V}^n_{\Delta_\alpha,c}$ can be constructed as residues of $(S^{\,n}_{\alpha,Q}{})^{-1}$
\begin{align}
c_{J}\,\xi_{\textrm{null}}^n = \lim\limits_{\alpha \to\alpha_{r,s}}(\alpha - \alpha_{r,s})\sum\limits_{|I|=n}  \left[(S^{\,n}_{\alpha,Q}{})^{-1}\right]_{JI}\, {\sf L}_{-I}\,\nu_{\Delta_\alpha},
\end{align}
where $rs \le n$ and $c_J$ are (in the generic case non-zero) numbers. Consequently, if we define
\begin{align}
\label{universal:singular:operator:Virasoro}
\widehat{A}_J^{\,n}(\alpha) = \omega_n(\alpha,Q)\sum\limits_{|I|=n}  \left[(S^{\,n}_{\alpha,Q}{})^{-1}\right]_{JI}\, L(\alpha)_{-I},
\end{align}
where
\begin{equation}
\omega_n(\alpha,Q) = -(-2)^n\prod\limits_{rs \le n} (\alpha - \alpha_{r,s}),
\end{equation}
then  for $r,s>0$ and $rs \le n$  the operators (\ref{universal:singular:operator:Virasoro}) are (non-trivial) endomorphisms of ${\cal H}_{\alpha}$ satisfying
\[
\lim\limits_{\alpha \to\alpha_{r,s}}\widehat{A}_J^{\,n}(\alpha)\mu_\alpha = 0.
\]
The simplest examples are
\begin{align}
\label{Virasoro:null:vectors:examples}
\nonumber
\widehat{A}_1^{\,1}(\alpha) & = L(\alpha)_{-1},
\\[4pt]
\nonumber
\widehat{A}_{2}^{\,2}(\alpha) & = L(\alpha)_{-1}^2 - 4\alpha^2L(\alpha)_{-2},
\\[4pt]
\nonumber
\widehat{A}_{1,1}^{\,2}(\alpha) & = (Q-2\alpha)L(\alpha)_{-1}^2 + 2\alpha L(\alpha)_{-2},
\\[-7pt]
\\[-7pt]
\nonumber
\widehat{A}_{3}^{\,3}(\alpha) & = L(\alpha)_{-1}\widehat{A}^{\,2}_{2}(\alpha) - \omega_2(\alpha,Q)L(\alpha)_{-3},
\\[4pt]
\nonumber
\widehat{A}^{\,3}_{1,2}(\alpha) & = (\alpha-Q)L(\alpha)_{-1}\widehat{A}_{2}^{\,2}(\alpha) - \omega_2(\alpha,Q)L(\alpha)_{-3},
\\[4pt]
\nonumber
\widehat{A}^{\,3}_{1,1,1}(\alpha)
& =
L(\alpha)_{-1}\left(\big(1-\alpha(\alpha-Q)\big)\widehat{A}_{1,1}^{\,2}(\alpha)- (\alpha -Q)\widehat{A}_{2}^{\,2}(\alpha)\right)
+
\omega_2(\alpha,Q)L(\alpha)_{-3}.
\end{align}
These are expressions of the form mentioned in (\ref{A2-vector-intro}). They were independently identified in \cite{Manabe:2015kbj} by matrix model techniques.

We now extend the Heisenberg algebra (\ref{Heisenberg}) by the operator ${\sf q}$ satisfying
\begin{equation}
\label{Heisenberg:2}
\left[{\sf a}_m,{\sf q}\right] = \frac12\delta_{m,0},\qquad m \in {\mathbb Z}.
\end{equation}
Then by
\[
\left[{\sf a}_0,{\rm e}^{2\alpha{\sf q}}\right] = \alpha\,{\rm e}^{2\alpha{\sf q}},
\hskip 1cm
\left[{\sf a}_m,{\rm e}^{2\alpha{\sf q}}\right] = 0,\hskip 5mm m \neq 0,
\]
and (\ref{charged:vacuumI}) we see that ${\rm e}^{2\alpha{\sf q}}$ can be regarded as a linear map
\be
\label{translation:in:alpha}
{\rm e}^{2\alpha{\sf q}}:\; {\cal H}_{\alpha'}  \to  {\cal H}_{\alpha +\alpha'},
\qquad
{\rm e}^{2\alpha{\sf q}}\,{\sf a}_{-I}\ket{\alpha'}  =  {\sf a}_{-I}\ket{\alpha + \alpha'}.
\ee
It is useful to assemble  operators ${\sf a}_m$ and ${\sf q}$ into a local bosonic field, defined by
\begin{equation}
\label{scalar:field}
\phi(x) = \phi_>(x) + \phi_<(x),\qquad
\phi_>(x) = {\sf a}_0\log x - \sum\limits_{m=1}^\infty\frac{{\sf a}_m}{m}\, x^{-m},
\qquad \phi_<(x) = {\sf q} + \sum\limits_{m=1}^\infty\frac{{\sf a}_{-m}}{m}\, x^{m}.
\end{equation}
We then introduce the energy-momentum tensor
\begin{equation}
\label{T:Virasoro}
T(x)\  = \ :\!\partial\phi(x)\partial\phi(x)\!: + Q\partial^2\phi(x) \ = \ \sum\limits_{m\in {\mathbb Z}} \frac{L_m}{x^{m+2}},
\end{equation}
which we also write as $T(x)=T_-(x)+T_+(x)$ so that $T_+(x)\ket{0}=0$, where
\begin{equation}
\label{Tplus:definition}
T_-(x) = \sum_{m=-\infty}^{-2}\frac{L_m}{x^{m+2}}, \qquad
T_+(x) = \sum_{m=-1}^{\infty}\frac{L_m}{x^{m+2}}.
\end{equation}
The modes of the energy-momentum tensor $L_m = \oint_0 \frac{dx}{2\pi i}\ x^{m + 1}\, T(x)$ are given explicitly by
\begin{equation}
\begin{split}
L_0 & =
2\sum\limits_{n=1}^\infty{\sf a}_{-n}{\sf a}_n +
{\sf a}_0\left({\sf a}_0 - Q\right) \\
L_m & =
\sum\limits_{n\neq 0,m}\!{\sf a}_{m-n}{\sf a}_n +\left(2{\sf a}_0 - (m+1)Q\right){\sf a}_m, \hskip 1cm m \neq 0,
\end{split}
\end{equation}
and they define a natural extension of the background charge representation of the Virasoro algebra (\ref{Virasoro:alebra:rep}) to the space
\[
{\cal H} = \int\limits_{\oplus} {\cal H}_{\alpha}\ d\alpha,
\]
since
\[
\forall\, \xi \in {\cal H}_{\alpha}\;\ \forall\, m\in {\mathbb Z}:\qquad L_m\xi = L(\alpha)_m\xi.
\]
The pairing $(\;\cdot\;,\;\cdot\;)_{\alpha,Q}$ can be naturally extended to the hermitian form  $(\; \cdot \; ,\; \cdot \; )_Q$ on ${\cal H}$ by requiring
that for any $\xi \in {\cal H}_{\alpha'}, \zeta \in {\cal H}_{\alpha}:$
\[
(\xi,\zeta)_Q
\; = \;
\left\{
\begin{array}{ccl}
0 &\;\; \textrm{for}\;& \alpha+\alpha'\neq Q,   \\[4pt]
(\xi,\zeta)_{\alpha,Q} &\;\;\textrm{for}\;& \alpha+\alpha' = Q.
\end{array}
\right.
\]
In what follows we are interested in calculating products $(\xi,\zeta)_Q$ for $\xi\in {\cal H}_{Q-\alpha}$ and $\zeta\in {\cal H}_{\alpha}$, which can be written as $\xi = {\cal O}_\xi^\dag\,\mu_Q, \zeta = {\cal O}_\zeta\,\mu_0$ for some operators
\[
{\cal O}_\xi: \; {\cal H}_{\alpha} \to {\cal H}_{0}, \qquad
{\cal O}_\zeta: \; {\cal H}_{0} \to {\cal H}_{\alpha}.
\]
It is then convenient to use the standard %in physical literature,
bra-ket notation and write $\bra{0}{\cal O}_\xi{\cal O}_\zeta\ket{0}$ to denote
\[
\big({\cal O}_\xi^\dag\,\mu_Q, {\cal O}_\zeta\,\mu_0\big)_Q = \big(\mu_Q, {\cal O}_\xi{\cal O}_\zeta\,\mu_0\big)_Q.
\]

%******************************************************************************************
%******************************************************************************************

\subsection{From CFT to $\alpha/\beta$ deformed eigenvalue integrals...}

In the previous subsection we introduced ingredients necessary to construct eigenvalue integrals and quantum curves. We conduct this construction in the rest of this section. First, using $\phi(x)$ we introduce the normal ordered exponential fields
\be
\mathsf{E}^{\alpha}(x) = {\rm e}^{2\alpha\phi_<(x)}\,{\rm e}^{2\alpha\phi_>(x)},   \label{E-vertexop}
\ee
which can be viewed as linear maps from ${\cal H}_{\alpha'}$ to ${\cal H}_{\alpha+\alpha'}$. They are primary fields with respect to the Virasoro algebra, i.e. they satisfy commutation relations of the form
\begin{equation}
\label{commutators:LE}
\left[L_m,\mathsf{E}^{\alpha}(x)\right] = x^m\left((m+1)\Delta_\alpha +x\partial_x\right)\mathsf{E}^{\alpha}(x),
\end{equation}
where $\Delta_\alpha = \alpha(\alpha - Q)$, see (\ref{cQrel}). It follows that
\begin{equation}
\label{TplusE:commutator}
\left[T_+(y),\mathsf{E}^{\alpha}(x)\right]
=
\left(\frac{\Delta_\alpha}{(y-x)^2} + \frac{1}{y-x}\frac{\partial}{\partial x}\right)\mathsf{E}^{\alpha}(x).
\end{equation}
In the parametrization $Q = \beta^{-1/2} - \beta^{1/2}$ %for $\beta \in {\mathbb R}_+$
we have $\Delta_{-\sqrt{\beta}} = \Delta_{1/\sqrt{\beta}} = 1$, and consequently
\begin{equation}
\label{commutators:TP:with:E}
\left[T_+(y),\mathsf{E}^{-\sqrt{\beta}}(x)\right] = \frac{\partial}{\partial x}\left(\frac{\mathsf{E}^{-\sqrt{\beta}}(x)}{y-x}\right),
\qquad  \left[T_+(y),\mathsf{E}^{1/\sqrt{\beta}}(x)\right] = \frac{\partial}{\partial x}\left(\frac{\mathsf{E}^{1/\sqrt{\beta}}(x)}{y-x}\right).
\end{equation}
Notice that for $|z_1| < |z_2|$ we have
\be
\left[\phi_>(z_1),\phi_<(z_2)\right] = \frac12\log z_1 - \frac12\sum\limits_{m =1}^{\infty}\frac{1}{m}\left(\frac{z_2}{z_1}\right)^m = \frac12\log(z_1-z_2).
\nonumber
\ee
Therefore, in terms of the Vandermonde determinant $\Delta(\vect{z}) = \prod_{1 \le a < b \le N} (z_a-z_b)$, we get
\begin{equation}
\label{oredering:of:screening:charges}
\prod\limits_{a=1}^N\mathsf{E}^{-\sqrt{\beta}}(z_a)\ket{0}
= \hskip -3pt\prod\limits_{1 \le a < b \le N}\hskip -5pt {\rm e}^{4\beta[\phi_>(z_a),\phi_<(z_b)]}\
{\rm e}^{-2\sqrt{\beta}\sum\limits_{a=1}^N\phi_<(z_a)}\ket{0}
= \Delta(\vect{z})^{2\beta}\
{\rm e}^{-2\sqrt{\beta}\sum\limits_{a=1}^N\phi_<(z_a)}\ket{0}.
\end{equation}
In addition, we introduce the coherent ``bra'' state
\be
\bra{V_{N\sqrt{\beta},\vect{\scriptstyle t}}} = \bra{0}{\rm e}^{2N\sqrt{\beta}{\sf q}}\prod\limits_{m=0}^{\infty}{\rm e}^{\frac{1}{\hbar}t_m{\sf a}_m}. \label{bra-Virasoro}
\ee
This state satisfies the relation
\begin{equation}
\label{exponent:on:coherent:state}
\bra{V_{N\sqrt{\beta},\vect{\scriptstyle t}}}{\rm e}^{-2\sqrt{\beta}\sum\limits_{a=1}^N\phi_<(z_a)}
= {\rm e}^{-\frac{\sqrt{\beta}}{\hbar}\sum\limits_{a=1}^NV(z_a)}\bra{V_{0,\vect{\scriptstyle t}}},
\qquad V(z) = \sum\limits_{m=0}^\infty t_m z^m.
\end{equation}

Combining the above ingredients, it follows that the $\beta$-deformed eigenvalue integral (\ref{ZV-intro}) can be represented as an integrated CFT expectation value (\ref{ZV-CFT-intro})
\be
Z = \int\!d^N\!z\ \bra{V_{N\sqrt{\beta},\vect{\scriptstyle t}}}\prod\limits_{a=1}^N\left.\left.\mathsf{E}^{-\sqrt{\beta}}(z_a)\right|0\right\rangle
 = \int\!d^N\!z\ \Delta(\vect{z})^{2\beta}\,{\rm e}^{-\frac{\sqrt{\beta}}{\hbar}\sum\limits_{a=1}^NV(z_a)}.
\ee
From the conformal field theory perspective, the relation referred to as the loop equation in matrix model formalism follows from (\ref{commutators:TP:with:E}) and the equality $T_+(y)\ket{0} = 0$
\begin{align}
\nonumber
\label{loop:equation:Virasoro}
\Left\langle T_+(y)\Right\rangle
& \equiv
\int\!d^N\!z\ \bra{V_{N\sqrt{\beta},\vect{\scriptstyle t}}}T_+(y)\prod\limits_{a=1}^N\left.\left.\mathsf{E}^{-\sqrt{\beta}}(z_a)\right|0\right\rangle = \\
& =
\int\!d^N\!z
\sum\limits_{b=1}^N\frac{\partial}{\partial z_a}
\left(
\frac{1}{y-z_a}\bra{V_{N\sqrt{\beta},\vect{\scriptstyle t}}}\prod\limits_{a=1}^N\left.\left.\mathsf{E}^{-\sqrt{\beta}}(z_a)\right|0\right\rangle
\right) = \\
\nonumber
& =
\int\!d^N\!z
\sum\limits_{b=1}^N\frac{\partial}{\partial z_a}
\left(
\frac{1}{y-z_a}\Delta(\vect{z})^{2\beta}\,{\rm e}^{-\frac{\sqrt{\beta}}{\hbar}\sum\limits_{a=1}^NV(z_a)}
\right) \; = \; 0.
\end{align}
It is also immediate to construct the wave-function (or the $\alpha/\beta$ deformed integral)
\be
\begin{split}
\label{Virasoro}
\widehat{\chi}_\alpha(x) & =
\int\!d^N\!z\ \bra{V_{N\sqrt{\beta}-\alpha/\hbar,\vect{\scriptstyle t}}}\mathsf{E}^{\frac{\alpha}{\hbar}}(x)\prod\limits_{a=1}^N\left.\left.\mathsf{E}^{-\sqrt{\beta}}(z_a)\right|0\right\rangle = \\
& = {\rm e}^{\frac{\alpha}{\hbar^2}V(x)}\int\!d^N\!z\ \prod\limits_{a=1}^N(x-z_a)^{-\frac{2\alpha\sqrt\beta}{\hbar}}\,\Delta(\vect{z})^{2\beta}\,{\rm e}^{-\frac{\sqrt{\beta}}{\hbar}\sum\limits_{a=1}^NV(z_a)}.
\end{split}
\ee
This wave-function is supposed to be annihilated by appropriately constructed quantum curves, for appropriate values of $\alpha$.

%******************************************************************************************
%******************************************************************************************

\subsection{...and to quantum curves}   \label{ssec-Virasoro-quantum-curves}

Once we introduced the wave-function (\ref{Virasoro}), we identify now the corresponding quantum curves, and show that they have the structure of singular vectors and can be written in terms of the representation of the Virasoro algebra in (\ref{hat:L:minus:n-intro}). In fact, the singular vector structure of quantum curves follows automatically from the conformal field theory construction of the wave-function -- this is the main advantage of the conformal field theory approach presented in this paper. The only non-trivial aspect is to derive the explicit representation of the Virasoro generators (\ref{hat:L:minus:n-intro}), which is the main aim of this section. This representation can be obtained from correlation functions of the form
\begin{equation}
\label{T:multipoint}
\bra{V_{N\sqrt{\beta}-\alpha,\vect{\scriptstyle t}}}T(y_1)\cdots T(y_l)\mathsf{E}^{\frac{\alpha}{\hbar}}(x)\prod\limits_{a=1}^N\left.\left.\mathsf{E}^{-\sqrt{\beta}}(z_a)\right|0\right\rangle,
\end{equation}
by defining
\begin{equation}
\label{rep:around:x}
{\cal L}_{-i_1}(x)\cdots {\cal L}_{-i_l}(x)\mathsf{E}^{\frac{\alpha}{\hbar}}(x)
= \frac{1}{(2\pi i)^l} \oint\limits_{x}\frac{dy_1}{(y_1-x)^{i_1-1}}\cdots\frac{dy_l}{(y_l-x)^{i_l-1}}\:
T(y_1)\cdots T(y_l)\mathsf{E}^{\frac{\alpha}{\hbar}}(x).
\end{equation}
The fact that ${\cal L}_m(x)$ satisfy the Virasoro algebra can be easily checked by using the OPE
\[
T(y_1)T(y_2) = \frac{c/2}{(y_1-y_2)^2} + \frac{T(y_2)}{(y_1 - y_2)^2} + \frac{\partial T(y_2)}{y_1-y_2}+\ldots,
\]
and the standard contour manipulation.

Calculation of the multipoint correlation function (\ref{T:multipoint}), even if tedious and difficult to present in a closed form for arbitrary $l$, is conceptually straightforward. First, using the commutation relation for $T_{\pm}(y)$ (defined in (\ref{Tplus:definition}))
\be
\begin{split}
\left[T_+(y_1),T_-(y_2)\right] = & \, \frac{c}{2}\frac{1}{(y_1-y_2)^4} + \left(\frac{2}{(y_1-y_2)^2} + \frac{1}{y_1-y_2}\frac{\partial}{\partial y_2}\right)T_-(y_2)   + \\
& + \left(\frac{2}{(y_2-y_1)^2} + \frac{1}{y_2-y_1}\frac{\partial}{\partial y_1}\right)T_+(y_1),
\end{split}
\ee
we ``normal order'' the product $T(y_1)\cdots T(y_l)$ and reduce the calculation to the situation where each $T_-(y_i)$ appears to the left of all $T_+(y_j).$ From (\ref{TplusE:commutator}) and (\ref{commutators:TP:with:E}) it then follows
\be
T_+(y)\ket{x,\vect{z}} =
\left(\frac{\Delta_{\frac{\alpha}{\hbar}}}{(y-x)^2} + \frac{1}{y-x}\frac{\partial}{\partial x}\right)\ket{x,\vect{z}}
+ \sum\limits_{b=1}^N\frac{\partial}{\partial z_b} \left(\frac{1}{y-z_b}\ket{x,\vect{z}} \right),    \label{T+yxz}
\ee
where we introduced a shorthand notation
\begin{align}
\label{ket:x:z}
\ket{x,\vect{z}} = \mathsf{E}^{\frac{\alpha}{\hbar}}(x)\prod\limits_{a=1}^N\mathsf{E}^{-\sqrt{\beta}}(z_a)\ket{0}.
\end{align}
The integration contour in all the $z_a$ variables has to be chosen in such a way that integrals of derivatives are zero. Consequently, the second term in (\ref{T+yxz}) does not contribute to the eigenvalue integral and will be omitted in what follows. We thus have
\begin{equation}
\label{Tplus:on:the :state}
T_+(y_1)\cdots T_+(y_k)\ket{x,\vect{z}}
=
\left(\frac{\Delta_{\frac{\alpha}{\hbar}}}{(y_k-x)^2} + \frac{1}{y_k-x}\frac{\partial}{\partial x}\right)
\cdots \left(\frac{\Delta_{\frac{\alpha}{\hbar}}}{(y_1-x)^2} + \frac{1}{y_1-x}\frac{\partial}{\partial x}\right)
\ket{x,\vect{z}} + \ldots
\end{equation}
where $\ldots$ at the end of this expression denotes terms which vanish upon the $z_a$ integration. Furthermore, from (\ref{oredering:of:screening:charges}) and the commutation relations of $\phi_>(x)$ and $\phi_<(z_a)$ fields
we get
\begin{align}
\ket{x,\vect{z}} =
\prod\limits_{b=1}^N(x-z_b)^{-\frac{2\alpha\sqrt\beta}{\hbar}}\,\Delta(\vect{z})^{2\beta}\,
{\rm e}^{\frac{2\alpha}{\hbar}\phi_<(x)}\prod\limits_{a=1}^N\left.\left.{\rm e}^{-2\sqrt{\beta}\phi_<(z_a)}\right|0\right\rangle,
\end{align}
so that our final task in calculating the correlation function (\ref{T:multipoint}) is to compute correlators of the form
\begin{equation}
\bra{V_{N\sqrt{\beta}-\alpha/\hbar,\vect{\scriptstyle t}}}T_-(y_1)\cdots T_-(y_l)\,{\rm e}^{\frac{2\alpha}{\hbar}\phi_<(x)}\prod\limits_{a=1}^N\left.\left.{\rm e}^{-2\sqrt{\beta}\phi_<(z_a)}\right|0\right\rangle.
\nonumber
\end{equation}

Since for $m,n < 0$
\[
\left[L_{m},{\sf q}\right] = {\sf a}_m,
\hskip 1cm
\left[L_{m},{\sf a}_n\right] = -n{\sf a}_{m+n},
\]
we have
\begin{equation}
\label{commutaton:TM:varphi}
\left[T_-(y),\phi_<(x)\right]
\; = \;
\sum\limits_{n=0}^\infty\sum\limits_{m=2}^{\infty}{\sf a}_{-m-n}x^ny^{m-2}
\; = \;
\frac{\partial\phi_<(y) - \partial\phi_<(x)}{y-x},
\end{equation}
and
\begin{equation}
\label{commutaton:TM:exponent}
\left[T_-(y),{\rm e}^{2\alpha\phi_<(x)}\right]
\; = \;
2\alpha\frac{\partial\phi_<(y) - \partial\phi_<(x)}{y-x}\,{\rm e}^{2\alpha\phi_<(x)}.
\end{equation}
Using
\begin{align}
\label{Tminus:on:vacuum}
T_-(y)\ket{0} = \left((\partial\phi_<(y))^2 + Q \partial^2\phi_<(y)\right)\ket{0},
\end{align}
we finally get
\begin{align}
&T_-(y)\,{\rm e}^{\frac{2\alpha}{\hbar}\phi_<(x)}\prod\limits_{a=1}^N\left.\left.{\rm e}^{-2\sqrt{\beta}\phi_<(z_a)}\right|0\right\rangle
\; = \; \bigg( \frac{2\alpha}{\hbar}\frac{\partial\phi_<(y) - \partial\phi_<(x)}{y-x}  + \\
& -  2\sqrt\beta\sum\limits_{b=1}^N \frac{\partial\phi_<(y) - \partial\phi_<(z_b)}{y-z_b}
+ (\partial\phi_<(y))^2 + Q \partial^2\phi_<(y)
\bigg)
{\rm e}^{\frac{2\alpha}{\hbar}\phi_<(x)}\prod\limits_{a=1}^N\left.\left.{\rm e}^{-2\sqrt{\beta}\phi_<(z_a)}\right|0\right\rangle. \nonumber
\end{align}
It should be clear how to express -- using equations (\ref{commutaton:TM:varphi})-(\ref{Tminus:on:vacuum}) -- states of the
form
\[
T_-(y_1)\cdots T_-(y_k)\,{\rm e}^{\frac{2\alpha}{\hbar}\phi_<(x)}\prod\limits_{a=1}^N\left.\left.{\rm e}^{-2\sqrt{\beta}\phi_<(z_a)}\right|0\right\rangle
\]
entirely in terms of (commuting with each other) exponents and derivatives of fields $\phi_<(x)$ and $\phi_<(z_a).$ Our computation of the correlation function
 (\ref{T:multipoint}) is now completed by noticing that
\begin{align}
\bra{V_{N\sqrt{\beta}-\alpha/\hbar,\vect{\scriptstyle t}}}{\rm e}^{\frac{2\alpha}{\hbar}\phi_<(x)}\prod\limits_{a=1}^N{\rm e}^{-2\sqrt{\beta}\phi_<(z_a)} =
{\rm e}^{\frac{\alpha}{\hbar^2}V(x)}\,{\rm e}^{-\frac{\sqrt{\beta}}{\hbar}\sum_{a=1}^NV(z_a)}
\bra{V_{0,\vect{\scriptstyle t}}}
\end{align}
and
\begin{align}
\bra{V_{0,\vect{\scriptstyle t}}}\partial^n\phi_<(x) = \frac{V^{(n)}(x)}{2\hbar}\bra{V_{0,\vect{\scriptstyle t}}}, \hskip 5mm n > 0.
\end{align}

In the simplest case of $l = 1$, the expectation value (\ref{T:multipoint}) takes form
\begin{align}
\label{single:T:correlator}
& \bra{V_{N\sqrt{\beta}-\alpha/\hbar,\vect{\scriptstyle t}}}T(y)\mathsf{E}^{\frac{\alpha}{\hbar}}(x)\prod\limits_{a=1}^N\left.\left.\mathsf{E}^{-\sqrt{\beta}}(z_a)\right|0\right\rangle =  \nonumber \\
& =  \Big( \frac{\Delta_\frac{\alpha}{\hbar}}{(y-x)^2} + \frac{1}{y-x}\frac{\partial}{\partial x}
+ \frac{\alpha}{\hbar^2}\frac{V'(y) -V'(x)}{y-x} - \frac{\sqrt{\beta}}{\hbar}\sum\limits_{a=1}^N\frac{V'(y)-V'(z_a)}{y-z_a}+  \\
&\quad  + \Big(\frac{V'(y)}{2\hbar}\Big)^2 + \frac{QV''(y)}{2\hbar} \Big) \Psi_\alpha(x,\vect{z})
+ \sum\limits_{a=1}^N\frac{\partial}{\partial z_a}\Big(\frac{\Psi_\alpha(x,\vect{z})}{y-z_a}\Big),
\nonumber
\end{align}
where $\Psi_\alpha(x,\vect{z})$ is a shorthand notation for
\begin{equation}
\label{Psi:explicit}
\Psi_\alpha(x,\vect{z})
=
\prod\limits_{a=1}^N(x-z_a)^{-\frac{2\alpha\sqrt\beta}{\hbar}}\,\Delta(\vect{z})^{2\beta}\,
{\rm e}^{\frac{\alpha}{\hbar^2}V(x)}\,{\rm e}^{-\frac{\sqrt{\beta}}{\hbar}\sum_{a=1}^NV(z_a)}.
\end{equation}
Noticing that
\begin{align}
\frac{V'(y)-V'(x)}{y-x}=
\sum\limits_{n=0}^{\infty}y^m \hskip -5pt \sum\limits_{m=n+2}^{\infty} \hskip -5pt m t_m  x^{m-n-2},
\nonumber
\end{align}
and using the formula
\begin{equation}
x^m\,{\rm e}^{\frac{\alpha}{\hbar^2}V(x)}
= \frac{\hbar^2}{\alpha}\partial_{t_m}\,{\rm e}^{\frac{\alpha}{\hbar^2}V(x)}
\nonumber
\end{equation}
(and, similarly, for $x \to z_a$ and $\alpha \to -\hbar\sqrt\beta$) we get the identity
\begin{align}
\Big(\frac{\alpha}{\hbar^2}\frac{V'(y) -V'(x)}{y-x}- \frac{\sqrt{\beta}}{\hbar}\sum\limits_{a=1}^N\frac{V'(x)-V'(z_a)}{x-z_a}\Big) \Psi_\alpha(x,\vect{z}) =
\frac{1}{\hbar^2}\widehat{f}_t(y)\, \Psi_\alpha(x,\vect{z}),
\end{align}
where
\begin{align}
\label{operator:f:definition}
\widehat{f}_t(y) = \hbar^2\sum\limits_{n=0}^{\infty}y^m \hskip -5pt \sum\limits_{m=n+2}^{\infty} \hskip -5pt m t_m \partial_{t_{m-n-2}}.
\end{align}
Our final representation of the Virasoro algebra generators in a form of differential operators acting on the wave-function $\widehat\chi_\alpha(x)$, defined as
\be
\label{L:hats:definition}
\int\! d^N\!z\, \bra{V_{N\sqrt{\beta}-\alpha/\hbar,\vect{\scriptstyle t}}}T(y)\mathsf{E}^{\frac{\alpha}{\hbar}}(x)\prod\limits_{a=1}^N\left.\left.\mathsf{E}^{-\sqrt{\beta}}(z_a)\right|0\right\rangle
= \sum\limits_{m=-\infty}^{\infty}\frac{{\widehat L}_m \widehat\chi_\alpha(x)}{(y-x)^{m+2}},
\ee
takes form advertised in (\ref{hat:L:minus:n-intro})
\be
\begin{split}
&\qquad\qquad\qquad\qquad\qquad\qquad {\widehat L}_{0} = \Delta_{\frac{\alpha}{\hbar}}, \qquad {\widehat L}_{-1} = \partial_x, \\
&{\widehat L}_{-n} = \frac{1}{\hbar^2(n-2)!} \Big(\frac14\partial^{n-2}_x\left(V'(x)\right)^2 + \frac{Q\hbar}{2}\partial_x^nV(x)
+ \partial_x^{n-2}\widehat{f}_t(x) \Big),\quad \textrm{for}\  n \ge 2.  \label{hat:L:minus:n}
\end{split}
\ee

Furthermore, it follows that quantum curves have the structure of singular vectors given in (\ref{Virasoro:null:vectors:examples}), with the above representation of Virasoro generators ${\widehat L}_{-n}$. For example, the form of $\widehat{A}_{2}^{\,2}(\alpha)$
in (\ref{Virasoro:null:vectors:examples}) yields
\be
\Big(\partial^2_x   -\frac{\alpha^2}{\hbar^4} \Big( \big(V'(x)\big)^2 + 2Q\hbar V''(x)
+ 4\widehat{f}_t(x) \Big)\Big) \widehat\chi_\alpha(x) = 0,
\ee
for $\alpha = 0$, $\frac{\hbar\sqrt{\beta}}{2}$, or $-\frac{\hbar}{2\sqrt{\beta}}$ (recall that in expressions for quantum curves we include an additional factor of $\hbar$ in the degenerate momenta), while
the form of $\widehat{A}_{3}^{\,3}(\alpha)$ leads to the equation
\begin{align}
\Big(\partial^3_x  -4\frac{\alpha^2}{\hbar^2}\partial_x{\widehat L}_{-2}  + \frac{2\alpha^2(2\alpha(2\alpha+Q\hbar)-\hbar^2)}{\hbar^4} {\widehat L}_{-3}\Big) \widehat\chi_\alpha(x) = 0,
\end{align}
with ${\widehat L}_{-2}, {\widehat L}_{-3}$ given explicitly in (\ref{hat:L:minus:n}) and
$\alpha = 0$, $\frac{\hbar\sqrt{\beta}}{2}$, $-\frac{\hbar}{2\sqrt{\beta}}$, $\hbar\sqrt{\beta}$, or $-\frac{\hbar}{\sqrt{\beta}}$. These are examples of quantum curves that we already mentioned in (\ref{A2-intro}) and (\ref{A3-intro}), and it is straightforward to construct quantum curves at higher levels. In the following section we generalize the construction presented above to the supersymmetric case.

%******************************************************************************************
%******************************************************************************************
%******************************************************************************************
%******************************************************************************************

\section{Super-quantum curves in the Neveu-Schwarz sector}   \label{sec-NS}

In this section we derive super-quantum curves in the Neveu-Schwarz sector from the conformal field theory perspective, analogously to the derivation of Virasoro quantum curves in the previous section. While the form of these super-quantum curves have been postulated in \cite{Ciosmak:2016wpx}, the conformal field theory approach proves that they indeed have the structure of Neveu-Schwarz singular vectors.

%******************************************************************************************
%******************************************************************************************

\subsection{Background charge representation in the Neveu-Schwarz sector}

Similarly as in the Virasoro case, we start our consideration from the analysis of the background charge representation, this time of the Neveu-Schwarz algebra. First, we extend the Heisenberg algebra (\ref{Heisenberg}) by fermionic oscillators and define a superalgebra\footnote{Note a change of the normalization in the bosonic commutator as compared to the Virasoro case (\ref{Heisenberg}).}
\be
\label{free:boson:fermion:NS}
\left[{\sf a}_m,{\sf a}_n\right] \; = \; m\delta_{m+n,0},\qquad
\left\{\psi_k,\psi_l\right\} \; = \; \delta_{k+l,0}, \qquad m,n\in\mathbb{Z}, \  k,l\in\mathbb{Z}+\frac12.
\ee
We denote ${\cal H}^{{\sr NS}}_{\alpha} = {\cal H}_{\alpha}\otimes {\cal F}_{{\sr NS}}$, where ${\cal H}_{\alpha}$ is defined as in (\ref{Zgrading}), and ${\cal F}_{{\sr NS}}$ is a free vector space generated by negative modes $\psi_{-k}$ out of the fermionic Fock vacuum $\Omega_{{\sr NS}}$, defined by the condition $\psi_k\Omega_{{\sr NS}} = 0$ for $k > 0$. The hermitian pairing
\be
(\,\cdot\,,\,\cdot\,)_{\alpha,Q}:\ {\cal H}^{{\sr NS}}_{Q-\alpha}\times{\cal H}^{{\sr NS}}_{\alpha} \to {\mathbb C}
\nonumber
\ee
is defined by conditions (\ref{bilinear:form:Virasoro}) supplemented with $\psi_k^\dag = \psi_{-k}.$

Consider now the Neveu-Schwarz (NS for short) algebra
\begin{align}
\label{NS}
\nonumber
\left[{\sf L}_m,{\sf L}_n\right] & =  (m-n){\sf L}_{m+n} +\frac{c}{12}m\left(m^2-1\right)\delta_{m+n,0},
\\
\left[{\sf L}_m,{\sf G}_k\right] & = \frac{m-2k}{2}{\sf G}_{m+k},
\\
\nonumber
\left\{{\sf G}_k,{\sf G}_l\right\} & =  2{\sf L}_{k+l} + \frac{c}{3}\left(k^2 -\frac14\right)\delta_{k+l,0},
\end{align}
where $m, n \in {\mathbb Z}$ and $k,l\in {\mathbb Z}+\frac12$. The Verma module ${\cal V}^{{\sr NS}}_{\Delta,c}$ of the Neveu-Schwarz algebra is a free vector space generated by ${\sf L}_{-m}, m > 0$ and
${\sf G}_{-k}, k > 0$ out of the NS highest weight state $\nu_{\Delta}^{{\sr NS}}$ defined by
\begin{equation}
\label{NS:highest:weight:state}
{\sf L}_m\nu_{\Delta}^{{\sr NS}} = {\sf G}_k \nu_{\Delta}^{{\sr NS}} = 0,\; m,k > 0,
\qquad
{\sf L}_0\nu_{\Delta}^{{\sr NS}} = \Delta\,\nu_{\Delta}^{{\sr NS}}.
\end{equation}
Definitions of Schapovalov hermitian form, as well as singular and null vectors in ${\cal V}^{{\sr NS}}_{\Delta,c},$ are obvious modifications of the corresponding notions
for the Virasoro Verma module.

We now fix\footnote{Note a change in the definition of $\Delta_\alpha$ as compared to the Virasoro case (\ref{cQrel}).}
\be
c = \frac32 - 3Q^2, \qquad  \Delta = \frac12\alpha(\alpha-Q) \equiv \Delta_\alpha.
\ee
Analogously as in the Virasoro case (\ref{Virasoro:alebra:rep}), the background charge representation of the NS algebra is defined by the map
\begin{align}
\nonumber
\sigma^{{\sr NS}}_{\alpha, Q}\ :\
{\rm End}\left({\cal V}^{{\sr NS}}_{\Delta, c}\right) \ni {\sf L}_m, {\sf G}_k\ \to \ L(\alpha)_m, G(\alpha)_k\ \in {\rm End}\left({\cal H}^{{\sr NS}}_{\alpha}\right),
\end{align}
where
\begin{align}
\label{NS:alebra:rep}
\nonumber
L(\alpha)_0 & = \sum\limits_{m=1}^\infty {\sf a}_{-m}{\sf a}_m + \sum\limits_{k=\frac12}^\infty k\psi_{-k}\psi_k + \frac12\alpha\left(\alpha - Q\right),
\\
L(\alpha)_n & = \frac12\sum\limits_{m\neq 0,n} {\sf a}_{n-m}{\sf a}_m + \frac12\sum\limits_{k\in{\mathbb Z}+\frac12}k\psi_{n-k}\psi_k +\frac12\left(2\alpha - (n+1)Q\right){\sf a}_n,
\hskip .5cm n \neq 0,
\\
\nonumber
G(\alpha)_k & = \sum\limits_{m\neq 0}{\sf a}_{m}\psi_{k-m} + \left(\alpha - \left(k+{\textstyle\frac12}\right)Q\right)\psi_k.
\end{align}
Define a matrix ${\cal S}$ of the transition map between canonical bases in ${\cal V}^{{\sr NS}}_{\Delta,c}$ and ${\cal H}^{{\sr NS}}_{\alpha}$
by the formula
\be
L(\alpha)_{-I}G(\alpha)_{-K}\, \mu_\alpha^{{\sr NS}} =
\sum\limits_{|J| +|L| = p }\hskip -5pt  \left[{\cal S}^{\,p}_{\alpha,Q}\right]_{IK,JL}\,{\sf a}_{-J}\psi_{-L}\,\mu_\alpha^{{\sr NS}},
\qquad
p = |I| + |K|,
\ee
where
\[
\psi_{-L} \equiv \psi_{-l_1}\cdots\psi_{-l_m}, \quad 0 < l_1 < \ldots < l_m, \quad |L| = \sum\limits_{i=1}^m l_i,
\]
and similarly for the multiindices $K$ in $G(\alpha)_{-K}$. As shown in \cite{Kato:1987qda}, the matrix $\big({\cal S}^{\,p}_{\alpha,Q}\big)^{-1}$ has (simple for generic values of $\alpha$) poles at
\be
\alpha = \alpha_{r,s} = \frac{(r-1)\beta^{-\frac12} - (s-1)\beta^{\frac12}}{2}, \quad r,s \in {\mathbb Z}_{>0}, \quad \frac{rs}{2} \le p,
\quad r + s \in 2{\mathbb Z}.
\ee
As in the Virasoro case we can construct null vectors of the NS algebra as residues of this matrix
\be
c_{JL}\,\xi_{{\sr NS}\, \textrm{null}}^p =
\lim\limits_{\alpha \to\alpha_{r,s}}(\alpha - \alpha_{r,s})\sum\limits_{|I|+|K|=p}  \left[\left({\cal S}^{\,p}_{\alpha,Q}\right)^{-1}\right]_{JL,IK}\,
{\sf L}_{-I}{\sf G}_{-K}\,\nu^{{\sr NS}}_{\Delta_\alpha},
\ee
and the endomorphisms of ${\cal H}^{{\sr NS}}_{\alpha}$ vanishing at $\alpha = \alpha_{r,s}$ as
\be
\widehat{A}^p_{JL}(\alpha) =
\omega_p^{{\sr NS}}(\alpha,Q) \sum\limits_{|I|+|K|=p}  \left[\left({\cal S}^{\,p}_{\alpha,Q}\right)^{-1}\right]_{JL,IK}\,
{L}(\alpha)_{-I}{G}(\alpha)_{-K},
\ee
where
\be
\omega_p^{{\sr NS}}(\alpha,Q) = \prod\limits_{^{1\le rs \le 2p}_{\hskip 2pt r+s \in 2 {\mathbb Z}}}\hskip -5pt \left(\alpha - \alpha_{r,s}\right).
\ee
The simplest examples are
\begin{align}
\label{NS:singular:opeartors:examples}
\nonumber
\widehat{A}^{1/2}_{1/2}(\alpha) & = G(\alpha)_{-\frac12},
\\
\nonumber
\widehat{A}^{3/2}_{3/2}(\alpha) & = \alpha^2 G(\alpha)_{-\frac32} -G(\alpha)_{-\frac12}L(\alpha)_{-1},
\\
\nonumber
\widehat{A}^{3/2}_{1,1/2}(\alpha) & = (\alpha + Q)G(\alpha)_{-\frac12}L(\alpha)_{-1}-\alpha G(\alpha)_{-\frac32},
\\[-8.5pt]
\\[-8.5pt]
\nonumber
2\widehat{A}^{2}_{2}(\alpha) & = (\alpha+Q) G(\alpha)_{-\frac12}\widehat{A}^{3/2}_{3/2}(\alpha) + \omega_{3/2}^{{\sr NS}}(\alpha,Q)G(\alpha)_{-\frac32}G(\alpha)_{-\frac12},
\\
\nonumber
2\widehat{A}^{2}_{1,1}(\alpha)
& =
(2\alpha+Q)G(\alpha)_{-\frac12}\widehat{A}^{3/2}_{1,1/2}(\alpha) + G(\alpha)_{-\frac12}\widehat{A}^{3/2}_{3/2}(\alpha) - (\alpha(\alpha+Q)-1)G(\alpha)_{-\frac32}G(\alpha)_{-\frac12},
\\
\nonumber
\widehat{A}^2_{3/2,1/2}(\alpha) & = \frac12(\alpha(\alpha+Q)-1)G(\alpha)_{-\frac32}G(\alpha)_{-\frac12} - \frac12G(\alpha)_{-\frac12}\widehat{A}^{3/2}_{3/2}(\alpha),
\end{align}
and an example at the level $p = \frac52$
\begin{align}
\nonumber
2\widehat{A}^{5/2}_{5/2}(\alpha)
=&\, G(\alpha)_{-\frac12} \left(\alpha \widehat{A}^{2}_{1,1}(\alpha) +\left(\alpha^2 + Q\alpha -2\right) \widehat{A}^{2}_{2}(\alpha) - \alpha \left(\alpha^2 + Q\alpha -2\right) \widehat{A}^2_{3/2,1/2}(\alpha)\right)   \\
& -  2\alpha\, \omega_2^{{\sr NS}}(\alpha,Q)G(\alpha)_{-\frac32}L(\alpha)_{-1}.  \label{NS:singular:opeartors:examples-52}
\end{align}

As in the Virasoro case, we also add to the algebra (\ref{free:boson:fermion:NS}) the operator ${\sf q}$ satisfying
\begin{equation}
\label{commutator:an:q:NS}
[{\sf a}_m,{\sf q}] = \delta_{m,0},
\end{equation}
and consider ${\rm e}^{\alpha' {\sf q}}$ as a map
\be
{\cal H}^{{\sr NS}}_{\alpha} \to {\cal H}^{{\sr NS}}_{\alpha+\alpha'},\qquad
{\rm e}^{\alpha' {\sf q}}\,\mu^{{\sr NS}}_\alpha = \mu^{{\sr NS}}_{\alpha+\alpha'}.
\ee
Finally we consider the bosonic field $\phi(x)$ defined by the formula (\ref{scalar:field}), together with the local fermion field
\begin{equation}
\label{free:fermion}
\psi(x) = \psi_>(x) + \psi_<(x),\qquad \psi_>(x) = \sum\limits_{k=\frac12}^{\infty} \psi_k\,x^{-k-\frac12},\qquad
\psi_<(x) = \sum\limits_{k=\frac12}^{\infty} \psi_{-k}\, x^{k-\frac12},
\end{equation}
and construct the energy-momentum tensor $T(x)$ and its partner spin $3/2$ field $S(x)$
\be
\begin{split}
\label{TG:NS}
T(x) & =  \frac12:\!\partial\phi(x)\partial\phi(x)\!: + \frac12 :\!\partial\psi(x)\psi(x)\!: + \frac{Q}{2}\partial^2\phi(x), \\
S(x) & =   \psi(x)\partial\phi(x) + Q\partial\psi(x).
\end{split}
\ee
Analogously to the bosonic case we denote
\be
\begin{split}
T_+(x) &= \sum_{m=-1}^{\infty} \frac{L_m}{x^{m+2}}, \qquad  T_-(x)  = T(x) - T_+(x) = \sum_{m=2}^\infty L_{-m}x^{m-2},\\
S_+(x) &= \sum_{k=-\frac12}^\infty \frac{G_k}{x^{k+\frac32}},  \qquad
S_-(x) = S(x) - S_+(x) = \sum_{k=\frac32}^\infty G_{-k} x^{k-\frac32},    \label{TpmSpm-NS}
\end{split}
\ee
so that $T_+\ket{0}=S_+\ket{0}=0$, where $\ket{0}=\mu_{0}^{{\sr NS}}$. The modes of these fields
\be
L_m = \oint\limits_0 \frac{dx}{2\pi i}\ x^{m + 1}\, T(x),   \qquad
G_k = \oint\limits_0 \frac{dx}{2\pi i}\ x^{k + \frac12}\, S(x),
\ee
have the explicit form
\be
\begin{split}
\label{NS:realization}
L_0 & =  \sum\limits_{m=1}^\infty {\sf a}_{-m}{\sf a}_m + \sum\limits_{k=\frac12}^\infty k\psi_{-k}\psi_k + \frac12{\sf a_0}\left({\sf a}_0 - Q\right),  \\
L_n & =  \frac12\sum\limits_{m \in {\mathbb Z}} {\sf a}_{n-m}{\sf a}_m + \frac12\sum\limits_{k\in{\mathbb Z}+\frac12}k\psi_{n-k}\psi_k -\frac12Q(n+1){\sf a}_n,  \quad n\neq 0, \\
G_k & =  \sum\limits_{m\in {\mathbb Z}}{\sf a}_{m}\psi_{k-m} - Q\left(k+{\textstyle\frac12}\right)\psi_k.
\end{split}
\ee
For every $\xi \in {\cal H}^{{{\sr NS}}}_{\alpha}$ these modes satisfy
\[
L_n\xi = L(\alpha)_n\xi, \qquad G_k\xi = G(\alpha)_k\xi,
\]
and they provide natural extensions of the operators $L(\alpha)_m$ and $G(\alpha)_k$ to the space
\[
{\cal H}^{\sr NS} = \int\limits_{\oplus} {\cal H}^{\sr NS}_{\alpha}\ d\alpha.
\]

%******************************************************************************************
%******************************************************************************************

\subsection{$\alpha/\beta$ deformed eigenvalue integrals in the Neveu-Schwarz sector...}

With the ingredients introduced in the previous section, we can now construct an expectation value representing a super-eigenvalue model in the Neveu-Schwarz sector. Again it is useful to define the normal ordered exponential
\be
\mathsf{E}^{\alpha}(x) = {\rm e}^{\alpha\phi_<(x)}\,{\rm e}^{\alpha\phi_>(x)},
\ee
which in the present case is defined without the factor 2 in the exponent due to different (as compared to the Virasoro case) normalization of the Heisenberg algebra (\ref{free:boson:fermion:NS}). We have
\be
\begin{split}
\label{commutators:TG:E}
\left[L_m,\mathsf{E}^{\alpha}(x)\right] & =  x^m\left(x\partial_x + (m+1)\Delta_\alpha\right)\mathsf{E}^{\alpha}(x), \qquad \Delta_\alpha = \frac12\alpha(\alpha-Q),  \\
\left[G_k,\mathsf{E}^{\alpha}(x)\right] & =  \alpha x^{k+\frac12}\,\psi(x)\mathsf{E}^{\alpha}(x).
\end{split}
\ee
We also define a superfield
\begin{equation}
\label{NS:superfield}
\Phi^{\alpha}(x,\theta) \; = \; {\rm e}^{-\theta G_{-1/2}}\,\mathsf{E}^{\alpha}(x)\,{\rm e}^{\theta G_{-1/2}} \; = \; \left(1+\alpha\psi(x)\theta\right)\mathsf{E}^{\alpha}(x)
\; = \; {\rm e}^{\alpha\psi(x)\theta}\,\mathsf{E}^{\alpha}(x),
\end{equation}
and in what follows sometimes use the notation
\begin{equation}
\label{Phi:m}
\Phi^{\alpha}_<(z,\theta) = {\rm e}^{\alpha(\phi_<(z) + \psi_<(z)\theta)}.
\end{equation}

Using the Neveu-Schwarz algebra (\ref{NS}) as well as the Jacobi identity
\begin{align}
%\label{Jacoby:1}
\left\{G_k,\left[G_l,\mathsf{E}^\alpha(x)\right]\right\}
+
\left\{G_l,\left[G_k,\mathsf{E}^\alpha(x)\right]\right\} =
\left[\left\{G_k,G_l\right\},\mathsf{E}^\alpha(x)\right]
\nonumber
\end{align}
we get
\begin{align}
\label{anticommutator:G:psi:E}
\left\{G_k,\psi(x)\mathsf{E}^{\alpha}(x)\right\} =
\frac{1}{\alpha} x^{k-\frac12}\left(x\partial_x + 2\Delta_\alpha\left(k+{\textstyle\frac12}\right)\right)\mathsf{E}^\alpha(x),
\end{align}
while the Jacobi identity
\begin{align}
%\label{Jacoby:2}
\left[L_m,\left[G_k,\mathsf{E}^\alpha(x)\right]\right] =
\left[\left[L_m,G_k\right],\mathsf{E}^\alpha(x)\right]
+
\left[G_k,\left[L_m,\mathsf{E}^\alpha(x)\right]\right]
\nonumber
\end{align}
gives
\begin{align}
\label{commutator:L:psi:E}
\left[L_m,\psi(x)\mathsf{E}^\alpha(x)\right] =
x^m\left(x\partial_x + (m+1)\left(\Delta_\alpha + {\textstyle\frac12}\right)\right)\psi(x)\mathsf{E}^\alpha(x).
\end{align}
(Note that the formulae (\ref{anticommutator:G:psi:E}) and (\ref{commutator:L:psi:E}) can be also obtained directly using (\ref{NS:realization}) and the commutation relations (\ref{free:boson:fermion:NS}) and (\ref{commutator:an:q:NS}).)
It follows that
\be
\begin{split}
\label{superfield:LG:commutators-NS}
\left[L_m,\Phi^{\alpha}(x,\theta)\right]
& = x^m\left(x\partial_x + (m+1)\left(\Delta_\alpha + {\textstyle\frac12}\theta\partial_\theta\right)\right)\Phi^{\alpha}(x,\theta),
\\
\left[G_k,\Phi^{\alpha}(x,\theta)\right]
& = x^{k-\frac12}
\Big(\theta\left(x\partial_x + 2\Delta_\alpha\left(k+{\textstyle\frac12}\right)\right)-x\partial_\theta \Big)\Phi^{\alpha}(x,\theta),
\end{split}
\ee
and in the particular case of $\alpha = -\sqrt{\beta}$, so that $\Delta_{-\sqrt{\beta}} = \frac12\sqrt{\beta}(Q+\sqrt{\beta})= \frac12$, we get
\be
\begin{split}
\label{screening:charges:LG:commutators}
\left[L_m,\Phi^{-\sqrt\beta}(x,\theta)\right]
& =  \left(\partial_x x - {\textstyle\frac12}(m+1)\partial_\theta\theta\right)\left(x^{m}\Phi^{-\sqrt{\beta}}(x,\theta)\right), \\
\left[G_k,\Phi^{-\sqrt\beta}(x,\theta)\right]
& = \Big(\theta\partial_x-\partial_\theta\Big)\left(x^{k+\frac12}\Phi^{-\sqrt\beta}(x,\theta)\right).
\end{split}
\ee
From (\ref{superfield:LG:commutators-NS}) it also follows that
\be
\begin{split}
\left[T_+(y),\Phi^{\alpha}(x,\theta)\right]
& = \frac{\Delta_{\alpha}+ \frac12\theta\partial_\theta}{(y-x)^2}\Phi^{\alpha}(x,\theta) + \frac{1}{y-x}\partial_x \Phi^{\alpha}(x,\theta),  \\
\left[S_+(y),\Phi^{\alpha}(x,\theta)\right]
& = \frac{2\Delta_{\alpha}\theta}{(y-x)^2} \Phi^{\alpha}(\theta,x)
+\frac{1}{y-x}\left(\theta\partial_x-\partial_\theta\right) \Phi^{\alpha}(\theta,x),  \label{TSplus-Phi-NS}
\end{split}
\ee
and in particular
\be
\begin{split}
\label{commutator:Tp:Sp:Phi}
\left[T_+(y),\Phi^{-\sqrt\beta}(x,\theta)\right] & =
\left(\partial_x - \frac12\partial_\theta \frac{\theta}{y-x}\right)\frac{\Phi^{-\sqrt\beta}(x,\theta)}{y-x},
\\
\left[S_+(y),\Phi^{-\sqrt\beta}(x,\theta)\right] & =
\Big(\theta\partial_x-\partial_\theta\Big)\frac{\Phi^{-\sqrt\beta}(x,\theta)}{y-x}.
\end{split}
\ee

Since
\be
[\theta \psi_>(x),\theta'\psi_<(x')] = -\theta\theta'\{\psi_>(x),\psi_<(x')\} \; = \; -\frac{\theta\theta'}{x-x'},
\ee
we get
\be
\begin{split}
\Phi^{\alpha}(x,\theta)\Phi^{\alpha'}(x',\theta') & =
{\rm e}^{\alpha\alpha'[\phi_>(x),\phi_<(x')]}\,{\rm e}^{\alpha\alpha'[\theta \psi_>(x),\theta'\psi_<(x')]}\,
:\!\Phi^{\alpha}(x,\theta)\Phi^{\alpha'}(x',\theta')\!: \\
& =  (x-x')^{\alpha\alpha'}\left(1-\frac{\alpha\alpha'\theta\theta'}{x-x'}\right):\!\Phi^{\alpha}(x,\theta)\Phi^{\alpha'}(x',\theta')\!: \\
& =
(x-x'-\theta\theta')^{\alpha\alpha'}:\!\Phi^{\alpha}(x,\theta)\Phi^{\alpha'}(x',\theta')\!:.
\end{split}
\ee
Consequently, if we define the NS screening charge operator as
\begin{align}
\label{screening:charge:NS}
\mathsf{Q}_{{\sr NS}} = \int\! dz\,d\theta\ \Phi^{-\sqrt{\beta}}(z,\theta),
\end{align}
then, using the notation (\ref{Phi:m}), we get
\begin{align}
\mathsf{Q}_{{\sr NS}}^N\ket{0} = \int\! d^N\!z\, d^N\!\theta\ \Delta_{{\sr NS}}(\vect{z},\vect{\theta})^{\beta}\,\prod\limits_{a=1}^N\Phi^{-\sqrt\beta}_<(z_a,\theta_a)\ket{0},
\end{align}
where $d^N\!z\, d^N\!\theta = \prod_{a=1}^Ndz_a\, d\theta_a$ and
\begin{align}
\Delta_{{\sr NS}}(\vect{z},\vect{\theta}) = \prod\limits_{1\le a < b \le N}\left(z_a-z_b-\theta_a\theta_b\right).   \label{Vandermonde-NS}
\end{align}
We also introduce a general coherent ``bra'' state in the NS sector
\begin{align}
\label{coherent:bra:NS}
\left\langle V_{N\sqrt{\beta},\hbox{{\boldmath $\scriptstyle t$},{\boldmath $\scriptstyle\xi$}}}\right| =
\bra{0}{\rm e}^{N\sqrt{\beta}\,{\sf q}}\prod\limits_{m=0}^{\infty}{\rm e}^{\frac{1}{\hbar}\left(t_m{\sf a}_{m}+\xi_{m+1/2}\psi_{m+1/2}\right)},
\end{align}
where
\[
\{\xi_k,\xi_l\} = \{\xi_k,\theta_a\} = \{\xi_k,\psi_l\} = 0.
\]

Combining the above ingredients we find that the Neveu-Schwarz $\beta$-deformed super-eigenvalue integral is represented by the following expectation value
\begin{equation}
Z =  \left\langle\left.\left. V_{N\sqrt{\beta},\hbox{{\boldmath $\scriptstyle t$},{\boldmath $\scriptstyle \xi$}}}\right|\mathsf{Q}_{{\sr NS}}^N\right|0\right\rangle
= \int\! d^N\!z\, d^N\!\theta\ \Delta_{{\sr NS}}(\vect{z},\vect{\theta})^{\beta}\,
{\rm e}^{-\frac{\sqrt\beta}{\hbar}\sum_{a=1}^NV(z_a,\theta_a)},   \label{Z-NS-1}
\end{equation}
where
\be
V(z,\theta) = V_{\sr B}(z) + V_{\sr F}(z)\theta, \qquad
V_{\sr B}(z) \; = \; \sum\limits_{m=0}^\infty t_m z^m, \qquad
V_{\sr F}(z) \; = \;  \sum\limits_{m=0}^\infty \xi_{m+1/2} z^m.   \label{Z-NS-2}
\ee
The loop equations for this model from CFT perspective follow from the relations $T_+(x)\ket{0} = S_+(x)\ket{0} = 0$; using
(\ref{commutator:Tp:Sp:Phi}) they take a familiar form
\begin{align}
\label{loop:equations:NS}
\Left\langle T_+(x)\Right\rangle
& \equiv
\left\langle\left.\left. V_{N\sqrt{\beta},\hbox{{\boldmath $\scriptstyle t$},{\boldmath $\scriptstyle \xi$}}}\right|T_+(x)\mathsf{Q}_{{\sr NS}}^N\right|0\right\rangle = \\
\nonumber
& =  \int\! d^N\!z\, d^N\!\theta\
\sum\limits_{a=1}^N \left(\partial_{z_a} - \frac12\partial_{\theta_a}\frac{\theta_a}{x-z_a}\right)
\left(\frac{1}{x-z_a}\Delta_{{\sr NS}}(\vect{z},\vect{\theta})^{\beta}\,
{\rm e}^{-\frac{\sqrt\beta}{\hbar}\sum_{b=1}^NV(z_b,\theta_b)}\right) = 0,  \\
\Left\langle S_+(x)\Right\rangle
& \equiv \left\langle\left.\left. V_{N\sqrt{\beta},\hbox{{\boldmath $\scriptstyle t$},{\boldmath $\scriptstyle \xi$}}}\right|S_+(x)\mathsf{Q}_{{\sr NS}}^N\right|0\right\rangle = \\
\nonumber
& = \int\! d^N\!z\, d^N\!\theta\
\sum\limits_{a=1}^N \Big(\partial_{\theta_a} - \theta_a\partial_{z_a}\Big)
\left( \frac{1}{x-z_a} \Delta_{{\sr NS}}(\vect{z},\vect{\theta})^{\beta}\,
{\rm e}^{-\frac{\sqrt\beta}{\hbar}\sum_{b=1}^NV(z_b,\theta_b)} \right) = 0.
\end{align}

Furthermore, the Neveu-Schwarz wave-function -- also referred to as the $\alpha/\beta$ eigenvalue integral -- is defined as the following expectation value
\begin{equation}
\label{wave:function:NS}
\widehat\chi_\alpha(x,\theta) =
\left\langle\left.\left. V_{N\sqrt{\beta}-\alpha/\hbar,\hbox{{\boldmath $\scriptstyle t$},{\boldmath $\scriptstyle\xi$}}}\right|\Phi^{\frac{\alpha}{\hbar}}(x,\theta)\mathsf{Q}_{{\sr NS}}^N\right|0\right\rangle.
\end{equation}
An explicit form of this expression defines a super-eigenvalue model considered in \cite{Ciosmak:2016wpx}
\begin{equation}
\widehat\chi_\alpha(x,\theta)
=
{\rm e}^{\frac{\alpha}{\hbar^2}V(x,\theta)}
\int\! d^N\!z\, d^N\!\theta\
\prod\limits_{a=1}^N
(x-z_a-\theta\theta_a)^{-\frac{\alpha\sqrt{\beta}}{\hbar}}
\Delta_{{\sr NS}}(\vect{z},\vect{\theta})^{\beta}\,
{\rm e}^{-\frac{\sqrt\beta}{\hbar}\sum_{b=1}^NV(z_b,\theta_b)}.    \label{wave:function:NS-2}
\end{equation}

%******************************************************************************************
%******************************************************************************************

\subsection{...and Neveu-Schwarz super-quantum curves}   \label{ssec-NS-quantum-curves}

In order to derive super-quantum curves, i.e. differential equations satisfied by the wave-function (\ref{wave:function:NS}), we need to compute correlation functions of the form
\[
\left\langle\left.\left. V_{N\sqrt{\beta}-\alpha/\hbar,\hbox{{\boldmath $\scriptstyle t$},{\boldmath $\scriptstyle\xi$}}}\right|
T(y_1)\cdots T(y_m)S(w_1)\cdots S(w_n)\Phi^{\frac{\alpha}{\hbar}}(x,\theta)\mathsf{Q}_{{\sr NS}}^N\right|0\right\rangle.
\]
Since this calculation, even if in principle straightforward, is rather lengthy and a closed formula for arbitrary $m$ and $n$ is not known, we restrict ourselves  to the simplest cases of $m = 1, n = 0$ and $m= 0, n = 1.$ First, we need to identify the following representation of the Neveu-Schwarz algebra
\be
\begin{split}
{\widehat  L}_{-n}\widehat\chi_\alpha(x,\theta) & =  \oint\limits_x \frac{dy}{2\pi i}\frac{1}{(y-x)^{n-1}}
\left\langle\left.\left. V_{N\sqrt{\beta}-\alpha/\hbar,\hbox{{\boldmath $\scriptstyle t$},{\boldmath $\scriptstyle\xi$}}}\right|T(y)\Phi^{\frac{\alpha}{\hbar}}(x,\theta)\mathsf{Q}_{{\sr NS}}^N\right|0\right\rangle,    \\
{\widehat G}_{-k}\widehat\chi_\alpha(x,\theta) & =  \oint\limits_x \frac{dy}{2\pi i}\frac{1}{(y-x)^{k-\frac12}}
\left\langle\left.\left. V_{N\sqrt{\beta}-\alpha/\hbar,\hbox{{\boldmath $\scriptstyle t$},{\boldmath $\scriptstyle\xi$}}}\right|S(y)\Phi^{\frac{\alpha}{\hbar}}(x,\theta)\mathsf{Q}_{{\sr NS}}^N\right|0\right\rangle.
\end{split}
\ee
To evaluate these expressions, first note that
\be
\begin{split}
T_-(y)\ket{0} & =  \frac12\left(\partial\phi_<(y)\partial\phi_<(y) + \partial\psi_<(y)\psi_<(y) + Q \partial^2\phi_<(y)\right)\ket{0},  \\
S_-(y)\ket{0} & =  \left(\psi_<(y)\partial\phi_<(y) + Q \partial\psi_<(y)\right)\ket{0}.
\end{split}
\ee
Using definitions (\ref{NS:realization}), (\ref{Phi:m}) and the commutation relations (\ref{free:boson:fermion:NS}) and (\ref{commutator:an:q:NS}), we find
\begin{align}
\left[T_-(y),\Phi^{\frac{\alpha}{\hbar}}_<(x,\theta)\right] & =
{\frac{\alpha}{\hbar}}\frac{\partial\phi_<(y)-\partial\phi_<(x)}{y-x}\Phi^{\frac{\alpha}{\hbar}}_<(x,\theta)
\\ \nonumber
& + {\frac{\alpha}{2\hbar}}\left(\frac{\partial\psi_<(y) - \partial\psi_<(x)}{y-x} + \frac{\psi_<(y)-\psi_<(x) - (y-x)\partial\psi_<(x)}{(y-x)^2}\right)\!\theta\, \Phi^{\frac{\alpha}{\hbar}}_<(x,\theta),
\\
\left[S_-(y),\Phi^{\frac{\alpha}{\hbar}}_<(x,\theta)\right]
& = {\frac{\alpha}{\hbar}}\left(\frac{\psi_<(y)-\psi_<(x)}{y-x} + \theta\frac{\partial\phi_<(y)-\partial\phi_<(x)}{y-x}\right)\Phi^{\frac{\alpha}{\hbar}}_<(x,\theta),
\end{align}
and similarly for the commutators $\left[T_-(y),\Phi^{-\sqrt{\beta}}_<(z_a,\theta_a)\right]$ and $\left[S_-(y),\Phi^{-\sqrt{\beta}}_<(z_a,\theta_a)\right].$

Since
\be
\begin{split}
\Big\langle V_{N\sqrt{\beta}-\alpha/\hbar,\hbox{{\boldmath $\scriptstyle t$},{\boldmath $\scriptstyle\xi$}}}\Big|\partial\phi_<(y)
& =
\Big\langle V_{N\sqrt{\beta}-\alpha/\hbar,\hbox{{\boldmath $\scriptstyle t$},{\boldmath $\scriptstyle\xi$}}}\Big| \frac{V'_{\sr B}(y)}{\hbar},   \\
\Big\langle V_{N\sqrt{\beta}-\alpha/\hbar,\hbox{{\boldmath $\scriptstyle t$},{\boldmath $\scriptstyle\xi$}}}\Big|\psi_<(y)
& =
\Big\langle V_{N\sqrt{\beta}-\alpha/\hbar,\hbox{{\boldmath $\scriptstyle t$},{\boldmath $\scriptstyle\xi$}}}\Big| \frac{V_{\sr F}(y)}{\hbar},
\end{split}
\ee
we have
\begin{align}
& \frac{\alpha}{\hbar^2}
\left(\frac{V'_{\sr B}(y)-V'_{\sr B}(x)}{y-x}
+\frac12\left(\frac{V'_{\sr F}(y)-V'_{\sr F}(x)}{y-x}
+  \frac{V_{\sr F}(y) - V_{\sr F}(x) -(y-x)V'_{\sr F}(x)}{(y-x)^2}\right)\theta
\right) {\rm e}^{\frac{\alpha}{\hbar^2}V(x,\theta)} = \nonumber \\
& = \;
\frac{1}{\hbar^2}\widehat{f}(y)\,{\rm e}^{\frac{\alpha}{\hbar^2} V(x,\theta)},
\end{align}
where
\be
%\begin{split}
\widehat{f}(y) =
\hbar^2\sum\limits_{n=0}^\infty y^n \hskip -5pt \sum\limits_{m=n+2}^\infty
\left(m t_m \partial_{t_{m-n-2}}+\left(m-\frac{n+1}{2}\right)\xi_{m+1/2}\partial_{\xi_{m-n-3/2}}\right).
%\widehat{f}(y) = & \widehat{f}_t(y) + \widehat{f}_\xi(y),  \\
%&\widehat{f}_t(y)  =  \hbar^2\sum\limits_{n=0}^\infty y^n \hskip -5pt \sum\limits_{m=n+2}^\infty m t_m \partial_{t_{m-n-2}},  \\
%&\widehat{f}_\xi(y) =  \hbar^2\sum\limits_{n=0}^\infty y^n \hskip -5pt \sum\limits_{m = n+2}^\infty \left(m-\frac{n+1}{2}\right)\xi_{m+1/2}\partial_{\xi_{m-n-3/2}}.
%\end{split}
\ee
Similarly
\be
\frac{\alpha}{\hbar^2} \left(\frac{V_{\sr F}(y)-V_{\sr F}(x)}{y-x}
+ \frac{V'_{\sr B}(y)-V'_{\sr B}(x)}{y-x}\theta
\right) {\rm e}^{\frac{\alpha}{\hbar^2}V(x,\theta)}
= \frac{1}{\hbar^2}\widehat{h}(y)\,{\rm e}^{\frac{\alpha}{\hbar^2}V(x,\theta)},
\ee
where
\be
%\begin{split}
\widehat{h}(y) = \hbar^2\sum_{n=0}^\infty y^n \hskip -5pt \sum_{m=n+1}^\infty
\left(\xi_{m+1/2} \partial_{t_{m-n-1}}+mt_{m}\partial_{\xi_{m-n-3/2}}\right).
%\widehat{h}(y)  = &  \widehat{h}_t(y) + \widehat{h}_\xi(y),   \\
%&\widehat{h}_t(y)  =  \hbar^2\sum_{n=0}^\infty y^n    \sum_{m=n+1}^\infty \xi_{m+1/2} \partial_{t_{m-n-1}},  \\
%&\widehat{h}_\xi(y)  =  \hbar^2\sum_{n=0}^\infty y^n    \sum_{m = n+2}^\infty mt_{m}\partial_{\xi_{m-n-3/2}}.
%\end{split}
\ee
Combining the above ingredients, and using (\ref{TSplus-Phi-NS}) and
(\ref{commutator:Tp:Sp:Phi}) which in particular contributes as surface terms, we get
\be
\begin{split}
&\left\langle\left.\left. V_{N\sqrt{\beta}-\alpha/\hbar,\hbox{{\boldmath $\scriptstyle t$},{\boldmath $\scriptstyle\xi$}}}\right| T(y)\Phi^{\frac{\alpha}{\hbar}}(x,\theta)\mathsf{Q}_{{\sr NS}}^N\right|0\right\rangle
=  \left(\frac{\Delta_{\frac{\alpha}{\hbar}}+ \frac12\theta\partial_\theta}{(y-x)^2} + \frac{1}{y-x}\partial_x\right)\widehat\chi_\alpha(x,\theta) + \\
& \qquad + \frac{1}{2\hbar^2}\left(
\left(V'_{\sr B}(y)\right)^2
+ V'_{\sr F}(y)V_{\sr F}(y)
+ Q\hbar V''_{\sr B}(y)
+2 \widehat{f}(y)
\right)\widehat\chi_\alpha(x,\theta),   \label{vev-T-NS}
\end{split}
\ee
and
\be
\begin{split}
&\left\langle\left.\left. V_{N\sqrt{\beta}-\alpha/\hbar,\hbox{{\boldmath $\scriptstyle t$},{\boldmath $\scriptstyle\xi$}}}\right|S(y)\Phi^{\frac{\alpha}{\hbar}}(x,\theta)\mathsf{Q}_{{\sr NS}}^N\right|0\right\rangle
= \left(\frac{2\theta\Delta_{\frac{\alpha}{\hbar}}}{(y-x)^2} +\frac{1}{y-x}\left(\theta\partial_x-\partial_\theta\right)\right)\widehat\chi_\alpha(x,\theta) + \\
& \qquad + \frac{1}{\hbar^2}\left(V_{\sr F}(y)V'_{\sr B}(y) + Q\hbar V'_{\sr F}(y)+ \widehat{h}(y)\right)\widehat\chi_\alpha(x,\theta). \label{vev-S-NS}
\end{split}
\ee
Expanding (\ref{vev-T-NS}) and (\ref{vev-S-NS}) in powers of $y$, we find that the representation of the Neveu-Schwarz algebra on the wave-function $\widehat\chi_\alpha(x,\theta)$ takes form
\be
\begin{split}
\label{L:final:representation}
& \qquad\qquad\qquad\qquad\qquad {\widehat L}_0 = \Delta_{\frac{\alpha}{\hbar}}+ \frac12\theta\partial_\theta, \qquad {\widehat L}_{-1} = \partial_x, \\
{\widehat L}_{-n} & =
\frac{1}{2\hbar^2 (n-2)!}\Big(\partial_x^{n-2}\left(V'_{\sr B}(x)\right)^2 + \partial_x^{n-2}\left(V'_{\sr F}(x)V_{\sr F}(x)\right)
+ Q\hbar \partial_x^nV_{\sr B}(x)+ 2\partial_x^{n-2}\widehat{f}(x)\Big),
\end{split}
\ee
and
\be
\begin{split}
\label{G:final:representation}
& \qquad\qquad\qquad  {\widehat G}_{\frac12} = 2\theta\Delta_{\frac{\alpha}{\hbar}}, \qquad
{\widehat G}_{-\frac12} =  \theta\partial_x-\partial_\theta,   \\
{\widehat G}_{-k}
& = \frac{1}{\hbar^2 \left(k-{\textstyle \frac32}\right)!} \Big( \partial_x^{k-\frac32}\left(V_{\sr F}(x)V'_{\sr B}(x)\right)
+ Q\hbar\partial_x^{k-\frac12}V_{\sr F}(x) +\partial_x^{k-\frac32}\widehat{h}(x) \Big).
\end{split}
\ee

From the above construction it automatically follows that super-quantum curves in the Neveu-Schwarz sector take form of singular Neveu-Schwarz vectors, such as those in examples (\ref{NS:singular:opeartors:examples}) and (\ref{NS:singular:opeartors:examples-52}), expressed in terms of the above generators of the Neveu-Schwarz algebra (\ref{L:final:representation}) and (\ref{G:final:representation}). In this way we reproduce -- and prove in general -- the results found in \cite{Ciosmak:2016wpx} (using the matrix model formalism).

%******************************************************************************************
%******************************************************************************************
%******************************************************************************************
%******************************************************************************************

\section{Ramond sector and Ramond-NS super-quantum curves}   \label{sec-Ramond}

In previous sections, using conformal field theory techniques, we proved that quantum curves associated to Virasoro and Neveu-Schwarz algebras, found in \cite{Manabe:2015kbj,Ciosmak:2016wpx}, indeed have structure of singular vectors for these algebras. In this section we generalize such an approach to the Ramond sector of the super-Virasoro algebra. In this case taking the viewpoint of conformal field theory has two important advantages. First of all, it enables to define the corresponding eigenvalue model, whose form is not obvious to identify a priori. Second, similarly as in the previous sections, it proves in general that Ramond quantum curves take form of singular vectors of the Ramond algebra.

When considering the Ramond sector, certain subtleties must be taken into account. Similarly as in earlier sections, the wave-function $\widehat{\chi}_{\alpha}(x)$ is identified as an expectation value of a certain $x$-dependent operator, evaluated in between two reference states. Such an expectation value can be defined in two general ways. First, in order to have a well-defined wave-function, we can choose the two reference states (represented, schematically, by the bra $\langle R|$ and the ket $|R\rangle$) from the Ramond sector, and the $x$-dependent operator $NS(x)$ to be of the Neveu-Schwarz type, schematically
\be
\widehat{\chi}_{\alpha}(x) \sim \langle R|\, NS(x)\, |R\rangle.     \label{chi-R-NS-R}
\ee
As the $x$-dependent part in this expression is encoded in the operator in the Neveu-Schwarz sector, it follows that quantum curves -- i.e. differential equations satisfied by $\widehat{\chi}_{\alpha}(x)$ -- have the structure of the Neveu-Schwarz singular vectors, such as those given in (\ref{NS:singular:opeartors:examples}) and (\ref{NS:singular:opeartors:examples-52}). Nonetheless, the form of super-Virasoro generators $\widehat{L}_n$ and $\widehat{G}_{n+1/2}$ in terms of which these quantum curves are expressed is now different than in section \ref{sec-NS}, and it encodes properties of the underlying Ramond algebra. We call such quantum curves as Ramond-NS super-quantum curves.

There is also the second possibility -- we can choose the $x$-dependent operator to be of the Ramond type, and then one of the reference states to be from the NS sector, and the other state from the Ramond sector, schematically
\be
\widehat{\chi}_{\alpha}(x) \sim \langle NS|\, R(x)\, |R\rangle.     \label{chi-NS-R-R}
\ee
In this case the corresponding quantum curves indeed have the structure of Ramond singular vectors, and we call such curves as Ramond-R super-quantum curves.

The construction of the above two types of Ramond quantum curves is a generalization of considerations in sections \ref{sec-Virasoro} and \ref{sec-NS} -- however details of such constructions are not completely obvious.
%The first advantage of this construction is that it naturally enables to identify the super-eigenvalue model in the Ramond sector, whose form is not obvious a priori.
In what follows, in sections \ref{ssec-background-R} and \ref{ssec-wavefunction-Ramond} we present the construction of Ramond-NS quantum curves, and in section \ref{ssec-Ramond-NS-Penner} we discuss the case of multi-Penner potential, which gives rise to one particular example of Ramond-NS super-quantum curves. Subsequently, in section \ref{sec-Ramond-R}, we present a construction of Ramond-R quantum curves. For brevity, in the Ramond-R case we essentially restrict the analysis to the Penner-like potential, and demonstrate that in this case quantum curves take form of a supersymmetric generalization of BPZ equations, which provides an independent check of our approach. Moreover, independently of conformal field theory analysis, in section \ref{sec-matrix} we derive Ramond-NS quantum curves from the super-eigenvalue model perspective and using matrix model techniques. Similarly, in section \ref{sec-matrix-Ramond-R} we rederive Ramond-R quantum curves using eigenvalue model techniques. This proves that CFT and matrix model methods lead to the same results, and enables to compare advantages of each of those approaches.

%******************************************************************************************
%******************************************************************************************

\subsection{Background charge representation and singular vectors}  \label{ssec-background-R}

The oscillator algebra in the Ramond sector takes the same form as in the Neveu-Schwarz sector (\ref{free:boson:fermion:NS}), however now with all indices integer
\be
\label{free:boson:fermion:R}
\left[{\sf a}_m,{\sf a}_n\right] \; = \; m\delta_{m+n,0},\qquad
\left\{\psi_m,\psi_n\right\} \; = \; \delta_{n+n,0}, \qquad m,n\in\mathbb{Z}.
\ee
There are two vacuum states in the Ramond sector, which we denote by $\ket{0,\pm}$, and which are defined by equations
\begin{equation}
{\sf a}_m\ket{0,\pm} = 0, \;\; m \ge 0, \qquad \psi_m\ket{0,\pm} = 0, \;\; m > 0, \qquad \psi_0\ket{0,\pm} = \frac{1}{\sqrt 2}\ket{0,\mp}.
\end{equation}
The super-Virasoro algebra, which now we refer to as the Ramond algebra, takes the same form as in (\ref{NS}), however now with all indices integer. We consider the following free field realization of this Ramond algebra
\begin{align}
\label{Ramond:realization}
\nonumber
L_0 & = \sum\limits_{m=1}^\infty {\sf a}_{-m}{\sf a}_m + \sum\limits_{m=1}^\infty m\psi_{-m}\psi_m + \frac12{\sf a_0}\left({\sf a}_0 - Q\right) + \frac{1}{16},
\\
L_n & = \frac12\sum\limits_{m\neq 0,n} {\sf a}_{n-m}{\sf a}_m + \frac12\sum\limits_{m\in{\mathbb Z}}m\psi_{n-m}\psi_m +\frac12\left(2{\sf a}_0 - (n+1)Q\right){\sf a}_n,\hskip .5cm n\neq 0,
\\
\nonumber
G_n & = \sum\limits_{m\in {\mathbb Z}}{\sf a}_{n-m}\psi_m - Q\left(n+{\textstyle\frac12}\right)\psi_n.
\end{align}
One can check that the following relations hold
\begin{equation}
\label{positive:GL:on:vacuum}
L_m\ket{0,\pm} = G_m\ket{0,\pm} = 0,\hskip 5mm m > 0,
\end{equation}
and
\begin{equation}
\label{zero:GL:on:vacuum}
L_0\ket{0,\pm} = \frac{1}{16}\ket{0,\pm}, \qquad G_0\ket{0,\pm} = -\frac{Q}{2}\psi_0\ket{0,\pm} = -\frac{Q}{2\sqrt{2}}\ket{0,\mp}.
\end{equation}
Due to the presence of the term $\frac12\psi_{-1}\psi_0$ in $L_{-1}$, there is no state annihilated by $L_{-1}$ in the Ramond sector.

By considering a transition map between the oscillator algebra and the Ramond version of the super-Virasoro algebra one can derive general expressions for singular vectors in the Ramond sector, analogously to considerations in previous sections. At the level zero the operator which gives a null vector for $\alpha = Q/2$ (while acting on the Ramond highest weight vector $\ket{\Delta_\alpha,\pm}$ with
$\Delta_\alpha = \frac12 \alpha(\alpha-Q) + \frac{1}{16}$) or the null field (while acting on the Ramond primary field ${\sf R}^\alpha_{\pm}(x)$) is given by
\[
{\widehat A}^{\,0}_{(\cdot|0)} = G_{0}.
\]
At the level 1 the operators which give null vectors/fields for  $\alpha = Q/2$, $\frac12\sqrt{\beta}$, or $-\frac{1}{2\sqrt\beta}$ may be presented in the form
\begin{align}
\begin{split}
{\widehat A}^{\,1}_{(\cdot|1)} & =  4\alpha(2\alpha-Q)G_{-1} - 8L_{-1}G_0, \\
{\widehat A}^{\,1}_{(1|0)} & =  4(2\alpha + Q)L_{-1}G_0 - (2\alpha-Q)G_{-1},\\
{\widehat A}^{\,1}_{(\cdot|1,0)} & =  8\alpha G_{-1}G_0 - 2(2\alpha-Q)L_{-1},\\
{\widehat A}^{\,1}_{(1|\cdot)} & =  2(2\alpha-Q)(2\alpha +Q)L_{-1} - 4 G_{-1}G_0. \label{singular-Ramond}
\end{split}
\end{align}
As our last example consider operators
\begin{align}
\begin{split}
\widehat{A}^{\,2}_{(1,1|\cdot)}
& =
L_{-1}\left(\left(\alpha + \textstyle{\frac32}Q\right)\widehat{A}^{\,1}_{(1|\cdot)} + \textstyle{\frac32} \widehat{A}^{\,1}_{(\cdot|1,0)}\right)
-
G_{-1}\left(\textstyle{\frac92}\widehat{A}^{\,1}_{(1|0)} + \left(\alpha + \textstyle{\frac32}Q\right)\widehat{A}^{\,1}_{(\cdot|1)}\right),
\\
\widehat{A}^{\,2}_{(2|\cdot)}
& =
G_{-1}\left(\textstyle{\frac32}\alpha\widehat{A}^{\,1}_{(1|0)} + \left(\alpha^2 + \textstyle{\frac32}Q\alpha - \textstyle{\frac34}\right)\widehat{A}^{\,1}_{(\cdot|1)}\right)
-
L_{-1}\left(\textstyle{\frac32}\widehat{A}^{\,1}_{(1|\cdot)} + 2\alpha \widehat{A}^{\,1}_{(\cdot|1,0)}\right),
\\
\widehat{A}^{\,2}_{(1|1,0)}
& =
G_{-1}\left(3\alpha\widehat{A}^{\,1}_{(1|0)} +\textstyle{\frac32}\widehat{A}^{\,1}_{(\cdot|1)}\right)
-
L_{-1}\left(\textstyle{\frac32}\widehat{A}^{\,1}_{(1|\cdot)} + 2\alpha \widehat{A}^{\,1}_{(\cdot|1,0)}\right),
\\
\widehat{A}^{\,2}_{(\cdot|2,0)}
& =
L_{-1}\left(\alpha\widehat{A}^{\,1}_{(1|\cdot)} +  \left(2\alpha^2 + Q\alpha - \textstyle{\frac32}\right) \widehat{A}^{\,1}_{(\cdot|1,0)}\right)
-
G_{-1}\left(2\alpha^2\widehat{A}^{\,1}_{(1|0)} + \alpha\widehat{A}^{\,1}_{(\cdot|1)}\right).
\nonumber
\end{split}
\end{align}
For $\alpha = Q/2$, $\frac12\sqrt{\beta}$, $-\frac{1}{2\sqrt\beta}$, $\frac32\sqrt{\beta}$, or $-\frac{3}{2\sqrt\beta}$ they give rise to the null vectors at level 2.

%******************************************************************************************
%******************************************************************************************

\subsection{Ramond-NS eigenvalue model}  \label{ssec-R-eigenvaluemodel}

We introduce now various fields and operators relevant in the Ramond sector. First, we consider the bosonic field (\ref{scalar:field}) and define a fermionic one
\begin{equation}
\psi(x)=\psi_>(x)+\psi_<(x),\qquad
\psi_>(x) = \sum\limits_{m=0}^{\infty}\psi_m x^{-m-\frac12},
\qquad
\psi_<(x) = %\sum\limits_{m=-\infty}^{-1}\psi_m x^{-m-\frac12} =
\sum\limits_{m=1}^{\infty}\psi_{-m}x^{m-\frac12},
\end{equation}
as well as the corresponding energy-momentum tensor and its superpartner
\be
\begin{split}
\label{TG:NS-Ramond}
T(x) & =  \frac12:\!\partial\phi(x)\partial\phi(x)\!: + \frac12 :\!\partial\psi(x)\psi(x)\!: + \frac{Q}{2}\partial^2\phi(x), \\
S(x) & =   \psi(x)\partial\phi(x) + Q\partial\psi(x).
\end{split}
\ee
These fields have mode decompositions
\begin{gather}
T(x)=T_+(x)+T_-(x),\qquad S(x)=S_+(x)+S_-(x), \nonumber \\
T_-(x) = \sum_{m=1}^{\infty} L_{-m} x^{m-2}, \qquad  T_+(x) = \sum_{m=0}^{\infty} \frac{L_m}{x^{m+2}},   \label{TS-Ramond}  \\
S_-(x) = \sum_{m=1}^{\infty} G_{-m} x^{m-\frac32}, \qquad S_+(x) = \sum_{m=0}^{\infty} \frac{G_m}{x^{m+\frac32}},  \nonumber
\end{gather}
with modes $L_m = \oint_0 \frac{dx}{2\pi i}\ x^{m + 1} T(x)$ and $G_k = \oint_0 \frac{dx}{2\pi i}\ x^{k + \frac12} S(x)$ given in (\ref{Ramond:realization}). Note that $T_+(x)\ket{0,\pm}=\frac{1}{16x^{2}}\ket{0,\pm}$, $S_+(x)\ket{0,\pm}=-\frac{Q}{2x^{3/2}}\psi_0\ket{0,\pm}$, and contrary to the definition in the Neveu-Schwarz sector (\ref{TpmSpm-NS}), now $T_-(x)$ and $S_-(x)$ contain singular terms at $x=0$ (for modes labeled by  $m=1$). Furthermore, we introduce the exponential (super)fields
\begin{align}
\begin{split}
\mathsf{E}^{\alpha}(x) &= {\rm e}^{\alpha\phi_<(x)}\,{\rm e}^{\alpha\phi_>(x)}, \\
\Phi^{\alpha}(x,\theta) &= \big(1+\alpha\psi(x)\theta)\mathsf{E}^{\alpha}(x) \; = \;   {\rm e}^{\alpha\psi(x)\theta}\mathsf{E}^{\alpha}(x). \label{Phi-Ramond}
\end{split}
\end{align}
In what follows we also use the notation
\begin{equation}
\label{Phi:m-Ramond}
\Phi^{\alpha}_<(z,\theta) = {\rm e}^{\alpha(\phi_<(z) + \psi_<(z)\theta)},\qquad
\Phi^{\alpha}_{\le}(x,\theta) = {\rm e}^{-\alpha \frac{\theta}{\sqrt{x}}\psi_0} \Phi^{\alpha}_<(x,\theta).
\end{equation}
Note that the following commutation relations hold
\be
\begin{split}
\label{superfield:LG:commutators-Ramond}
\left[L_m,\Phi^{\alpha}(x,\theta)\right]
& = x^m\left(x\partial_x + (m+1)\left(\Delta_\alpha + {\textstyle\frac12} - {\textstyle\frac12}\partial_\theta\theta\right)\right)\Phi^{\alpha}(x,\theta),
\\
\left[G_m,\Phi^{\alpha}(x,\theta)\right]
& = x^{m-\frac12} \Big(\theta\left(x\partial_x + 2\Delta_\alpha\left(m+{\textstyle\frac12}\right)\right)-x\partial_\theta
\Big)\Phi^{\alpha}(x,\theta),\quad \Delta_\alpha=\frac12\alpha(\alpha-Q),
\end{split}
\ee
which in the special case $\alpha = -\sqrt{\beta}$, so that $\Delta_{-\sqrt{\beta}} = \frac12\sqrt{\beta}(Q+\sqrt{\beta})= \frac12$, take form
\be
\begin{split}
\label{screening:charges:LG:commutators-Ramond}
\left[L_m,\Phi^{-\sqrt\beta}(x,\theta)\right]
& =  \left(\partial_x x - {\textstyle\frac12}(m+1)\partial_\theta\theta\right)\left(x^{m}\Phi^{-\sqrt{\beta}}(x,\theta)\right), \\
\left[G_m,\Phi^{-\sqrt\beta}(x,\theta)\right]
& = \Big(\theta\partial_x-\partial_\theta\Big)\left(x^{m+\frac12}\Phi^{-\sqrt\beta}(x,\theta)\right).
\end{split}
\ee
From (\ref{superfield:LG:commutators-Ramond}) we also find
\be
\begin{split}
\left[T_{+}(y),\Phi^{\alpha}(x,\theta)\right]
&=
\partial_x\frac{x}{y(y-x)}\Phi^{\alpha}(x,\theta)
+\frac{\Delta_{\alpha}-\frac12-\frac12\partial_{\theta}\theta}{(y-x)^2}\Phi^{\alpha}(x,\theta), \\
\left[S_{+}(y),\Phi^{\alpha}(x,\theta)\right]
&=
\left(\theta\partial_x-\partial_{\theta}\right)
\sqrt{\frac{x}{y}}\frac{1}{y-x}\Phi^{\alpha}(x,\theta)
+\frac{(\Delta_{\alpha}-\frac12)(y+x)\theta}{\sqrt{yx}(y-x)^2}\Phi^{\alpha}(x,\theta),
\label{TS_comm_screen_R_sp}
\end{split}
\ee
and in particular
\be
\begin{split}
\left[T_{+}(y),\Phi^{-\sqrt{\beta}}(x,\theta)\right]
&=
\left(\partial_x\frac{x}{y(y-x)}
-\frac{\partial_{\theta}\theta}{2(y-x)^2}\right)\Phi^{-\sqrt{\beta}}(x,\theta), \\
\left[S_{+}(y),\Phi^{-\sqrt{\beta}}(x,\theta)\right]
&=
\left(\theta\partial_x-\partial_{\theta}\right)
\sqrt{\frac{x}{y}}\frac{1}{y-x}\Phi^{-\sqrt{\beta}}(x,\theta).
\label{TS_comm_screen_R}
\end{split}
\ee

Since
\be
\begin{split}
{\rm e}^{\psi_>(x)\theta}\,{\rm e}^{\psi_<(x')\theta'}
& = \left(1-\{\psi_>(x),\psi_<(x')\}\theta\theta'\right)\,{\rm e}^{\psi_<(x')\theta'}\,{\rm e}^{\psi_>(x)\theta} = \\
& = \left(1-\frac{\theta\theta'}{\sqrt{xx'}}\frac{x'}{x-x'}\right){\rm e}^{\psi_<(x')\theta'}\,{\rm e}^{\psi_>(x)\theta},
\end{split}
\ee
we get the identity
\be
\prod\limits_{a=1}^N\Phi^{-\sqrt{\beta}}(z_a,\theta_a)\ket{0,\pm}
= \hskip -10pt \prod\limits_{1\le a < b \le N}\hskip -5pt \left(z_a-z_b-\theta_a\theta_b\sqrt{\frac{z_b}{z_a}}\right)^\beta \prod\limits_{a=1}^N{\rm e}^{\sqrt{\beta}\frac{\theta_a}{ \sqrt{z_a}} \psi_0}
\prod\limits_{a=1}^N \Phi^{-\sqrt{\beta}}_<(z_a,\theta_a)\ket{0,\pm},
\ee
which is relevant for the eigenvalue model of the Ramond-NS type. To simplify this relation, note that we have
\[
\prod\limits_{a=1}^N{\rm e}^{\sqrt{\beta}\frac{\theta_a}{ \sqrt{z_a}} \psi_0}=
\prod\limits_{a=1}^N \left(1+\sqrt{\beta}\frac{\theta_a}{ \sqrt{z_a}}\psi_0\right) =
\bigg(1+\sqrt{\beta}\sum\limits_{a=1}^N
\frac{\theta_a}{\sqrt{z_a}}\psi_0\bigg)
\exp\bigg(-{\textstyle\frac{\beta}{2}}\hskip -10pt\sum\limits_{1 \le a<b\le N}\frac{\theta_a\theta_b}{ \sqrt{z_az_b}}\bigg),
\]
so that we get
\begin{align}
& \left(z_a-z_b-\theta_a\theta_b\sqrt{\frac{z_b}{z_a}}\right)^\beta\exp\left(-\frac{\beta}{2}\frac{\theta_a\theta_b}{\sqrt{z_az_b}}\right) = (z_a-z_b)^{\beta}\left(1-\beta\sqrt{\frac{z_b}{z_a}}\frac{\theta_a\theta_b}{z_a-z_b}\right)\left(1-\frac{\beta}{2}\frac{\theta_a\theta_b}{\sqrt{z_az_b}}\right)  \nonumber \\
& = (z_a-z_b)^{\beta}\left(1-\frac{\beta}{2}\frac{\theta_a\theta_b}{\sqrt{z_az_b}}\frac{z_a+z_b}{z_a-z_b}\right)
= \left(z_a-z_b-\frac{z_a+z_b}{2\sqrt{z_az_b}}\,\theta_a\theta_b\right)^\beta.
\end{align}
Then it follows that
\be
\prod_{a=1}^N\Phi^{-\sqrt{\beta}}(z_a,\theta_a)\ket{0,\pm}
= {\rm e}^{-\psi_0\sqrt{\beta}\sum\limits_{a=1}^{N} \frac{\theta_a}{\sqrt{z_a}}} \Delta_{\sr R}(\vect{z},\vect{\theta})^\beta \prod_{a=1}^N \Phi^{-\sqrt{\beta}}_<(z_a,\theta_a)\ket{0,\pm},
\ee
where we introduced a (Ramond-NS) version of the Vandermonde determinant
\begin{equation}
\Delta_{\sr R}(\vect{z},\vect{\theta}) = \prod_{1\le a < b \le N}  \left(z_a-z_b-\frac{z_a+z_b}{2\sqrt{z_az_b}}\,\theta_a\theta_b\right).   \label{Vandermonde-Ramond}
\end{equation}

In order to define the Ramond-NS model it is natural to define the following coherent ``bra'' states
\begin{align}
\label{coherent:bra:R}
\left\langle V^{\pm}_{N\sqrt{\beta},\hbox{{\boldmath $t$},{\boldmath $\xi$}}}\right| =
\bra{0,\pm}{\rm e}^{N\sqrt{\beta}\,{\sf q}}\,{\rm e}^{\frac{2}{\hbar}\xi_0\psi_0}\prod\limits_{m=0}^{\infty}{\rm e}^{\frac{1}{\hbar}\left(t_m{\sf a}_{m}+\xi_{m+1}\psi_{m+1}\right)}.
\end{align}
Since $\left<0,+\, |\, 0,-\right>=0$ (in our choice of normalization), we get
\be
\bra{0,+}{\rm e}^{\frac{2}{\hbar}\xi_0\psi_0}{\rm e}^{-\psi_0\sqrt{\beta}\sum_{a=1}^{N} \frac{\theta_a}{\sqrt{z_a}}}\ket{0,+}
= 1-2\psi_0^2\frac{\sqrt{\beta}}{\hbar}\sum_{a=1}^{N} \frac{\xi_0\theta_a}{\sqrt{z_a}}
= {\rm e}^{-\frac{\sqrt{\beta}}{\hbar}\sum_{a=1}^{N} \frac{\xi_0\theta_a}{\sqrt{z_a}}},
\ee
and then
\be
\label{eq:2:24}
\left\langle V^+_{N\sqrt{\beta},\hbox{{\boldmath $t$},{\boldmath $\xi$}}}\right|\prod\limits_{a=1}^N\left.\left.\Phi^{-\sqrt{\beta}}(z_a,\theta_a)\right|0,+\right\rangle
= \Delta_{\sr R}(\vect{z},\vect{\theta})^\beta\, {\rm e}^{-\frac{\sqrt{\beta}}{\hbar}\sum_{a=1}^N\! V_{\sr R}(z_a,\theta_a)},
\ee
where
\be
V_{\sr R}(z,\theta)  =  V_{\sr B}(z) + V_{\sr F}(z)\frac{\theta}{\sqrt{z}},
\qquad V_{\sr B}(z) = \sum\limits_{m=0}^\infty t_m z^m,
\qquad V_{\sr F}(z) = \sum\limits_{m=0}^\infty \xi_m z^{m}.    \label{VRztheta}
\ee
The partition function of the Ramond-NS type is now represented as the integrated expectation value (\ref{eq:2:24})
\be
Z  = \left.\left.\left\langle V^+_{N\sqrt{\beta},\hbox{{\boldmath $t$},{\boldmath $\xi$}}}\right|\mathsf{Q}_{{\sr R}}^N\right|0,+\right\rangle
= \int\! d^N\!z\, d^N\!\theta\
\Delta_{\sr R}(\vect{z},\vect{\theta})^\beta\,
{\rm e}^{-\frac{\sqrt{\beta}}{\hbar}\sum_{a=1}^N\! V_{\sr R}(z_a,\theta_a)},
\label{Z-Ramond}
\ee
where we defined the Ramond screening charge operator by
\begin{align}
\label{screening:charge:R}
\mathsf{Q}_{{\sr R}} = \int\! dz\,d\theta\ \Phi^{-\sqrt{\beta}}(z,\theta).
\end{align}
This partition function (\ref{Z-Ramond}) is regarded as the definition of the super-eigenvalue model for the Ramond-NS sector.\footnote{
Note that, since we defined two ``bra'' states $\langle V^{\pm}_{N\sqrt{\beta},\hbox{{\boldmath $t$},{\boldmath $\xi$}}} |$  in (\ref{coherent:bra:R}), and there are two ``ket'' Ramond vacua $|0,\pm\rangle$, one could in principle consider three additional correlators analogous to (\ref{eq:2:24}). However, since
\[
\bra{0,+}{\rm e}^{\frac{2}{\hbar}\xi_0\psi_0}{\rm e}^{-\psi_0\sqrt{\beta}\sum_{a=1}^{N} \frac{\theta_a}{\sqrt{z_a}}}\ket{0,-}
=
\frac{1}{\sqrt 2}\Big(\frac{2}{\hbar}\xi_0 + \sqrt{\beta}\sum_{a=1}^{N} \frac{\theta_a}{\sqrt{z_a}}\Big),
\]
we see that $\langle V^{\pm}_{N\sqrt{\beta},\hbox{{\boldmath $t$},{\boldmath $\xi$}}} |\prod\limits_{a=1}^N \Phi^{-\sqrt{\beta}}(z_a,\theta_a) | 0,\mp \rangle$ (with opposite Ramond vacua chosen on both sides of the correlator) is no longer of the desired form, as in the right hand side of (\ref{eq:2:24}). In what follows we therefore consider only the correlator (\ref{eq:2:24}), with both $\ket{0,+}$ Ramond vacua in both in and out states (the case of both $\ket{0,-}$ vacua is equivalent and does not need to be considered separately).}

In what follows, for an operator $\mathcal{O}$, we also use the notation
\be
\Left<\mathcal{O}\Right> =
\left.\left.\left\langle V^+_{N\sqrt{\beta},\hbox{{\boldmath $t$},{\boldmath $\xi$}}}\right|\mathcal{O}\: \mathsf{Q}_{{\sr R}}^N \right|0,+\right\rangle.   \label{vev-R}
\ee
In particular the partition function (\ref{Z-Ramond}) can be written as $Z=\Left< 1 \Right>$. Using (\ref{TS_comm_screen_R}) with the relations (\ref{positive:GL:on:vacuum}) and (\ref{zero:GL:on:vacuum}) we find
\begin{align}
&
\Left< S_+(x)\Right>+
\frac{Q}{2\hbar x^{3/2}}\left(\xi_0-\frac{\hbar^2}{2}\partial_{\xi_0}\right)Z=
\label{s_p_loop_r}%(\ref{s_p_loop_r})
\\
&=
\int\! d^N\!z\, d^N\!\theta\
\sum\limits_{a=1}^N \left(\theta_a\partial_{z_a} - \partial_{\theta_a}\right)
\left(\sqrt{\frac{z_a}{x}}\frac{1}{x-z_a}\Delta_{\sr R}(\vect{z},\vect{\theta})^{\beta}\,
\mathrm{e}^{-\frac{\sqrt{\beta}}{\hbar}\sum_{b=1}^NV(z_b,{\theta}_b)}\right)
=0,
\nonumber\\
&
\Left< T_+(x)\Right>-\frac{1}{16x^2}\:Z=
\label{t_p_loop_r}%(\ref{t_p_loop_r})
\\
&=
\int\! d^N\!z\, d^N\!\theta\
\sum\limits_{a=1}^N \left(\partial_{z_a} - \frac{x}{2}\partial_{\theta_a}\frac{\theta_a}{z_a(x-z_a)}\right)
\left(\frac{z_a}{x(x-z_a)}\Delta_{\sr R}(\vect{z},\vect{\theta})^{\beta}\,
\mathrm{e}^{-\frac{\sqrt{\beta}}{\hbar}\sum_{b=1}^NV(z_b,{\theta}_b)}\right)
=0.
\nonumber
\end{align}
These equations are interpreted as loop equations in the matrix model language in section \ref{sub-sec-loop-matrix}. We now introduce a representation of the Ramond algebra in terms of operators acting on the partition function as
\be
g_n\,Z =
\Left< G_n \Right>,\qquad
%\int\!  d^N\!z\, d^N\!\theta \
%\left\langle V^+_{N\sqrt{\beta},\hbox{{\boldmath $t$},{\boldmath $\xi$}}}\right|G_n\prod\limits_{a=1}^N\left.\left.\Phi^{-\sqrt{\beta}}(z_a,\theta_a)\right|0,+\right\rangle,  \\
\ell_n\,Z =
\Left< L_n \Right>.   %\label{GnZLnZ}
%\int\!  d^N\!z\, d^N\!\theta \
%\left\langle V^+_{N\sqrt{\beta},\hbox{{\boldmath $t$},{\boldmath $\xi$}}}\right|L_n\prod\limits_{a=1}^N\left.\left.\Phi^{-\sqrt{\beta}}(z_a,\theta_a)\right|0,+\right\rangle.
\ee
Then from the loop equations (\ref{t_p_loop_r}) and (\ref{s_p_loop_r}) we find super-Virasoro constraints
\begin{equation}
g_{n}Z=-\frac{Q}{2\hbar}\delta_{n,0}\Big(\xi_0-\frac{\hbar^2}{2}\partial_{\xi_0}\Big)Z,
\qquad \ell_{n}Z=\frac{1}{16}\delta_{n,0}Z, \qquad n\ge 0,
\label{s_vir_const-LH}
\end{equation}
where
\begin{align}
\begin{split}
g_{n}&=\sum_{m=1}^{\infty}mt_m\partial_{\xi_{m+n}}
+\sum_{m=0}^{\infty}\xi_m\partial_{t_{m+n}}
+\frac{\hbar^2}{2}\partial_{\xi_0}\partial_{t_n}
+\hbar^2\sum_{m=1}^{n}\partial_{\xi_m}\partial_{t_{n-m}} + \\
&\quad -Q\hbar\Big(n+\frac{1}{2}\Big)\partial_{\xi_n} -\frac{Q}{2\hbar}\delta_{n,0}\Big(\xi_0-\frac{\hbar^2}{2}\partial_{\xi_0}\Big),  \label{gn-Ramond-LH}
\end{split}
\end{align}
and
\begin{align}
\begin{split}
\ell_{n}&=
\sum_{m=1}^{\infty}mt_m\partial_{t_{m+n}}
+\sum_{m=0}^{\infty}\Big(m+\frac{n}{2}\Big)\xi_m\partial_{\xi_{m+n}}
+\frac{\hbar^2}{2}\sum_{m=0}^{n}\partial_{t_{n-m}}\partial_{t_{m}} + \\
&\quad +\frac{\hbar^2}{4}n\partial_{\xi_{0}}\partial_{\xi_{n}} +\frac{\hbar^2}{2}\sum_{m=1}^{n-1}m\partial_{\xi_{m-n}}\partial_{\xi_{m}} -\frac{Q\hbar}{2}(n+1)\partial_{t_n}+\frac{1}{16}\delta_{n,0}.   \label{ln-Ramond-LH}
\end{split}
\end{align}

%******************************************************************************************
%******************************************************************************************

\subsection{Wave-function and Ramond-NS super-quantum curves}   \label{ssec-wavefunction-Ramond}

We define the wave-function in the Ramond-NS sector, representing the expression (\ref{chi-R-NS-R}), as the expectation value %(recall (\ref{vev-R}))
\be
\widehat\chi_\alpha(x,\theta) = \Left< \Phi^{\frac{\alpha}{\hbar}}(x,\theta)\Right>
= \left.\left.\left\langle V^+_{N\sqrt{\beta}-\frac{\alpha}{\hbar},\hbox{{\boldmath $t$},{\boldmath $\xi$}}}\right|\Phi^{\frac{\alpha}{\hbar}}(x,\theta)
\mathsf{Q}_{{\sr R}}^N
\right|0,+\right\rangle.  \label{chi-alpha-Ramond}
\ee
We write down this expression more explicitly (after the replacement $\theta \to \eta\sqrt{x}$) in (\ref{chi_hat_w}), where we analyze the eigenvalue model viewpoint.
%It is also useful to consider a decomposition of the wave-function into the bosonic and fermionic components
%\be
%\widehat\chi_\alpha(x,\theta) = \widehat\chi_{{\sr B},\alpha}(x) +  \widehat\chi_{{\sr F},\alpha}(x)\theta.   \label{chiR-BF}
%\ee

One our goal is to find a representation of the Ramond algebra acting on $\widehat\chi_\alpha(x,\theta)$, as well as on components $\widehat\chi_{{\sr B},\alpha}(x)$ and $\widehat\chi_{{\sr F},\alpha}(x)$. To this end we compute expectation values of operators defined in (\ref{TS-Ramond}). We find
\be
\begin{split}
& \Left< \sqrt{y}S(y)\Phi^{\frac{\alpha}{\hbar}}(x,\theta)\Right> =
\Big(\frac{\sqrt{x}}{y-x}\left(\theta\partial_x-\partial_\theta\right)  + \frac{\Delta_{\frac{\alpha}{\hbar}}\theta(y+x)}{\sqrt{x}(y-x)^2}\Big)
\widehat\chi_\alpha(x,\theta) +  \\
& \qquad \qquad +  \frac{1}{\hbar^2} \Big(\widehat h(y)  + \Big(V'_{\sr B}(y)- \frac{Q\hbar}{2 y}\Big)\Big( V_{\sr F}(y) - \frac12\hbar^2\partial_{\xi_0}\Big) + Q\hbar V'_{\sr F}(y)  \Big)\widehat\chi_\alpha(x,\theta),      \label{vev-S-Ramond}
\end{split}
\ee
where
\be
\widehat h(y) = \hbar^2  \sum_{n=0}^\infty y^n \hskip -5pt \sum_{m=n+1}^\infty
\big(\xi_m \partial_{t_{m-n-1}} + m t_m \partial_{\xi_{m-n-1}}\big),    \label{h-hat-Ramond}
\ee
and similarly
\be
\begin{split}
& \Left< yT(y)\Phi^{\frac{\alpha}{\hbar}}(x,\theta)\Right> =
\Big(\frac{x}{y-x}\partial_x + \frac{y\big(\Delta_{\frac{\alpha}{\hbar}} + {\textstyle\frac12}\theta \partial_\theta\big)}{(y-x)^2} + \frac{1}{16y}\Big)  \widehat\chi_\alpha(x,\theta) +  \\
&   \qquad \qquad + \frac{1}{\hbar^2} \Big(  {\widehat f}(y) + \frac12
\Big(y\left(V'_{\sr B}(y)\right)^2 + Qy V''_{\sr B}(y) +
V'_{\sr F}(y)\Big(V_{\sr F}(y)-\frac{\hbar^2}{2}\partial_{\xi_0}\Big) \Big) \Big)
\widehat\chi_\alpha(x,\theta),   \label{vev-T-Ramond}
\end{split}
\ee
where
\be
{\widehat f}(y) = \hbar^2\sum_{n=0}^\infty y^n \hskip -5pt \sum_{m=n+1}^\infty \Big(mt_m\partial_{t_{m-n-1}} + \Big(m-\frac{n+1}{2}\Big)\xi_m\partial_{\xi_{m-n-1}}\Big). \label{f-hat-Ramond}
\ee
Computations leading to the above results are straightforward albeit technically involved, therefore we present them separately in appendices \ref{ssec-S-Ramond} and \ref{ssec-T-Ramond}. However, at this point we stress an important subtlety concerning the character of the expansion of the field $S(y)$ in powers of $y$. The crucial information about this expansion is encoded in the first term on the right hand side of (\ref{vev-S-Ramond}), for which we have
\be
\frac{1}{\sqrt{y}}\left(\frac{\sqrt{x}}{y-x}\left(\theta\partial_x-\partial_\theta\right)  + \frac{\Delta_{\frac{\alpha}{\hbar}}\theta(y+x)}{\sqrt{x}(y-x)^2}\right) = \frac{2\Delta_{\frac{\alpha}{\hbar}}\theta}{(y-x)^2} + \left(\frac{1}{y-x} - \frac{1}{2x}\right)\left(\theta\partial_x-\partial_\theta\right) + \frac{\Delta_{\frac{\alpha}{\hbar}}\theta}{4x^2} + \ldots  \label{Sy-expand}
\ee
with the dots denoting terms vanishing for $y \to x$. On one hand, there is a square root singularity in $y$ for $y\to 0$ or $y\to\infty$, which is a manifestation of the fact that at these points we inserted Ramond vacua. On the other hand, for $y\to x$ the above expression has (as follows from its right hand side) the second order pole, which means that the operator inserted at $x$ is identified as a Neveu-Schwarz operator. Furthermore, from the presence of this second order pole, and since in general the expansion of the supercurrent is of the form $S(y) = \sum_k G_k(x) (y-x)^{-k-3/2}$, we deduce that the summation variable $k$ must take half-integer values, and therefore the modes $G_k$ are relevant for the Neveu-Schwarz sector. Expanding then (\ref{vev-S-Ramond}) around $y=x$ and taking advantage of (\ref{Sy-expand}) we identify the representation of the Ramond algebra acting on the wave-function
\begin{align}
\nonumber
\widehat{G}_{\frac12} \widehat\chi_\alpha(x,\theta) & = \Left< G_{\frac12}\cdot\Phi^{\frac{\alpha}{\hbar}}(x,\theta)\Right>
= 2\Delta_{\frac{\alpha}{\hbar}}\theta\, \widehat\chi_\alpha(x,\theta),  \\
\widehat{G}_{-\frac12} \widehat\chi_\alpha(x,\theta) & = \Left< G_{-\frac12}\cdot\Phi^{\frac{\alpha}{\hbar}}(x,\theta)\Right>
= \left(\theta\partial_x-\partial_\theta\right)\widehat\chi_\alpha(x,\theta),  \\
\nonumber
\widehat{G}_{-\frac32} \widehat\chi_\alpha(x,\theta) & =\Left< G_{-\frac32}\cdot\Phi^{\frac{\alpha}{\hbar}}(x,\theta)\Right> =  \Big(\frac{\Delta_{\frac{\alpha}{\hbar}}\theta}{4x^2} - \frac{1}{2x}\left(\theta\partial_x-\partial_\theta\right)\Big)\widehat\chi_\alpha(x,\theta) +  \\
\nonumber
& \quad +  \frac{x^{-1/2}}{\hbar^2}\Big( \Big(V'_{\sr B}(x)- \frac{Q\hbar}{2 x}\Big)\Big(V_{\sr F}(x) - \frac12\hbar^2\partial_{\xi_0}\Big) + Q\hbar V'_{\sr F}(x)+ \widehat h(x) \Big)\widehat\chi_\alpha(x,\theta).
\end{align}
%For $\widehat\chi_\alpha(x,\theta) = \widehat\chi_{{\sr B},\alpha}(x) +  \widehat\chi_{{\sr F},\alpha}(x)\theta$, as defined in (\ref{chiR-BF}), we thus get
%\begin{align}
%\nonumber
%\widehat{\sf G}_{\frac12}\widehat\chi_{{\sr B},\alpha}(x) & = 0,  \\
%\widehat{\sf G}_{-\frac12}\widehat\chi_{{\sr B},\alpha}(x) & = \widehat\chi_{{\sr F},\alpha}(x),   \\
%\nonumber
%\widehat{\sf G}_{-\frac32}\widehat\chi_{{\sr B},\alpha}(x) & = -\frac{1}{2x}\widehat\chi_{{\sr F},\alpha}(x)  +  \\
%\nonumber
%& \quad +  \frac{x^{-1/2}}{\hbar^2}\Big( \Big(V'_{\sr B}(x)- \frac{Q\hbar}{2 x}\Big)\Big(V_{\sr F}(x) - \frac12\hbar^2\partial_{\xi_0}\Big) + Q\hbar V'_{\sr F}(x)+ \widehat h(x) \Big)\widehat\chi_{{\sr B},\alpha}(x),
%\end{align}
%and
%\begin{align}
%\nonumber
%\widehat{\sf G}_{\frac12}\widehat\chi_{{\sr F},\alpha}(x) & =  2\Delta_{\frac{\alpha}{\hbar}}\widehat\chi_{{\sr B},\alpha}(x),  \\
%\widehat{\sf G}_{-\frac12}\widehat\chi_{{\sr F},\alpha}(x) & = \partial_x\widehat\chi_{{\sr B},\alpha}(x),   \\
%\nonumber
%\widehat{\sf G}_{-\frac32}\widehat\chi_{{\sr F},\alpha}(x) & = \Big(\frac{\Delta_{\frac{\alpha}{\hbar}}}{4x^2} -\frac{1}{2x}\partial_x\Big)\widehat\chi_{{\sr B},\alpha}(x)  + \\
%\nonumber
%&  \quad +  \frac{x^{-1/2}}{\hbar^2}\Big( \Big(V'_{\sr B}(x)- \frac{Q\hbar}{2 x}\Big)\Big(V_{\sr F}(x) - \frac12\hbar^2\partial_{\xi_0}\Big) + Q\hbar V'_{\sr F}(x)+ \widehat h(x) \Big)\widehat\chi_{{\sr F},\alpha}(x).
%\end{align}
Similarly, expanding (\ref{vev-T-Ramond}) around $y=x$  we get in particular
\begin{align}
\nonumber
\widehat{L}_0 \widehat\chi_\alpha(x,\theta) & = \Left< L_0\cdot \Phi^{\frac{\alpha}{\hbar}}(x,\theta)\Right>
= \Big(\Delta_{\frac{\alpha}{\hbar}} + \frac12\theta \partial_\theta\Big)\widehat\chi_\alpha(x,\theta),   \\
\widehat{L}_{-1} \widehat\chi_\alpha(x,\theta) & = \Left< L_{-1}\cdot \Phi^{\frac{\alpha}{\hbar}}(x,\theta)\Right> =
\partial_x\widehat\chi_\alpha(x,\theta),  \\
\nonumber
\widehat{L}_{-2} \widehat\chi_\alpha(x,\theta) & = \Left< L_{-2}\cdot \Phi^{\frac{\alpha}{\hbar}}(x,\theta)\Right>
=  \Big(\frac{1}{16x^2} -\frac{1}{x}\partial_x\Big) \widehat\chi_\alpha(x,\theta)  + \\
\nonumber
+ \frac{1}{\hbar^2} &\left\{ \frac{1}{x}{\widehat f}(x)
+ \frac12
\left(\left(V'_{\sr B}(x)\right)^2 + Q\hbar V''_{\sr B}(x) +
\frac{1}{x}V'_{\sr F}(x)\left(V_{\sr F}(x)-\frac{\hbar^2}{2}\partial_{\xi_0}\right)
\right)
\right\}
\widehat\chi_\alpha(x,\theta).
\end{align}
%and for $\widehat\chi_\alpha(x,\theta) = \widehat\chi_{{\sr B},\alpha}(x) +  \widehat\chi_{{\sr F},\alpha}(x)\theta$, as defined in (\ref{chiR-BF}), we find
%\begin{align}
%\nonumber
%\widehat{\sf L}_{0}\widehat\chi_{{\sr B},\alpha}(x) & =  \Delta_{\frac{\alpha}{\hbar}} \chi_{{\sr B},\alpha}(x),
%\\
%\nonumber
%\widehat{\sf L}_{0}\widehat\chi_{{\sr F},\alpha}(x) & =  \left(\Delta_{\frac{\alpha}{\hbar}}+{\textstyle\frac12}\right) \chi_{{\sr F},\alpha}(x),
%\\
%\widehat{\sf L}_{-1}\widehat\chi_{{\sr B/F},\alpha}(x) & =  \partial_x\chi_{{\sr B/F},\alpha}(x),  \\
%\nonumber
%\widehat{\sf L}_{-2}\widehat\chi_{{\sr B/F},\alpha}(x)
%& =
%\Big(\frac{1}{16x^2} -\frac{1}{x}\partial_x\Big)\chi_{{\sr B/F},\alpha}(x) + \\
%\nonumber
%& +  \frac{1}{\hbar^2} \left\{ \frac{1}{x}{\widehat f}(x)
%+ \frac12
%\left(\left(V'_{\sr B}(x)\right)^2 + QV''_{\sr B}(x) +
%\frac{1}{x}V'_{\sr F}(x)\left(V_{\sr F}(x)-\frac{\hbar^2}{2}\partial_{\xi_0}\right)
%\right)
%\right\}
%\chi_{{\sr B/F},\alpha}(x).
%\end{align}

It then follows that super-quantum curves will take form of Neveu-Schwarz singular vectors (\ref{NS:singular:opeartors:examples}), expressed in terms of the representation of super-Virasoro generators given above. We present an explicit example of a super-quantum curve at level 3/2 in section \ref{ssec-superquantum-32}, and yet more specific example of the multi-Penner model in section \ref{ssec-Ramond-NS-Penner}.

%******************************************************************************************
%******************************************************************************************
%******************************************************************************************
%******************************************************************************************

\section{Ramond-R super-quantum curves}    \label{sec-Ramond-R}

In this section we introduce wave-functions and derive super-quantum curves of the Ramond-R type, which have the structure of Ramond singular vectors, such as those in (\ref{singular-Ramond}). To define such curves we need to consider wave-functions defined in terms of expectation values of $x$-dependent operators $R(x)$ of Ramond type, presented schematically in (\ref{chi-NS-R-R})
\be
\widehat{\chi}_{\alpha}(x) \sim \langle NS|\, R(x)\, |R\rangle.       \label{chi-NS-R-R-bis}
\ee
It is most natural to define such Ramond fields $R(x)$ by taking advantage of ``chiral spin fields'' (or ``twist fields'') $\sigma_{\pm}(x)$  discussed  in \cite{Poghosian:1996dw,Chorazkiewicz:2011zd}.
Therefore we first discuss properties of such fields, and subsequently define the whole correlator (\ref{chi-NS-R-R-bis}).

In fact, a definition of wave-functions and quantum curves of the Ramond-R type is much more involved than in previous examples. For this reason we do not discuss models with generic potentials in the Ramond-R sector, but focus our considerations on the Penner-like potentials. In principle, an arbitrary potential could be presented as a combination of various Penner-like potentials, so in this sense our results provide a basic building block of super-quantum curves corresponding to more general potentials. Furthermore, we show that the Ramond-R super-quantum curve with the Penner-like potential reduces to a supersymmetric version of the BPZ equation in conformal field theory, which is a nice confirmation of our formalism. We rederive these results in section \ref{sec-matrix-Ramond-R} using solely techniques of eigenvalue models, which proves that both approaches are equivalent.

%******************************************************************************************

\subsection{Chiral spin fields}

As indicated above, to start with we summarize properties of chiral spin fields. These fields have scaling dimension $\frac{1}{16}$ and can be defined by the OPE with the chiral fermion $\psi(z)$
\begin{align}
\label{OPE:psi:sigma}
\psi(z)\sigma_{\pm}(w) \sim {\rm e}^{\mp \frac{i\pi}{4}}\frac{\sigma_{\mp}(w)}{\sqrt{2(z-w)}},
\end{align}
and the braiding property
\begin{align}
\label{bariding:psi:sigma}
\sqrt{w-z}\,\sigma_{\pm}(w)\psi(z)
=
\pm i\sqrt{z-w}\,\psi(z)\sigma_{\pm}(w).
\end{align}
It follows from (\ref{OPE:psi:sigma}) that if we define the states $\ket{\sigma_\pm}$ via an action of the chiral spin fields on the Neveu-Schwarz vacuum $\ket{0}$
\[
\ket{\sigma_\pm} = \lim\limits_{w \to 0} \sigma_{\pm}(w)\ket{0},
\]
then
\[
\psi_0\ket{\sigma_\pm} = \frac{{\rm e}^{\mp\frac{i\pi}{4}}}{\sqrt 2}\ket{\sigma_\mp},
\hskip 1cm
\bra{\sigma_\pm}\psi_0 = \frac{{\rm e}^{\pm\frac{i\pi}{4}}}{\sqrt 2}\bra{\sigma_\mp}.
\]
These relations enable to identify the Ramond vacua as $\ket{\sigma_+} = \ket{0,+}$ and $\ket{\sigma_-} = {\rm e}^{\frac{i\pi}{4}}\ket{0,-}.$
The reason for introducing this phase shift is that the one-point correlation functions of the $\sigma_{\pm}$ fields are equal and thus can be both normalized to be unity
\begin{equation}
\label{sigma:normalization}
\big\langle 0\big|\sigma_{\pm}(1)\big|\,\sigma_\pm\,\big\rangle = 1.
\end{equation}

In order to determine the $x$-dependence of the correlators $\big\langle 0\big|\sigma_{\pm}(x)\big|\,\sigma_\pm\,\big\rangle$
we use the fact that with respect to the Virasoro algebra $\sigma_{\pm}(x)$ are primary fields of scaling dimension $\frac{1}{16}$
\begin{equation}
\label{sigmas:as:primary:fields}
\left[L_m,\sigma_{\pm}(x)\right] = x^m\left(x\partial_x + \frac{1}{16}(m+1)\right)\sigma_{\pm}(x).
\end{equation}
Therefore, since $\bra{0}L_1 = 0$ and $L_1\ket{\sigma_\pm} = 0$, we get from (\ref{sigmas:as:primary:fields}):
\begin{align}
0=
\big\langle 0\big|L_1\sigma_{\pm}(x)\big|\,\sigma_\pm\,\big\rangle
\; = \;
\big\langle 0\big|\left[L_1,\sigma_{\pm}(x)\right]\big|\,\sigma_\pm\,\big\rangle
\; = \;
x\left(x\partial_x + \frac18\right)\big\langle 0\big|\sigma_{\pm}(x)\big|\,\sigma_\pm\,\big\rangle
\nonumber
\end{align}
which, in view of (\ref{sigma:normalization}), yields
\begin{equation}
\label{one:point:sigma}
\big\langle 0\big|\sigma_{\pm}(x)\big|\,\sigma_\pm\,\big\rangle
= x^{-1/8}.
\end{equation}

To calculate the one-point correlation functions of the fermion field $\psi(x)$ in the presence of the spin fields let us consider the functions
\begin{align}
\label{sigma:psi:sigma:Ramond}
s_{\pm}(z)=
\sqrt{z(x-z)}\big\langle\,0\,\big|\sigma_{\mp}(x)\psi(z)\big|\,\sigma_\pm\,\big\rangle
\; = \;
\mp i \sqrt{z(z-x)}\big\langle\,0\,\big|\psi(z)\sigma_{\mp}(x)\big|\,\sigma_\pm\,\big\rangle.
\end{align}
By (\ref{OPE:psi:sigma}), $s_{\pm}(z)$ is a holomorphic function of $z$ with the only possible singularities (poles) at 0, $x$ and infinity. Since
\[
s_{\pm}(z) \; \sim \;
\left\{
\begin{array}{lcl}
{\rm e}^{\mp\frac{i\pi}{4}}\,\sqrt{\frac{x}{2}}\,\big\langle 0\big|\sigma_{\mp}(x)\big|\,\sigma_\mp\,\big\rangle & \; {\rm for}\; & z \to 0,
\\[8pt]
\mp i {\rm e}^{\pm\frac{i\pi}{4}}\,\sqrt{\frac{x}{2}}\,\big\langle 0\big|\sigma_{\pm}(x)\big|\,\sigma_\pm\,\big\rangle  & \; {\rm for}\; & z \to x,
\\[8pt]
\mp i \big\langle 0\big|\psi_{\frac12}\sigma_{\mp}(x)\big|\,\sigma_\pm\,\big\rangle & \; {\rm for}\; & z \to \infty,
\end{array}
\right.
\]
this function is actually a ($x$-dependent) constant
\[
s_{\pm}(z)
=
{\rm e}^{\mp\frac{i\pi}{4}}\,\sqrt{\frac{x}{2}}\,\big\langle 0\big|\sigma_{\mp}(x)\big|\,\sigma_\mp\,\big\rangle
=
{\rm e}^{\mp\frac{i\pi}{4}}\,\sqrt{\frac{x}{2}}\,\big\langle 0\big|\sigma_{\pm}(x)\big|\,\sigma_\pm\,\big\rangle.
\]
In view of  (\ref{one:point:sigma}) we obtain the one-point functions of the fermion field
\begin{align}
\begin{split}
\big\langle 0\big|\sigma_{\mp}(x)\psi(z)\big|\,\sigma_\pm\,\big\rangle
& =
\frac{{\rm e}^{\mp \frac{i\pi}{4}}}{\sqrt 2} \sqrt{\frac{x}{z}}\frac{1}{\sqrt{x-z}}\, x^{-1/8}, \label{one:point:sigma-bis} \\
\big\langle 0\big|\psi(z)\sigma_{\mp}(x)\big|\,\sigma_\pm\,\big\rangle
& =
\frac{{\rm e}^{\pm \frac{i\pi}{4}}}{\sqrt 2} \sqrt{\frac{x}{z}}\frac{1}{\sqrt{z-x}}\, x^{-1/8}.
\end{split}
\end{align}
Similar technique allows to calculate higher point correlation functions of the fermion fields. For example, for the two-point functions we get
\begin{align}
\begin{split}
\big\langle\,0\,\big|\sigma_{\pm}(x)\psi(w)\psi(z)\big|\,\sigma_\pm\,\big\rangle
& =
\frac12\left(\sqrt{\frac{z(x-w)}{w(x-z)}}+\sqrt{\frac{w(x-z)}{z(x-w)}}\right)\frac{x^{-1/8}}{w-z},
\\
\big\langle\,0\,\big|\psi(w)\sigma_{\pm}(x)\psi(z)\big|\,\sigma_\pm\,\big\rangle
& =
\pm\frac{i}{2}\left(\sqrt{\frac{z(w-x)}{w(x-z)}}-\sqrt{\frac{w(x-z)}{z(w-x)}}\right)\frac{x^{-1/8}}{w-z},
\\
\big\langle\,0\,\big|\psi(w)\psi(z)\sigma_{\pm}(x)\big|\,\sigma_\pm\,\big\rangle
& =
\frac12\left(\sqrt{\frac{z(w-x)}{w(z-x)}}+\sqrt{\frac{w(z-x)}{z(w-x)}}\right)\frac{x^{-1/8}}{w-z}.
\end{split}
\end{align}
The result for the higher point correlation functions can be summarized by the following version of the Wick theorem\footnote{The chiral fermion $\psi(z)$ and the fermionic variable $\theta$ anticommute $\{\psi(z), \theta\}=0$. We assign the commutation relation $[\sigma_+(x), \theta]=0$ between $\sigma_+(x)$ and $\theta$, and then, from the definition \eqref{OPE:psi:sigma} of the chiral spin fields, it follows that $\sigma_-(x)$ and $\theta$ anticommute $\{\sigma_-(x), \theta\}=0$.}
\begin{align}
\label{Wick}
\big\langle\,0\,\big|\sigma_{\pm}(x)\prod\limits_{a=1}^N{\rm e}^{\psi(z_a)\theta_a}\big|\,\sigma_\pm\,\big\rangle
& =
\big\langle 0\big|\sigma_{+}(x)\big|\,\sigma_+\,\big\rangle
\exp\Big( -\sum\limits_{a<b}\big\langle \sigma_{+}(x)\psi(z_a)\psi(z_b)\big\rangle\,\theta_a\theta_b \Big), \nonumber \\
\big\langle\,0\,\big|\sigma_{\pm}(x)\prod\limits_{a=1}^N{\rm e}^{\psi(z_a)\theta_a}\big|\,\sigma_\mp\,\big\rangle
& =
\mp\sum\limits_{a=1}^N \theta_a\big\langle\,0\,\big|\sigma_{\pm}(x)\psi(z_a)\big|\,\sigma_\mp\,\big\rangle  \times \\
& \qquad \qquad \qquad \times
\exp\Big(-\sum\limits_{a<b}\big\langle \sigma_{+}(x)\psi(z_a)\psi(z_b)\big\rangle\,\theta_a\theta_b \Big), \nonumber
\end{align}
where
\begin{equation}
\label{normalized:two:point:function}
\big\langle \sigma_{+}(x)\psi(z_a)\psi(z_b)\big\rangle
=
\frac{\big\langle\,0\,\big|\sigma_{+}(x)\psi(z_a)\psi(z_b)\big|\,\sigma_+\,\big\rangle}{\big\langle 0\big|\sigma_{+}(x)\big|\,\sigma_+\,\big\rangle}
= \frac{  \sqrt{\frac{z_a(x-z_b)}{z_b(x-z_a)}}+\sqrt{\frac{z_b(x-z_a)}{z_a(x-z_b)}} }{2(z_a-z_b)}.
\end{equation}
Modifications appearing in the case when some of the fermions are to the left and some to the right of the spin field $\sigma_\pm(x)$ are straightforward.

%******************************************************************************************

\subsection{Ramond-R wave-function}

The key in the current construction is the field denoted schematically $R(x)$ in (\ref{chi-NS-R-R-bis}). More precisely, we introduce a pair of such fields, which are Ramond chiral primary fields with conformal dimension $\frac12\alpha(\alpha - Q) + \frac{1}{16}\equiv \Delta_\alpha + \frac{1}{16}$, and which we define as
\begin{equation}
\label{Ramond:chiral:fields}
{\sf R}^\alpha_{\pm}(x) = {\sf E}^\alpha(x)\sigma_{\pm}(x),
\end{equation}
with ${\sf E}^\alpha(x)$ given in (\ref{Phi-Ramond}). Furthermore, we introduce a correlator, whose integrated form represents the Ramond-R eigenvalue model
\be
\label{PsiR:definition}
{\Psi}_{\pm}(x,\vect{z},\vect{\theta}) = \bra{\alpha_0}{\sf R}_{\pm}^{\frac{\alpha}{\hbar}}(x)\Phi^{\frac{\gamma}{\hbar}}(w,\eta)\prod\limits_{a=1}^N\Phi^{-\sqrt{\beta}}(z_a,\theta_a)\ket{\sigma_+}, \qquad
\frac{\alpha  +\gamma}{\hbar} - N\sqrt{\beta} = Q-\frac{\alpha_0}{\hbar}.
\ee
Let us clarify the structure of this correlator. First, the insertion of ${\sf R}^{\frac{\alpha}{\hbar}}_{\pm}(x)$ is analogous to a determinant-like insertion in a Virasoro (non-supersymmetric) matrix model, and introduces a dependence of the wave-function on $x$. Second, $\Phi^{\frac{\gamma}{\hbar}}(w,\eta)$ gives rise to the Penner-like potential (and it plays a role analogous to more general references states, such as (\ref{bra-Virasoro}) or (\ref{coherent:bra:NS}) in other models that we considered, which gave rise to more general potentials). Third, a series of fields $\Phi^{-\sqrt{\beta}}(z_a,\theta_a)$ introduces, as usual, the structure of the eigenvalue model and gives rise to the Vandermonde-like determinant. To simplify calculations, we assume the presence of a chiral field at infinity and define
\be
\bra{\alpha_0} = \bra{0}{\rm e}^{\frac{\alpha_0}{\hbar}{\sf q}}
=
\lim_{z\to\infty}z^{-\frac{\alpha_0}{\hbar}\left(\frac{\alpha_0}{\hbar}-Q\right)}\bra{0}\Phi^{\frac{\alpha_0}{\hbar}}(z,\theta).
\ee
Generalization of the above model to the multi-Penner case can be achieved by replacing in the correlator (\ref{PsiR:definition})
$\Phi^{\frac{\gamma}{\hbar}}(w,\eta)$ with $\prod_{i=1}^M \Phi^{\frac{\gamma_i}{\hbar}}(w_i,\eta_i)$ with $M > 1$, and more general potentials can be introduced by considering yet more general insertions.

Let us now evaluate the correlator (\ref{PsiR:definition}). Using (\ref{Wick}) and the normal ordering formula for the bosonic fields ${\sf E}$ we get its explicit form, which can be presented as
\be
{\Psi}_{\pm}(x,\vect{z},\vect{\theta}) =
\Theta_{\pm}(x)\, (x-w)^{\frac{\alpha\gamma}{\hbar^2}}\,
%\prod\limits_{a=1}^N(x-z_a)^{-\frac{\alpha\sqrt{\beta}}{\hbar}}\,
\Delta_{{\sr R},x}(\vect{z},\vect{\theta})^\beta\,
{\rm e}^{-\frac{\sqrt{\beta}}{\hbar}\sum_{a=1}^N
\left(V_{{\sr B},x}(z_a) +V_{{\sr F},x}(z_a)\theta_a\right)},   \label{Ramond-R-eigenvalue}
\ee
where
\begin{align}
\begin{split}
\Delta_{{\sr R},x}(\vect{z},\vect{\theta}) &=  \label{Vandermonde-Ramond-R}
\prod\limits_{a<b}(z_a-z_b)\, {\rm e}^{-\big\langle\sigma_{+}(x)\psi(z_a)\psi(z_b)\big\rangle\,\theta_a\theta_b} = \\
& =  \prod\limits_{a<b}
\left(z_a-z_b -\left(\sqrt{\frac{z_a(x-z_b)}{z_b(x-z_a)}}+\sqrt{\frac{z_b(x-z_a)}{z_a(x-z_b)}}\right)\frac{\theta_a\theta_b}{2}\right)
\end{split}
\end{align}
can be viewed as a Ramond version of the Vandermonde determinant, and the analog of the Penner potential takes form
\begin{equation}
V_{{\sr B},x}(z_a) + V_{{\sr F},x}(z_a)\theta_a,
\end{equation}
where
\begin{align}
\begin{split}
\label{Ramond:superpotential}
V_{{\sr B},x}(z) & =  \alpha\log(x-z)+\gamma\log(z-w),  \\
V_{{\sr F},x}(z) & =
-\gamma\big\langle\sigma_{+}(x)\psi(w)\psi(z)\big\rangle\eta  = \frac{\gamma\eta}{2(z-w)}\left(\sqrt{\frac{w(x-z)}{z(x-w)}}+ \sqrt{\frac{z(x-w)}{w(x-z)}}\right).
\end{split}
\end{align}
Furthermore, we also denoted
\begin{align}
\begin{split}
\Theta_+(x) & =  \big\langle 0\big|\sigma_{+}(x)\big|\,\sigma_{+}\,\big\rangle = x^{-1/8},  \\
\Theta_-(x) & =
\frac{\gamma\eta}{\hbar}\big\langle\,0\,\big|\sigma_{-}(x)\psi(w)\big|\,\sigma_{+}\,\big\rangle -\sqrt{\beta} \sum\limits_{a=1}^N \theta_a\big\langle\,0\,\big|\sigma_{-}(x)\psi(z_a)\big|\,\sigma_{+}\,\big\rangle = \\
&=
\frac{{\rm e}^{-\frac{i\pi}{4}}}{\sqrt{2}}\left(
\frac{\gamma\eta}{\hbar}\frac{\sqrt{x}}{\sqrt{w(x-w)}}
-\sqrt{\beta}\sum\limits_{a=1}^N \frac{\sqrt{x}\theta_a}{\sqrt{z_a(x-z_a)}}
\right)x^{-1/8},
\label{ramond_theta_def}
\end{split}
\end{align}
where the explicit forms of $\Theta_+(x)$ and $\Theta_-(x)$ are derived in (\ref{one:point:sigma}) and (\ref{one:point:sigma-bis}), respectively.
Introducing yet another fermionic variable $\xi$ we can combine ${\Psi}_{\pm}(x,\vect{z},\vect{\theta})$ into a single super-function
\be
{\Psi}(x,\xi,\vect{z},\vect{\theta}) = {\Psi}_{+}(x,\vect{z},\vect{\theta}) + \frac{\sqrt{2}}{\hbar}{\rm e}^{\frac{i\pi}{4}}\,\xi\,{\Psi}_{-}(x,\vect{z},\vect{\theta}). \label{psi-Ramond-R-1}
\ee
This allows to express the Ramond-R wave-function for the one-Penner potential $\chi^{\sr R}_{\alpha}(x,\xi)$ which we discuss in section \ref{sec-matrix-Ramond-R} in the form
\be
\chi^{\sr R}_{\alpha}(x,\xi) = x^{1/8} (x-w)^{-\frac{\alpha\gamma}{\hbar^2}}
{\rm e}^{-\frac{\gamma\xi\eta}{\hbar^2}\frac{\sqrt{x}}{\sqrt{w(x-w)}}}\,\int\!d^N\!z\, d^N\!\theta\ {\Psi}(x,\xi,\vect{z},\vect{\theta}). \label{psi-Ramond-R-2}
\ee

%******************************************************************************************

\subsection{Ramond-R super-quantum curves}\label{subsec_rr_sq}

Analogously as in other cases, super-quantum curves of the Ramond-R type are differential equations satisfied by integrated form of $\Psi_{\pm}(x,\vect{z},\vect{\theta})$. In general they can be obtained by constructing null vectors in the Ramond Verma module. As an example, we consider correlation functions of the field
(cf.\ the first equation in (\ref{singular-Ramond}))
\begin{equation}
\label{null:operator}
\left(\frac{\alpha}{\hbar}\left(\frac{2\alpha}{\hbar}-Q\right)G_{-1} - 2L_{-1}G_0\right)\cdot {\sf R}_\pm^{\frac{\alpha}{\hbar}}(x).
\end{equation}
Such a correlator should vanish for $\alpha = \frac{\hbar Q}{2}$, $\alpha = \frac{\hbar\sqrt{\beta}}{2}$, and $\alpha = -\frac{\hbar}{2\sqrt\beta}.$ The second and the third case leads to a pair of the first order differential equations derived below.

We first denote
\begin{align}
\begin{split}
s_{\pm}(y) & =  \sqrt{y}\sqrt{y-x}\bra{\alpha_0}S(y) {\sf R}_\pm^{\frac{\alpha}{\hbar}}(x)\Phi^{\frac{\gamma}{\hbar}}(w,\eta)
\prod\limits_{a=1}^N\Phi^{-\sqrt{\beta}}(z_a,\theta_a)\ket{\sigma_+} =\\
& = \mp i\sqrt{y}\sqrt{x-y}\bra{\alpha_0} {\sf R}_\pm^{\frac{\alpha}{\hbar}}(x)S(y)\Phi^{\frac{\gamma}{\hbar}}(w,\eta)
\prod\limits_{a=1}^N\Phi^{-\sqrt{\beta}}(z_a,\theta_a)\ket{\sigma_+}.
\end{split}
\end{align}

Since, as it follows from the mode expansion of the current $S(y)$ around the fields and states present in the correlation functions $s_{\pm}(y)$
as well as the commutation formulae (\ref{TS_comm_screen_R_sp}) and (\ref{TS_comm_screen_R}),
\begin{align}
\begin{split}
\bra{\alpha_0}S(y) &  \sim  y^{-2},
\\
\sqrt{y-x}S(y)\,{\sf R}_\pm^{\frac{\alpha}{\hbar}}(x)
& \sim
\frac{1}{y-x} G_0\cdot {\sf R}_\pm^{\frac{\alpha}{\hbar}}(x)
= \frac{\frac{\alpha}{\hbar}-Q/2}{y-x}\frac{{\rm e}^{\mp\frac{i\pi}{4}}}{\sqrt{2}}\, {\sf R}_\mp^{\frac{\alpha}{\hbar}}(x),
\\
S(y)\Phi^{\frac{\gamma}{\hbar}}(w,\eta)
& \sim
\left(\frac{2\Delta_{\frac{\gamma}{\hbar}}\eta}{(y-w)^2} + \frac{\eta\partial_w - \partial_\eta}{(y-w)}\right)\Phi^{\frac{\gamma}{\hbar}}(w,\eta),
\\
S(y)\Phi^{-\sqrt{\beta}}(z_a,\theta_a)
& \sim  \left(\frac{\theta_a}{(y-z_a)^2} + \frac{\theta_a\partial_{z_a} - \partial_{\theta_a}}{(y-z_a)}\right)\Phi^{-\sqrt{\beta}}(z_a,\theta_a)
 =  \left(\theta_a\partial_{z_a} - \partial_{\theta_a}\right) \frac{\Phi^{-\sqrt{\beta}}(z_a,\theta_a)}{y-z_a},
\\
\sqrt{y}\,S(y)\ket{\sigma_+} & \sim \frac{1}{y} G_0\ket{\sigma_+} \; = \; -\frac{Q}{2}\frac{{\rm e}^{-\frac{i\pi}{4}}}{\sqrt{2}\,y}\,\ket{\sigma_-},
\end{split}
\end{align}
we see that $s_\pm(y)$ are meromorphic functions, vanishing for $y \to \infty$ and having first order poles with
locations and residues which can be read off
from the equations above. $s_\pm(y)$ can be thus written as a sum of the pole terms and, taking into account the equalities
\begin{align}
\begin{split}
\bra{\alpha_0}{\sf R}_{-}^{\frac{\alpha}{\hbar}}(x)\Phi^{\frac{\gamma}{\hbar}}(w,\eta)\prod\limits_{a=1}^N\Phi^{-\sqrt{\beta}}(z_a,\theta_a)\ket{\sigma_-}
& =
{\Psi}_{+}(x,\vect{z},\vect{\theta}),
\\
\bra{\alpha_0}{\sf R}_{+}^{\frac{\alpha}{\hbar}}(x)\Phi^{\frac{\gamma}{\hbar}}(w,\eta)\prod\limits_{a=1}^N\Phi^{-\sqrt{\beta}}(z_a,\theta_a)\ket{\sigma_-}
& =
-i{\Psi}_{-}(x,\vect{z},\vect{\theta}),
\nonumber
\end{split}
\end{align}
which follow from (\ref{Wick}), we get
\begin{align}
&\bra{\alpha_0}S(y){\sf R}_\pm^{\frac{\alpha}{\hbar}}(x)\Phi^{\frac{\gamma}{\hbar}}(w,\eta)\prod\limits_{a=1}^N\Phi^{-\sqrt{\beta}}(z_a,\theta_a)\ket{\sigma_+} = \nonumber \\
& \quad =
\frac12\left( (\frac{2\alpha}{\hbar} - Q)\sqrt{\frac{x}{2y(y-x)^3}} \pm Q\sqrt{\frac{x}{2 y^3(y-x)}} \right) {\rm e}^{\mp\frac{i\pi}{4}}\, {\Psi}_{\mp}(x,\vect{z},\vect{\theta}) +
\nonumber\\
& \qquad - i\sqrt{\frac{w(x-w)}{y(y-x)}}\left(\frac{2\Delta_{\frac{\gamma}{\hbar}}\eta}{(y-w)^2}
+ \frac{1}{y-w}\left(\left(\frac{1}{w}-\frac{1}{x-w}\right)\Delta_{\frac{\gamma}{\hbar}}\eta + \eta\partial_w - \partial_\eta\right)\right) {\Psi}_{\pm}(x,\vect{z},\vect{\theta}) + \nonumber \\
& \qquad -  i\sum\limits_{a=1}^N \left(\theta_a\partial_{z_a} - \partial_{\theta_a}\right)\left(\sqrt{\frac{z_a(x-z_a)}{y(y-x)}}\frac{{\Psi}_{\pm}(x,\vect{z},\vect{\theta})}{y-z_a}\right).
\label{S:R:correlator}
\end{align}

Using (\ref{S:R:correlator}) and the definition
\be
S(y){\sf R}_\pm^{\alpha}(x) = \sum\limits_{m\in\mathbb Z} \frac{1}{(y-x)^{m+\frac32}}\,G_m\cdot {\sf R}_\pm^{\alpha}(x),
\ee
we get in particular
\be
\label{G0:on:R}
\bra{\alpha_0}G_0\cdot {\sf R}_\pm^{\frac{\alpha}{\hbar}}(x)\Phi^{\frac{\gamma}{\hbar}}(w,\eta)\prod\limits_{a=1}^N\Phi^{-\sqrt{\beta}}(z_a,\theta_a)\ket{\sigma_+}
= \frac{\frac{\alpha}{\hbar}-Q/2}{\sqrt{2}}{\rm e}^{\mp \frac{i\pi}{4}}\,\Psi_{\mp}(x,\vect{z},\vect{\theta}),
\ee
and
\begin{align}
\label{G:minus1:on:R}
& \bra{\alpha_0}G_{-1}\cdot {\sf R}_\pm^{\frac{\alpha}{\hbar}}(x)\Phi^{\frac{\gamma}{\hbar}}(w,\eta)\prod\limits_{a=1}^N\Phi^{-\sqrt{\beta}}(z_a,\theta_a)\ket{\sigma_+}
= -\frac{\frac{\alpha}{\hbar}-Q/2\mp Q}{2\sqrt{2}\,x} {\rm e}^{\mp \frac{i\pi}{4}}\,\Psi_{\mp}(x,\vect{z},\vect{\theta}) + \nonumber\\
& \quad - i\sqrt{\frac{w}{x(x-w)}}\left(\frac{\Delta_{\frac{\gamma}{\hbar}}\eta\,x}{w(x-w)} +
\eta\partial_w - \partial_\eta\right){\Psi}_{\pm}(x,\vect{z},\vect{\theta}) + \nonumber  \\
& \quad -
i\sum\limits_{a=1}^N \left(\theta_a\partial_{z_a} - \partial_{\theta_a}\right)\left(\sqrt{\frac{z_a(x-z_a)}{x}}\frac{{\Psi}_{\pm}(x,\vect{z},\vect{\theta})}{x-z_a}\right).
\end{align}
Defining finally
\begin{equation}
{\widehat \chi}^{\sr R}_{\pm,\alpha}(x) = \int\! d^N\!z\, d^N\!\theta\,
\bra{\alpha_0}{\sf R}_\pm^\frac{\alpha}{\hbar}(x)\Phi^{\frac{\gamma}{\hbar}}(w,\eta)\prod\limits_{a=1}^N\Phi^{-\sqrt{\beta}}(z_a,\theta_a)\ket{\sigma_+},
\end{equation}
and
\begin{equation}
{\widehat G}_m {\widehat \chi}^{\sr R}_{\pm,\alpha}(x) = \int\! d^N\!z\, d^N\!\theta\,\bra{\alpha_0}G_m\cdot
{\sf R}_\pm^{\frac{\alpha}{\hbar}}(x)\Phi^{\frac{\gamma}{\hbar}}(w,\eta)\prod\limits_{a=1}^N\Phi^{-\sqrt{\beta}}(z_a,\theta_a)\ket{\sigma_+},
\end{equation}
and using (\ref{null:operator}), (\ref{G0:on:R}) and (\ref{G:minus1:on:R}), we arrive at differential equations for the wave-functions
\begin{align}
\label{NVD:Ramond}
& \left(\frac{\partial}{\partial x} +\frac{1}{8x}\right){\widehat \chi}^{\sr R}_{+,\alpha}(x)
=
-{\rm e}^{\frac{i\pi}{4}}\,
\frac{\alpha}{\hbar}\sqrt{\frac{2w}{x(x-w)}}\left(\frac{\Delta_\frac{\gamma}{\hbar}\eta\, x}{w(x-w)} + \eta\partial_w - \partial_\eta\right)
{\widehat \chi}^{\sr R}_{-,\alpha}(x),
\\[2pt]
& \left(\frac{\partial}{\partial x} + \frac{1}{8x} - \frac{Q\alpha}{\hbar\,x}\right){\widehat \chi}^{\sr R}_{-,\alpha}(x)
=
{\rm e}^{-\frac{i\pi}{4}}\,
\frac{\alpha}{\hbar}\sqrt{\frac{2w}{x(x-w)}}\left(\frac{\Delta_\frac{\gamma}{\hbar}\eta\, x}{w(x-w)} + \eta\partial_w - \partial_\eta\right)
{\widehat \chi}^{\sr R}_{+,\alpha}(x),
\nonumber
\end{align}
which are valid for $\alpha = \frac{\hbar}{2}\sqrt{\beta}$ or $-\frac{\hbar}{2\sqrt\beta}$.

The equation (\ref{NVD:Ramond}) can be further simplified by using the scaling covariance (the unbroken subgroup of  the global $SL(2,{\mathbb C})$ covariance) of the correlation functions (\ref{PsiR:definition}). Indeed, since
\[
L_0\ket{\sigma_{\pm}} = \frac{1}{16}\ket{\sigma_{\pm}}, \qquad
\bra{\alpha_0}L_0 = \Delta_{\frac{\alpha_0}{\hbar}}\bra{\alpha_0},
%\qquad
%\Delta_{\alpha_0} = \frac12\alpha_0(\alpha_0-Q),
\]
then using commutation relations
\begin{align}
\begin{split}
\left[L_0,{\sf R}_\pm^{\alpha}(x)\right]
& =
\left(x\partial_x + \left(\Delta_{\alpha} + {\textstyle\frac{1}{16}}\right)\right){\sf R}_\pm^{\alpha}(x),
\\
\left[L_0,\Phi^{\gamma}(w,\eta)\right]
& =
\left(w\partial_w + \left(\Delta_{\gamma} + {\textstyle\frac12}\eta\partial_\eta\right)\right)\Phi^{\gamma}(x,\eta),
\\
\left[L_0,\Phi^{-\sqrt\beta}(z_a,\theta_a)\right]
& =  \left(\partial_{z_a} z_a - {\textstyle\frac12}\partial_{\theta_a}\theta_a\right)\Phi^{-\sqrt{\beta}}(z_a,\theta_a),
\nonumber
\end{split}
\end{align}
we get
\begin{align}
\Big(x\partial_x + w\partial_w + {\textstyle\frac{\eta}{2}}\partial_\eta
+
\sum\limits_{a=1}^N\!\left(\partial_{z_a} z_a - {\textstyle\frac12}\partial_{\theta_a}\theta_a\right)\!\!
\Big)\Psi_{\pm}(x,\vect{z},\vect{\theta})
=
\left(\!\Delta_{\frac{\alpha_0}{\hbar}}-\Delta_{\frac{\alpha}{\hbar}} - \Delta_{\frac{\gamma}{\hbar}} - {\textstyle\frac{1}{8}}\right)\!\Psi_{\pm}(x,\vect{z},\vect{\theta}),
\nonumber
\end{align}
and consequently
\begin{align}
\label{L0:Ward:identity:Ramond}
\left(x\partial_x + w\partial_w + {\textstyle\frac12}\eta\partial_\eta\right)
{\widehat \chi}^{\sr R}_{\pm,\alpha}(x)
=
\left(\Delta_{\frac{\alpha_0}{\hbar}}-\Delta_{\frac{\alpha}{\hbar}} - \Delta_{\frac{\gamma}{\hbar}} - {\textstyle\frac{1}{8}}\right)
{\widehat \chi}^{\sr R}_{\pm,\alpha}(x).
\end{align}
Using (\ref{L0:Ward:identity:Ramond}) and defining
\begin{equation}
{\widehat \chi}^{\sr R}_{\pm,\alpha}(x)
=
x^{-\frac18}\left(f^{\sr R}_{\pm,\alpha}(x) \mp \eta g^{\sr R}_{\pm,\alpha}(x)\right),
\hskip 1cm
g^{\sr R}_{\pm,\alpha}(x) = \mp\, x^{\frac18}\partial_\eta{\widehat \chi}^{\sr R}_{\pm,\alpha}(x),
\label{fg_ramond}
\end{equation}
we can reduce partial differential equations (\ref{NVD:Ramond}) to two pairs of coupled ordinary differential equations
\begin{align}
\label{BPZ:R:1}
\frac{\partial f^{\sr R}_{+,\alpha}(x)}{\partial x}
& =
{\rm e}^{\frac{i\pi}{4}}\,
\frac{\alpha}{\hbar}\sqrt{\frac{2w}{x(x-w)}}
g^{\sr R}_{-,\alpha}(x),    \\
\nonumber
\left(\frac{\partial}{\partial x} - \frac{Q\alpha}{\hbar\,x}\right)g^{\sr R}_{-,\alpha}(x)
& =
{\rm e}^{-\frac{i\pi}{4}}\,
\frac{\alpha}{\hbar}\sqrt{\frac{2w}{x(x-w)}}\Big(\frac{\Delta_{\frac{\alpha_0}{\hbar}}
-\Delta_{\frac{\alpha}{\hbar}}  -x\partial_x}{w}+\frac{\Delta_{\frac{\gamma}{\hbar}}}{x-w}\Big)
f^{\sr R}_{+,\alpha}(x),
\end{align}
and
\begin{align}
\label{BPZ:R:2}
\left(\frac{\partial}{\partial x} - \frac{Q\alpha}{\hbar\,x}\right)f^{\sr R}_{-,\alpha}(x)
& =
{\rm e}^{-\frac{i\pi}{4}}\,
\frac{\alpha}{\hbar}\sqrt{\frac{2w}{x(x-w)}}
g^{\sr R}_{+,\alpha}(x),    \\
\nonumber
\frac{\partial g^{\sr R}_{+,\alpha}(x)}{\partial x}
& =
{\rm e}^{\frac{i\pi}{4}}\,
\frac{\alpha}{\hbar}\sqrt{\frac{2w}{x(x-w)}}\Big(\frac{\Delta_{\frac{\alpha_0}{\hbar}}
-\Delta_{\frac{\alpha}{\hbar}} -x\partial_x}{w}+\frac{\Delta_{\frac{\gamma}{\hbar}}}{x-w}\Big)
f^{\sr R}_{-,\alpha}(x).
\end{align}
It is straightforward to derive from (\ref{BPZ:R:1}) and (\ref{BPZ:R:1}) a second order, ordinary differential equations satisfied by the functions
$f^{\sr R}_{\pm,\alpha}(x)$ and $g^{\sr R}_{\pm,\alpha}(x)$ (we discuss these equations in Section \ref{sec-matrix-Ramond-R}). These are Ramond-R super-quantum curve equations we have been after.  In the context of superconformal field theory these equations can be regarded as supersymmetric versions of BPZ equations, and they were  discussed and solved in \cite{Zamolodchikov:1988nm,Poghosian:1996dw}. The fact that we reproduce supersymmetric BPZ equations ensures that we have chosen a proper definition of the Ramond-R wave-functions (and the corresponding eigenvalue model), and it is a nice test of our formalism.

%******************************************************************************************
%******************************************************************************************
%******************************************************************************************
%******************************************************************************************

\section{Ramond-NS super-eigenvalue model and super-quantum curves}    \label{sec-matrix}

In this section we rederive super-quantum curves in the Ramond-NS sector, however in the formalism of matrix models (or rather eigenvalue models), analogously as in \cite{Manabe:2015kbj,Ciosmak:2016wpx}. This enables us to compare both approaches (i.e. conformal field theory and matrix models) and confirm that they lead to the same results. However, as an additional result, in this section we also derive classical curves, which from the matrix model viewpoint would be interpreted as spectral curves, and which describe equilibrium distribution of eigenvalues. The interpretation of such classical curves is more natural from matrix model perspective rather than conformal field theory.

Our starting point is the expression for the partition function (\ref{Z-Ramond}), which we proposed based on conformal field theory considerations, and which we interpret now as a super-eigenvalue model with $N$ bosonic and fermionic variables $z_a$ and $\theta_a$
\begin{equation}
Z= \int\! d^N\!z\, d^N\!\theta\ \Delta_{\sr R}(\vect{z},\vect{\theta})^{\beta}\mathrm{e}^{-\frac{\sqrt{\beta}}{\hbar}\sum_{a=1}^NV_{\sr R}(z_a,\theta_a)},
\label{matrix_defin}%(\ref{matrix_defin})
\end{equation}
with $d^N\!z\, d^N\!\theta=\prod_{a=1}^Ndz_a\,d\theta_a$, the Vandermonde-like determinant given in (\ref{Vandermonde-Ramond})
\begin{equation}
\Delta_{\sr R}(\vect{z},\vect{\theta})^{\beta}=
\prod_{1\le a<b\le N}\Big(z_a-z_b-\frac{1}{2}(z_a+z_b)\frac{\theta_a\theta_b}{\sqrt{z_az_b}}\Big)^{\beta},
\end{equation}
and the potential is given by (\ref{VRztheta})
\be
V_{\sr R}(x,\theta)=V_{\sr B}(x)+V_{\sr F}(x)\frac{\theta}{\sqrt{x}},\qquad V_{\sr B}(x)=\sum_{n=0}^{\infty}t_nx^n,\qquad
V_{\sr F}(x)=\sum_{n=0}^{\infty}\xi_{n}x^n,    \label{VR-VB-VF}
\ee
with bosonic times and fermionic times $t_n$ and $\xi_{n}$, such that $\{\theta_a, \xi_{n}\}=0$. In this section we also use an equivalent notation, with redefined anticommuting variable\footnote{This is the same trick used in \cite{Blondeau-Fournier:2016jth} to remove the half-integer powers of the variables $z_a$ in the construction of Ramond singular vectors.}
\be
\theta = \eta\sqrt{z}.   \label{theta-eta}
\ee
In terms of variables ${\eta}_a=\theta_a / \sqrt{z_a}$ defined in (\ref{theta-eta}), by $d^N\!\theta=\prod_{a=1}^{N}z_a^{-1/2}\,d^N\!{\eta}$ the partition function (\ref{matrix_defin}) takes form
\begin{equation}
Z= \int\! d^N\!z\,d^N\!{\eta}\ \Delta_{\mathsf{r}}(\vect{z},\vect{{\eta}})^{\beta}\mathrm{e}^{-\frac{\sqrt{\beta}}{\hbar}\sum_{a=1}^NV_{\mathsf{r}}(z_a,{\eta}_a)},
\label{matrix_def}%(\ref{matrix_def})
\end{equation}
where $V_{\mathsf{r}}(x,{\eta})=V_{\sr B}(x)+V_{\sr F}(x){\eta}$, and
\begin{equation}
\Delta_{\mathsf{r}}(\vect{z},\vect{{\eta}})^{\beta}=\prod_{a=1}^{N}z_a^{-1/2}\cdot
\prod_{1\le a<b\le N}\Big(z_a-z_b-\frac{1}{2}(z_a+z_b){\eta}_a{\eta}_b\Big)^{\beta}.
\end{equation}
We also decompose the partition function as
\begin{align}
\begin{split}
Z&=Z_0+\frac{\xi_0}{\hbar} Z_1,\\
Z_0&\equiv
Z\big|_{\xi_0=0}=
\int\! d^N\!z\,d^N\!{\eta}\ \Delta_{\mathsf{r}}(\vect{z},\vect{{\eta}})^{\beta}\mathrm{e}^{-\frac{\sqrt{\beta}}{\hbar}\sum_{a=1}^N
\left(V_{\sr B}(z_a)+V_{\sr F}^{(+)}(z_a){\eta}_a\right)},\\
Z_1&\equiv
\hbar\partial_{\xi_0}Z=
-\sqrt{\beta}
\int\! d^N\!z\,d^N\!{\eta}\ \Delta_{\mathsf{r}}(\vect{z},\vect{{\eta}})^{\beta}\left(\sum_{a=1}^N{\eta}_a\right)
\mathrm{e}^{-\frac{\sqrt{\beta}}{\hbar}\sum_{a=1}^N
\left(V_{\sr B}(z_a)+V_{\sr F}^{(+)}(z_a){\eta}_a\right)},
\label{matrix_decomp}%(\ref{matrix_decomp})
\end{split}
\end{align}
where
\begin{equation}
V_{\sr F}^{(+)}(x)=\sum_{n=1}^{\infty}\xi_{n}x^n.
\nonumber
\end{equation}
The partition functions $Z_0$ and $Z_1$ represent a two-dimensional Ramond vacuum. In this section the unnormalized expectation value of an operator $\mathcal{O}$ is denoted by
\begin{equation}
\Left<\mathcal{O}\Right> = \int\! d^N\!z\,d^N\!{\eta}\,
\mathcal{O} \Delta_{\mathsf{r}}(\vect{z},\vect{{\eta}})^{\beta}
\mathrm{e}^{-\frac{\sqrt{\beta}}{\hbar}\sum_{a=1}^NV_{\mathsf{r}}(z_a,{\eta}_a)}.
\label{def_unnorm_exp}%(\ref{def_unnorm_exp})
\end{equation}

%******************************************************************************************
%******************************************************************************************

\subsection{Free field realization}

In terms of times in the potential (\ref{VR-VB-VF}) the oscillator algebra (\ref{free:boson:fermion:R}) in the Ramond sector with (\ref{commutator:an:q:NS}) is realized as
\begin{align}
\begin{split}
&q=\frac{1}{\hbar}t_0,\quad {\sf a}_{-n}=\frac{1}{\hbar}t_n,\quad
{\sf a}_0=\hbar\partial_{t_0},\quad {\sf a}_n=\hbar\partial_{t_n},
\\
&\psi_0=\frac{1}{\hbar}\xi_0+\frac{\hbar}{2}\partial_{\xi_0},\quad
\psi_{-n}=\frac{1}{\hbar}\xi_n,\quad \psi_{n}=\hbar\partial_{\xi_n},
\qquad n\ge 1.
\end{split}
\end{align}
Here note that the derivatives of the partition function with respect to times can be represented as the following expectation values
\begin{equation}
\partial_{t_n}Z=-\frac{\sqrt{\beta}}{\hbar}\Left<\sum_{a=1}^Nz_a^n\Right>,\qquad
\partial_{\xi_n}Z=-\frac{\sqrt{\beta}}{\hbar}\Left<\sum_{a=1}^Nz_a^n{\eta}_a\Right>.
\label{id_mat_free}%(\ref{id_mat_free})
\end{equation}
Therefore we can introduce bosonic and fermionic quantum fields, whose negative modes -- normally represented by derivatives with respect of times -- can be written as
\begin{align}
\begin{split}
\phi(x)&
=\frac{1}{\hbar}\sum_{n=0}^{\infty}t_nx^n
-\sqrt{\beta}N\log x+\sqrt{\beta}\sum_{n=1}^{\infty}\sum_{a=1}^{N}\frac{z_a^n}{n}x^{-n}
=\frac{1}{\hbar}V_{\sr B}(x)-\sqrt{\beta}\sum_{a=1}^N\log(x-z_a),    \\
\sqrt{x}\psi(x)&
=\frac{1}{\hbar}\xi_0-\frac{\sqrt{\beta}}{2}\sum_{a=1}^{N}\eta_a
+\frac{1}{\hbar}\sum_{n=1}^{\infty}\xi_nx^n
-\sqrt{\beta}\sum_{n=1}^{\infty}\sum_{a=1}^{N}z_a^n\eta_a x^{-m}
=\frac{1}{\hbar}\widehat{V}_{\sr F}(x)
-\sqrt{\beta}\sum_{a=1}^N\frac{z_a{\eta}_a}{x-z_a},
\label{free_b_f}%(\ref{free_b_f})
\end{split}
\end{align}
where we have defined
\begin{equation}
\widehat{V}_{\sr F}(x)\equiv V_{\sr F}(x)-\frac{\sqrt{\beta}\hbar}{2}\sum_{a=1}^N{\eta}_a
=\widehat{\xi}_0+\sum_{n=1}^{\infty}\xi_{n}x^n,\qquad
\widehat{\xi}_0\equiv \xi_0-\frac{\sqrt{\beta}\hbar}{2}\sum_{a=1}^N{\eta}_a.
\label{vf_ref}%(\ref{vf_ref})
\end{equation}
These fields satisfy standard relations
\begin{align}
\phi(x_1)\phi(x_2)=\log(x_1-x_2)+\ldots,\qquad
\psi(x_1)\psi(x_2)=\frac{1}{x_1-x_2}+\ldots.
\nonumber
\end{align}
%Note that in (\ref{vf_ref}) the representation $-\hbar\partial_{\xi_0}/\sqrt{\beta}$ of $\Left<\sum_{a=1}^N {\eta}_a\Right>$ realizes the relation $\widehat{\xi}_0^2/\hbar^2=1/2$ of the zero mode.
The Ramond supercurrent $S(x)$ and the energy-momentum tensor $T(x)$ are constructed as
\begin{align}
S(x)&\equiv\sum_{n\in{\IZ}}g_nx^{-n-\frac32}= \psi(x)\partial_x\phi(x)+Q\partial_x\psi(x),
\label{s_current}%(\ref{s_current})
\\
T(x)&\equiv\sum_{n\in{\IZ}}\ell_nx^{-n-2}= \frac12:\partial_x\phi(x)\partial_x\phi(x):+\frac12:\partial_x\psi(x)\psi(x):
+\frac12Q\partial_x^2\phi(x),
\label{e_m_tensor}%(\ref{e_m_tensor})
\end{align}
where $Q=\beta^{-1/2}-\beta^{1/2}$ corresponds to the background charge in $\mathcal{N}=1$ super-Liouville field theory. We also write
\be
\begin{split}
S(x)&=S_+(x)+S_-(x),\qquad S_+(x)=\sum_{n=0}^{\infty}  g_n x^{-n-\frac32}, \\
T(x)&=T_+(x)+T_-(x),\qquad T_+(x)=\sum_{n=0}^{\infty} \ell_nx^{-n-2}.
\end{split}
\ee
The OPEs of these fields are given by
\begin{align}
\begin{split}
&
S(x_1)S(x_2)=\frac{2c}{3(x_1-x_2)^3}+\frac{2}{x_1-x_2}T(x_2)+\ldots,\\
&
T(x_1)S(x_2)=\frac{3}{2(x_1-x_2)^2}S(x_2)+\frac{1}{x_1-x_2}S'(x_2)+\ldots,\\
&
T(x_1)T(x_2)=\frac{c}{2(x_1-x_2)^4}+\frac{2}{(x_1-x_2)^2}T(x_2)+\frac{1}{x_1-x_2}T'(x_2)+\ldots,
\label{ope_s_em}%(\ref{ope_s_em})
\end{split}
\end{align}
where the central charge reads
\be
c= \frac{3}{2}-3Q^2.
\ee
The OPEs (\ref{ope_s_em}) imply that the modes $g_n$ and $\ell_n$ defined by the expansions (\ref{s_current}) and (\ref{e_m_tensor}) satisfy the super-Virasoro algebra (\ref{NS}). After some manipulations we also find
\begin{align}
\begin{split}
\sqrt{x}S_+(x)&= \frac{\beta}{2}\sum_{a,b=1}^N\frac{(x+z_a) {\eta}_a}{(x-z_a)(x-z_b)}
+\frac{\sqrt{\beta}Q}{2}\sum_{a=1}^N\frac{(x+z_a) {\eta}_a}{(x-z_a)^2}
-\frac{\sqrt{\beta}Q}{4x}\sum_{a=1}^N{\eta}_a + \\
&\quad -\frac{Q}{2\hbar x}\xi_0 -\frac{\sqrt{\beta}}{\hbar}\sum_{a=1}^N\frac{z_aV_{\sr B}'(z_a){\eta}_a+V_{\sr F}(z_a)}{x-z_a},
\label{s_p_x}%(\ref{s_p_x})
\end{split}
\\
\begin{split}
\sqrt{x}S_-(x)&= \frac{1}{\hbar^2}V_{\sr B}'(x)\widehat{V}_{\sr F}(x)+\frac{Q}{\hbar}V_{\sr F}'(x)
-\frac{Q}{2\hbar x}\left(V_{\sr F}(x)-\xi_0\right) +  \\
&\quad -\frac{\sqrt{\beta}}{\hbar}\sum_{a=1}^N\frac{V_{\sr F}(x)-V_{\sr F}(z_a)}{x-z_a}
-\frac{\sqrt{\beta}}{\hbar}\sum_{a=1}^N\frac{\big(V_{\sr B}'(x)-V_{\sr B}'(z_a)\big)z_a{\eta}_a}{x-z_a}, \label{s_m_x}%(\ref{s_m_x})
\end{split}
\end{align}
and similarly
\begin{align}
\begin{split}
xT_+(x)&= \frac{\beta}{2}\sum_{a,b=1}^N\frac{x}{(x-z_a)(x-z_b)}
+\frac{\sqrt{\beta}Q}{2}\sum_{a=1}^N\frac{x}{(x-z_a)^2}
-\frac{\beta}{4}\sum_{a,b=1}^N\frac{z_a{\eta}_a{\eta}_b}{(x-z_a)^2} + \\
&\quad +\frac{\beta}{2}\sum_{a,b=1}^N\frac{z_az_b{\eta}_a{\eta}_b}{(x-z_a)(x-z_b)^2}
-\frac{\sqrt{\beta}}{\hbar}\sum_{a=1}^N\frac{z_a\big(V_{\sr B}'(z_a)+V_{\sr F}'(z_a){\eta}_a\big)}{x-z_a} +  \\
&\quad -\frac{\sqrt{\beta}}{2\hbar}\sum_{a=1}^N\frac{z_aV_{\sr F}(z_a){\eta}_a}{(x-z_a)^2}+\frac{1}{16x}, \label{t_p_x}%(\ref{t_p_x})
\end{split}
\\
\begin{split}
xT_-(x)&= \frac{x}{2\hbar^2}V_{\sr B}'(x)^2+\frac{1}{2\hbar^2}V_{\sr F}'(x)\widehat{V}_{\sr F}(x) +\frac{Qx}{2\hbar}V_{\sr B}''(x)
-\frac{\sqrt{\beta}}{\hbar}\sum_{a=1}^N\frac{xV_{\sr B}'(x)-z_aV_{\sr B}'(z_a)}{x-z_a} + \\
&\quad -\frac{\sqrt{\beta}}{2\hbar}\sum_{a=1}^N\frac{\big(V_{\sr F}'(x)-V_{\sr F}'(z_a)\big)z_a{\eta}_a}{x-z_a}
-\frac{\sqrt{\beta}}{2\hbar}\sum_{a=1}^N\frac{z_aV_{\sr F}^{(2)}(x,z_a){\eta}_a}{(x-z_a)^2}, \label{t_m_x}%(\ref{t_m_x})
\end{split}
\end{align}
where we denote
\begin{equation}
V_{\sr F}^{(2)}(x,z_a)\equiv V_{\sr F}(x)-V_{\sr F}(z_a)-(x-z_a)V_{\sr F}'(z_a).
\label{vf2_def}%(\ref{vf2_def})
\end{equation}

%******************************************************************************************
%******************************************************************************************

\subsection{Loop equations and super-Virasoro constraints}\label{sub-sec-loop-matrix}

We can now determine loop equations for the super-eigenvalue model (\ref{matrix_def}). As usual, these equations follow from the invariance of the partition function under changes of integration variables. The partition function (\ref{matrix_def}) is invariant under
\begin{equation}
z_a\ \to\ z_a+\frac{z_a {\eta}_a \delta}{\sqrt{x}(x-z_a)},\qquad
{\eta}_a\ \to\ {\eta}_a+\frac{\delta}{\sqrt{x}(x-z_a)},
\label{f_shift}%(\ref{f_shift})
\end{equation}
with a fermionic constant $\delta$. We see that this invariance leads to the loop equation (\ref{s_p_loop_r}):
\begin{equation}
\int\! d^N\!z\,d^N\!{\eta}\
\sum_{a=1}^N\big({\eta}_a\partial_{z_a}z_a-\partial_{{\eta}_a}\big)
\bigg[\frac{1}{x-z_a}
\Delta_{\mathsf{r}}(\vect{z},\vect{{\eta}})^{\beta}
\mathrm{e}^{-\frac{\sqrt{\beta}}{\hbar}\sum_{a=1}^NV(z_a,{\eta}_a)}\bigg]=0,
\nonumber
\end{equation}
which can be written as
\begin{equation}
\Left<S_+(x)\Right>=
-\frac{Q}{2\hbar x^{3/2}}\left(\xi_0-\frac{\hbar^2}{2}\partial_{\xi_0}\right)Z,
\label{f_ward_id2}%(\ref{f_ward_id2})
\end{equation}
where the supercurrent $S_+(x)$ is given by (\ref{s_p_x}).
The partition function (\ref{matrix_def}) is also invariant under
\begin{equation}
z_a\ \to\ z_a+\frac{z_a \epsilon}{x(x-z_a)},\qquad
{\eta}_a\ \to\ {\eta}_a+\frac{z_a {\eta}_a \epsilon}{2x(x-z_a)^2},
\label{b_shift}%(\ref{b_shift})
\end{equation}
with an infinitesimal parameter $\epsilon$. We see that this leads to another loop equation (\ref{t_p_loop_r}):
\begin{equation}
\int\! d^N\!z\,d^N\!{\eta}\
\sum_{a=1}^N\Big(\partial_{z_a}-\partial_{{\eta}_a}\frac{{\eta}_a}{2(x-z_a)}\Big)
\bigg[\frac{z_a}{x-z_a}
\Delta_{\mathsf{r}}(\vect{z},\vect{{\eta}})^{\beta}
\mathrm{e}^{-\frac{\sqrt{\beta}}{\hbar}\sum_{a=1}^NV(z_a,{\eta}_a)}\bigg]=0,
\nonumber
\end{equation}
which can be written as
\begin{equation}
\Left<T_+(x)\Right>=\frac{1}{16x^2}\:Z,
\label{b_ward_id2}%(\ref{b_ward_id2})
\end{equation}
where the energy-momentum tensor $T_+(x)$ is given by (\ref{t_p_x}).
Therefore by the mode expansion and the realizations (\ref{id_mat_free}), these loop equations give super-Virasoro constraints (\ref{s_vir_const-LH}):
\begin{equation}
g_{n}Z=-\frac{Q}{2\hbar}\delta_{n,0}\Big(\xi_0-\frac{\hbar^2}{2}\partial_{\xi_0}\Big)Z,
\qquad \ell_{n}Z=\frac{1}{16}\delta_{n,0}Z, \qquad n\ge 0.
\end{equation}

%where
%\begin{align}
%\begin{split}
%\ell_{n}&= \sum_{m=1}^{\infty}mt_m\partial_{t_{m+n}} %+\sum_{m=0}^{\infty}\Big(m+\frac{n}{2}\Big)\xi_m\partial_{\xi_{m+n}}
%+\frac{\hbar^2}{2}\sum_{m=0}^{n}\partial_{t_{n-m}}\partial_{t_{m}} +  \\
%&\quad +\frac{\hbar^2}{4}n\partial_{\xi_{0}}\partial_{\xi_{n}} +\frac{\hbar^2}{2}\sum_{m=1}^{n-1}m\partial_{\xi_{m-n}}\partial_{\xi_{m}} -\frac{Q\hbar}{2}(n+1)\partial_{t_n}+\frac{1}{16}\delta_{n,0}.   \label{ln-Ramond}
%\end{split}
%\end{align}
%The realization of the super-Virasoro algebra in terms of operators (\ref{gn-Ramond}) and (\ref{ln-Ramond}) reproduces (\ref{gn-Ramond-LH}) and (\ref{ln-Ramond-LH}), while the above loop equations reproduce (\ref{s_vir_const-LH}).

Note that from the constraints (\ref{s_vir_const-LH}), by the decomposition (\ref{matrix_decomp}) of the partition function we obtain
\begin{align}
\begin{split}
& \widehat{g}_nZ_0-\frac{\xi_0}{\hbar}\widehat{g}_nZ_1\equiv
g_n Z=\left(\frac{Q}{4}Z_1-\frac{Q}{2\hbar}\xi_0Z_0\right)\delta_{n,0},\\
& \widehat{\ell}_nZ_0+\frac{\xi_0}{\hbar}\widehat{\ell}_nZ_1\equiv
\ell_n Z=\frac{1}{16}\left(Z_0+\frac{\xi_0}{\hbar}Z_1\right)\delta_{n,0},
\end{split}
\end{align}
where $\widehat{g}_n$ and $\widehat{\ell}_n$ are abstract operators acting on $Z_0$ and $Z_1$. In particular, the constraints
\begin{equation}
\widehat{g}_0Z_0=\frac{Q}{4}Z_1,\quad
\widehat{g}_0Z_1=\frac{Q}{2}Z_0,\quad
\widehat{\ell}_0Z_0=\frac{1}{16}Z_0,\quad
\widehat{\ell}_0Z_1=\frac{1}{16}Z_1,
\end{equation}
are consistent with the relation $\widehat{g}_0^2=\widehat{\ell}_0-c/24$ in the super-Virasoro algebra.

Also note, that while right hand sides of equations (\ref{f_ward_id2}) and (\ref{b_ward_id2}) are non-zero (and so might seem non-standard), this is only a consequence of our conventions. If the modes $L_0$ and $G_0$ would not be included in the definition of respectively $T_+(x)$ and $S_+(x)$, then (\ref{f_ward_id2}) and (\ref{b_ward_id2}) would have zero on the right hand side.

%******************************************************************************************
%******************************************************************************************

\subsection{Super-spectral curve} \label{ssec-spectralcurve}

In the analysis of matrix or eigenvalue models, a spectral curve is an algebraic curve that encodes equilibrium distribution of eigenvalues. In case of super-eigenvalue models one finds a supersymmetric spectral curve, defined in terms of supersymmetric algebraic equations. Such super-spectral curves in the Neveu-Schwarz sector have been derived in \cite{Ciosmak:2016wpx}. In this section we derive an analogous super-spectral curve in the Ramond sector, for the super-eigenvalue model defined by (\ref{matrix_defin}). To this end we analyze the loop equations (\ref{f_ward_id2}) and (\ref{b_ward_id2}) in the large $N$ limit
\begin{equation}
N\to \infty,\qquad \hbar\to 0,\qquad \textrm{with}\quad \mu\equiv\beta^{\frac12} \hbar N=\textrm{const}.
\label{large_N_lim}%(\ref{large_N_lim})
\end{equation}
We also use the notation
\begin{equation}
\widehat{\hbar}\equiv\big(\beta^{\frac12}-\beta^{-\frac12}\big)\hbar=-Q\hbar.
\end{equation}
By defining
\begin{equation}
Y_{\sr B}(x;\widehat{\hbar})\equiv Y_{\sr B}(x) = \lim_{\begin{subarray}{c}N\to\infty\\\widehat{\hbar}\,\textrm{fixed}\end{subarray}}\frac{\hbar}{Z}\Left<\partial_x\phi(x)\Right>,\qquad
Y_{\sr F}(x;\widehat{\hbar})\equiv Y_{\sr F}(x) = \lim_{\begin{subarray}{c}N\to\infty\\\widehat{\hbar}\,\textrm{fixed}\end{subarray}}\frac{\hbar}{Z}\Left<\psi(x)\Right>,
\label{def_y_bf}%(\ref{def_y_bf})
\end{equation}
the loop equations (\ref{f_ward_id2}) and (\ref{b_ward_id2}) yield
\begin{equation}
\sqrt{x}Y_{\sr B}(x)Y_{\sr F}(x)-V_{\sr B}'(x)\widetilde{V}_{\sr F}(x)+\widehat{\hbar}\Big(V_{\sr F}'(x)
-\frac{1}{2x}\widetilde{V}_{\sr F}(x)-\sqrt{x}Y_{\sr F}'(x)\Big)-h^{(0)}(x)=0,
\label{largeN_Ward_b}%(\ref{largeN_Ward_b})
\end{equation}
and
\begin{equation}
xY_{\sr B}(x)^2+xY_{\sr F}'(x)Y_{\sr F}(x)-xV_{\sr B}'(x)^2-V_{\sr F}'(x)\widetilde{V}_{\sr F}(x)+\widehat{\hbar}x\big(V_{\sr B}''(x)-Y_{\sr B}'(x)\big)-2f^{(0)}(x)=0,
\label{largeN_Ward_f}%(\ref{largeN_Ward_f})
\end{equation}
respectively. Here
\begin{equation}
\widetilde{V}_{\sr F}(x)\equiv \widetilde{V}_{\sr F}(x;\hbar)
=V_{\sr F}(x)+\frac{1}{2}
\lim_{\begin{subarray}{c}N\to\infty\\\widehat{\hbar}\,\textrm{fixed}\end{subarray}}\frac{\sqrt{\beta}\hbar}{Z}\Left<\sum_{a=1}^N{\eta}_a\Right>
=V_{\sr F}(x)-\frac{1}{2}
\lim_{\begin{subarray}{c}N\to\infty\\\widehat{\hbar}\,\textrm{fixed}\end{subarray}}\hbar\frac{Z_1}{Z_0},
\end{equation}
and
\begin{align}
h^{(0)}(x)& \equiv
h^{(0)}(x;\widehat{\hbar}) =
\nonumber\\
& = -\lim_{\begin{subarray}{c}N\to\infty\\\widehat{\hbar}\ \textrm{fixed}\end{subarray}}\frac{\sqrt{\beta}\hbar}{Z}\Left<
\sum_{a=1}^{N}\bigg(\frac{V_{\sr F}(x)-V_{\sr F}(z_a)}{x-z_a}+\frac{\big(xV_{\sr B}'(x)-z_aV_{\sr B}'(z_a)\big){\eta}_a}{x-z_a}\bigg)
\Right>,
\label{h_x_0_cl}%(\ref{h_x_0_cl})
\\
f^{(0)}(x)&\equiv
f^{(0)}(x;\widehat{\hbar}) =
\nonumber\\
&=
-\lim_{\begin{subarray}{c}N\to\infty\\\widehat{\hbar}\ \textrm{fixed}\end{subarray}}\frac{\sqrt{\beta}\hbar}{Z}\Left<
\sum_{a=1}^{N}\bigg(\frac{xV_{\sr B}'(x)-z_aV_{\sr B}'(z_a)}{x-z_a}
+\frac{\big(xV'_{\sr F}(x)-z_aV'_{\sr F}(z_a)\big){\eta}_a}{2(x-z_a)}\Right. +
\nonumber\\
&\hspace{10em}
\Left.+\frac{z_aV^{(2)}_{\sr F}(x,z_a){\eta}_a}{2(x-z_a)^2}\bigg)\Right>,
\label{f_x_0_cl}%(\ref{f_x_0_cl})
\end{align}
where $V^{(2)}_{\sr F}(x,z_a)$ is defined in (\ref{vf2_def}). For polynomial potentials, $h^{(0)}(x)$ and $f^{(0)}(x)$ are polynomials of $x$.
For $\widehat{\hbar}=0$, or in particular for $\beta=1$, denoting
\be
y_{\sr B}(x) = Y_{\sr B}(x;0), \qquad  y_{\sr F}(x) = Y_{\sr F}(x;0),   \label{yByF}
\ee
the loop equations (\ref{largeN_Ward_b}) and (\ref{largeN_Ward_f}) in the large $N$ limit yield a super-spectral curve
\begin{align}
\begin{cases}
& A_{\sr F}(x,y_{\sr B}|y_{\sr F})\equiv y_{\sr B}(x)y_{\sr F}(x)+G(x)=0,  \\
& A_{\sr B}(x,y_{\sr B}|y_{\sr F})\equiv y_{\sr B}(x)^2+y_{\sr F}'(x)y_{\sr F}(x)+2L(x)=0,
\label{b_f_sp_curve}%(\ref{b_f_sp_curve})
\end{cases}
\end{align}
where
\begin{align}
\begin{split}
G(x) & = -x^{-1/2}V_{\sr B}'(x)V_{\sr F}(x)-x^{-1/2}h^{(0)}(x;0),\\
L(x) & = -\frac12 V_{\sr B}'(x)^2-\frac12 x^{-1}V_{\sr F}'(x)\widetilde{V}_{\sr F}(x;0)-x^{-1}f^{(0)}(x;0).
\nonumber
\end{split}
\end{align}
The equation (\ref{b_f_sp_curve}) is a supersymmetric algebraic equation, which defines a supersymmetric algebraic curve. In matrix model interpretation this curve encodes equilibrium distribution of eigenvalues. We refer to this curve as the super-spectral curve.

%******************************************************************************************
%******************************************************************************************

\subsection{Wave-function and deformed currents}

The quantum curves that we are after are supposed to quantize the super-spectral curve (\ref{b_f_sp_curve}). They should take form of differential equations that annihilate the wave-function, which in the operator formalism we introduced in (\ref{chi-alpha-Ramond}). More explicitly, in the eigenvalue representation this wave-function takes form
\begin{equation}
\widehat{\chi}_{\alpha}(x,\sqrt{x}{\eta})=\Left<\mathrm{e}^{\frac{\alpha}{\hbar}\big(\phi(x)+\sqrt{x}\psi(x){\eta}\big)}\Right>,
\label{chi_hat_def}%(\ref{chi_hat_def})
\end{equation}
where ${\eta}$ is a fermionic variable with $\{{\eta}, {\eta}_a\}=\{{\eta}, \xi_{n}\}=0$, and $\alpha$ is a bosonic parameter that we refer to as the momentum. Using expressions (\ref{free_b_f}) we can further write
\begin{equation}
\widehat{\chi}_{\alpha}(x,\sqrt{x}{\eta})=\mathrm{e}^{\frac{\alpha}{\hbar^2}V_{\sr B}(x)+\frac{\alpha}{\hbar^2}V_{\sr F}(x){\eta}}\Left<\chi_{\alpha}^{\textrm{ins}}(x,\sqrt{x}{\eta})\Right>
\equiv
\mathrm{e}^{\frac{\alpha}{\hbar^2}V_{\sr B}(x)+\frac{\alpha}{\hbar^2}V_{\sr F}(x){\eta}}\chi_{\alpha}(x,\sqrt{x}{\eta}),
\label{chi_hat_w}%(\ref{chi_hat_w})
\end{equation}
where we have defined $\chi_{\alpha}(x,\sqrt{x}{\eta})$ as the unnormalized expectation value of the operator
\begin{align}
\chi_{\alpha}^{\textrm{ins}}(x,\sqrt{x}{\eta})&
=\mathrm{e}^{-\frac{\sqrt{\beta}}{\hbar}\sum_{a=1}^N
\alpha\left(\log(x-z_a)-\frac{1}{2}{\eta}{\eta}_a
-\frac{z_a}{x-z_a}{\eta}{\eta}_a\right)} =
\nonumber\\
&=\left(1+\frac{\sqrt{\beta}}{\hbar}\alpha{\eta}
\sum_{a=1}^N\left(\frac12{\eta}_a+
\frac{z_a{\eta}_a}{x-z_a}\right)\right)\prod_{a=1}^N(x-z_a)^{-\frac{\sqrt{\beta}}{\hbar}\alpha}.
\label{chi_ins_def}%(\ref{chi_ins_def})
\end{align}
The wave-function can be decomposed into bosonic and fermionic components as
\begin{align}
\widehat{\chi}_{\alpha}(x,\sqrt{x}{\eta})=\widehat{\chi}_{{\sr B},\alpha}(x)+\sqrt{x}\widehat{\chi}_{{\sr F},\alpha}(x)\eta.
\label{chi_bf_def}%(\ref{chi_bf_def})
\end{align}

%******************************************************************************************
%******************************************************************************************

%\subsection{Deformed currents and loop equations for the wave-function}

By analogy with a derivation of loop equations for the partition function, in this section we derive loop equations for the wave-function $\chi_{\alpha}(x,\sqrt{x}{\eta})$. This analysis is equivalent to the operator formalism presented in section \ref{sec-Ramond}. To proceed we regard the wave-function as an eigenvalue model with deformed potentials
\begin{align}
\begin{split}
\widetilde{V}_{\sr B}(y;x)&
=V_{\sr B}(y)+\alpha\log(x-y),\\
\widetilde{V}_{\sr F}(y;x,{\eta})&
=V_{\sr F}(y)-\frac{\alpha {\eta}}{2}-\frac{\alpha y{\eta}}{x-y},   \label{V-tilde-BF}
\end{split}
\end{align}
which replace $V_{\sr B}(y)$ and $V_{\sr F}(y)$ in the supercurrent $S(y)$ and the energy-momentum tensor $T(y)$. Note that the fermionic time $\xi_0$ is also deformed as $\xi_0 \to \widetilde{\xi}_0\equiv\xi_0 -\alpha {\eta}/2$.
From (\ref{s_p_x}) and (\ref{s_m_x}) we find the deformed super-current
\begin{equation}
S(y;x,{\eta})=S_+(y;x,{\eta})+S_-(y;x,{\eta}),
\nonumber\
\end{equation}
where
\begin{align}
\sqrt{y}S_+(y;x,{\eta})
& = \frac{\alpha\sqrt{\beta}}{\hbar}\sum_{a=1}^N
\frac{z_a {\eta}_a}{(x-z_a)(y-z_a)}
+\frac{Q\sqrt{\beta}}{2}\sum_{a=1}^N\frac{(y+z_a){\eta}_a}{(y-z_a)^2}
+\frac{\beta}{2}\sum_{a,b=1}^N\frac{(y+z_a){\eta}_a}{(y-z_a)(y-z_b)} +
\nonumber\\
&\ \ \
+\frac{\alpha{\eta}\sqrt{\beta}}{\hbar}\sum_{a=1}^N\frac{z_a}{(x-z_a)(y-z_a)}
+\frac{\alpha{\eta}\sqrt{\beta}}{2\hbar}\sum_{a=1}^N\frac{1}{y-z_a} +
\nonumber\\
&\ \ \ -\frac{\sqrt{\beta}}{\hbar}\sum_{a=1}^N\frac{yV_{\sr B}'(y){\eta}_a+V_{\sr F}(y)}{y-z_a}
-\frac{h(y)}{\hbar^2}
-\frac{Q}{2\hbar y}\Big(\widetilde{\xi}_0
+\frac{\sqrt{\beta}\hbar}{2}\sum_{a=1}^N{\eta}_a\Big),
\end{align}
and
\begin{align}
\sqrt{y}S_-(y;x,{\eta})&=
\frac{2\Delta_{\frac{\alpha}{\hbar}}x{\eta}}{(y-x)^2}+
\frac{1}{y-x}\Big[x{\eta}\Big(\partial_x+\frac{\alpha}{\hbar^2}V_{\sr B}'(y)\Big)-\Big(\partial_{{\eta}}-\frac{\alpha}{\hbar^2}V_{\sr F}(y)\Big)
+\Delta_{\frac{\alpha}{\hbar}}{\eta}\Big] +
\nonumber\\
&\ \ \
+\frac{1}{\hbar^2}\Big[V_{\sr B}'(y)\Big(V_{\sr F}(y)+\frac{\alpha{\eta}}{2}
+\frac{\sqrt{\beta}\hbar}{2}\sum_{a=1}^{N}{\eta}_a\Big)
+Q\hbar V_{\sr F}'(y) +
\nonumber\\
&\qquad\qquad
-\frac{Q\hbar}{2y}\big(V_{\sr F}(y)-\xi_0\big)
+h(y)\Big].
\label{s_def_m}%(\ref{s_def_m})
\end{align}
Here
\begin{align}
\Delta_{\frac{\alpha}{\hbar}}=\frac{\alpha}{2\hbar}\Big(\frac{\alpha}{\hbar}-Q\Big),
\end{align}
and
\begin{align}
h(y)=
-\sqrt{\beta}{\hbar}\sum_{a=1}^N
\bigg(\frac{V_{\sr F}(y)-V_{\sr F}(z_a)}{y-z_a}+
\frac{\big(yV_{\sr B}'(y)-z_aV_{\sr B}'(z_a)\big){\eta}_a}{y-z_a}\bigg).   \label{hy-R-NS}
\end{align}
As the operator acting on the partition function $Z$ or the wave-function $\chi_{\alpha}(x,\sqrt{x}{\eta})$, $h(y)$ is equivalently represented by the partial differential operator
\begin{align}
\widehat{h}(y)=\hbar^2\sum_{n=0}^{\infty}y^n \hskip -5pt
\sum_{m=n+1}^{\infty}\left(\xi_m\partial_{t_{m-n-1}}+mt_m\partial_{\xi_{m-n-1}}\right),
\label{h_hat_op}%(\ref{h_hat_op})
\end{align}
which we found independently in (\ref{h-hat-Ramond}). In the large $N$ limit (\ref{large_N_lim}), the expectation value of this operator reproduces $h^{(0)}(y)$ in (\ref{h_x_0_cl})
\begin{align}
\lim_{\begin{subarray}{c}N\to\infty\\\widehat{\hbar}\ \textrm{fixed}\end{subarray}}\frac{1}{Z}\Left<h(y)\Right>
=
\lim_{\begin{subarray}{c}N\to\infty\\\widehat{\hbar}\ \textrm{fixed}\end{subarray}}\frac{1}{Z}\widehat{h}(y)Z
=h^{(0)}(y).
\label{h_hat_classic}%(\ref{h_hat_classic})
\end{align}
By considering deformed potentials (\ref{V-tilde-BF}), instead of the loop equation (\ref{f_ward_id2}) for the partition function $Z$ we now obtain a loop equation for $\chi_{\alpha}(x,\sqrt{x}{\eta})$
\begin{equation}
\Left<S_+(y;x,{\eta})\chi_{\alpha}^{\textrm{ins}}(x,\sqrt{x}{\eta})\Right>=
-\frac{Q}{2\hbar y^{3/2}}\left(\xi_0-\frac{\alpha\eta}{2}-\frac{\hbar^2}{2}\partial_{\xi_0}\right)
\chi_{\alpha}(x,\sqrt{x}{\eta}).
\label{ward_wave_s}%(\ref{ward_wave_s})
\end{equation}

Similarly, from (\ref{t_p_x}) and (\ref{t_m_x}) we find the deformed energy-momentum tensor
\begin{equation}
T(y;x,{\eta})=T_+(y;x,{\eta})+T_-(y;x,{\eta}),
\nonumber\
\end{equation}
where
\begin{align}
yT_+(y;x,{\eta})&=
\frac{\alpha\sqrt{\beta}}{\hbar}\sum_{a=1}^N\frac{z_a}{(x-z_a)(y-z_a)}
+\frac{Q\sqrt{\beta}}{2}\sum_{a=1}^N\frac{y}{(y-z_a)^2}
+\frac{\beta}{2}\sum_{a,b=1}^N\frac{y}{(y-z_a)(y-z_b)} +
\nonumber\\
&\ \ \
-\frac{\beta}{4}\sum_{a,b=1}^N\frac{z_a{\eta}_a{\eta}_b}{(y-z_a)^2}
+\frac{\beta}{2}\sum_{a,b=1}^N\frac{z_az_b{\eta}_a{\eta}_b}{(y-z_a)(y-z_b)^2}
+\frac{\alpha{\eta}\sqrt{\beta}}{\hbar}\sum_{a=1}^N\frac{xz_a{\eta}_a}{(x-z_a)^2(y-z_a)} +
\nonumber\\
&\ \ \
+\frac{\alpha{\eta}\sqrt{\beta}}{2\hbar}\sum_{a=1}^N\frac{z_a^2{\eta}_a}{(x-z_a)(y-z_a)^2}
+\frac{\alpha{\eta}\sqrt{\beta}}{4\hbar}\sum_{a=1}^N\frac{z_a{\eta}_a}{(y-z_a)^2} +
\nonumber\\
&\ \ \
-\frac{\sqrt{\beta}}{2\hbar}\sum_{a=1}^N\frac{2yV_{\sr B}'(y)+yV_{\sr F}'(y){\eta}_a}{y-z_a}
-\frac{\sqrt{\beta}}{2\hbar}\sum_{a=1}^N\frac{V_{\sr F}(y)z_a{\eta}_a}{(y-z_a)^2}
-\frac{f(y)}{\hbar^2}+\frac{1}{16y},
\end{align}
and
\begin{align}
yT_-(y;x,{\eta})&=
\frac{1}{2(y-x)^2}\Big[2y\Delta_{\frac{\alpha}{\hbar}}+x{\eta}\Big(\partial_{{\eta}}-\frac{\alpha}{\hbar^2}V_{\sr F}(y)\Big)\Big]+
\frac{1}{y-x}\Big[x\partial_x+\frac{\alpha y}{\hbar^2}V_{\sr B}'(y)
-\frac{\alpha{\eta}y}{2\hbar^2}V_{\sr F}'(y)\Big] +
\nonumber\\
&\ \ \
+\frac{1}{\hbar^2}\Big[\frac{y}{2}V_{\sr B}'(y)^2+\frac12V_{\sr F}'(y)
\Big(V_{\sr F}(y)-\frac{\alpha{\eta}}{2}+\frac{\sqrt{\beta}\hbar}{2}\sum_{a=1}^N{\eta}_a\Big)
+\frac12Q\hbar yV_{\sr B}''(y)+f(y)\Big].
\label{t_def_m}%(\ref{t_def_m})
\end{align}
Here
\begin{align}
f(y)=
-\sqrt{\beta}{\hbar}\sum_{a=1}^N
\bigg(\frac{yV_{\sr B}'(y)-z_aV_{\sr B}'(z_a)}{y-z_a}+
\frac{\big(yV_{\sr F}'(y)-z_aV_{\sr F}'(z_a)\big){\eta}_a}{2(y-z_a)}
+\frac{V_{\sr F}^{(2)}(x,z_a)z_a{\eta}_a}{2(y-z_a)^2}\bigg),    \label{fy-R-NS}
\end{align}
where $V_{\sr F}^{(2)}(x,z_a)$ is defined in (\ref{vf2_def}). We can also represent $f(y)$ as a partial differential operator acting on the partition function $Z$ or the wave-function $\chi_{\alpha}(x,\sqrt{x}{\eta})$
\begin{align}
\widehat{f}(y)=\hbar^2\sum_{n=0}^{\infty}y^n \hskip -5pt
\sum_{m=n+1}^{\infty}\Big(mt_m\partial_{t_{m-n-1}}+\Big(m-\frac{n+1}{2}\Big)\xi_m\partial_{\xi_{m-n-1}}\Big),
\label{f_hat_op}%(\ref{f_hat_op})
\end{align}
and in the large $N$ limit (\ref{large_N_lim}) we reproduce $f^{(0)}(y)$ in (\ref{f_x_0_cl})
\begin{align}
\lim_{\begin{subarray}{c}N\to\infty\\\widehat{\hbar}\ \textrm{fixed}\end{subarray}}\frac{1}{Z}\Left<f(y)\Right>
=
\lim_{\begin{subarray}{c}N\to\infty\\\widehat{\hbar}\ \textrm{fixed}\end{subarray}}\frac{1}{Z}\widehat{f}(y)Z
=f^{(0)}(y).
\label{f_hat_classic}%(\ref{f_hat_classic})
\end{align}
Using the deformed potentials, instead of the loop equation (\ref{b_ward_id2}) for the partition function $Z$ we obtain now a loop equation for $\chi_{\alpha}(x,\sqrt{x}{\eta})$
\begin{equation}
\Left<T_+(y;x,{\eta})\chi_{\alpha}^{\textrm{ins}}(x,\sqrt{x}{\eta})\Right>=\frac{1}{16y^2}\chi_{\alpha}(x,\sqrt{x}{\eta}).
\label{ward_wave_t}%(\ref{ward_wave_t})
\end{equation}

%******************************************************************************************
%******************************************************************************************

\subsection{Building blocks of super-quantum curves}

We introduce now a representation of super-Virasoro operators $G_{n+\frac12}$ and $L_{n}$, acting on the wave-function $\chi_{\alpha}(x,\sqrt{x}{\eta})$ by
\begin{align}
G_{n+\frac12}\chi_{\alpha}(x,\sqrt{x}{\eta})&=
\oint_{y=x}\frac{dy}{2\pi i}(y-x)^{n+1}S(y;x,{\eta})\chi_{\alpha}(x,\sqrt{x}{\eta})
=
\nonumber\\
&= \oint_{y=x}\frac{dy}{2\pi i}(y-x)^{n+1}\Big[S_-(y;x,{\eta})
 -\frac{Q}{2\hbar y^{3/2}}\Big(\xi_0-\frac{\alpha\eta}{2}-\frac{\hbar^2}{2}\partial_{\xi_0}\Big)\Big]\chi_{\alpha}(x,\sqrt{x}{\eta}),
\nonumber \\
L_{n}\chi_{\alpha}(x,\sqrt{x}{\eta})&=
\oint_{y=x}\frac{dy}{2\pi i}(y-x)^{n+1}T(y;x,{\eta})\chi_{\alpha}(x,\sqrt{x}{\eta})
=
\nonumber\\
&= \oint_{y=x}\frac{dy}{2\pi i}(y-x)^{n+1}\Big[T_-(y;x,{\eta})
+\frac{1}{16y^2}\Big]\chi_{\alpha}(x,\sqrt{x}{\eta}), \nonumber
\end{align}
where we have used the loop equations (\ref{ward_wave_s}) and (\ref{ward_wave_t}).
From (\ref{s_def_m}) and (\ref{t_def_m}) we obtain for example
\begin{align}
\begin{split}
G_{-\frac12}&=x^{1/2}\eta\Big(\partial_x+\frac{\alpha}{\hbar^2}V_{\sr B}'(x)\Big)
-x^{-1/2}\Big(\partial_{\eta}-\frac{\alpha}{\hbar^2}V_{\sr F}(x)\Big),  \\
G_{-\frac32}&=\frac{x^{-3/2}}{4}\Delta_{\frac{\alpha}{\hbar}}\eta-
\frac{x^{-3/2}}{2}\Big(x\eta\Big(\partial_x+\frac{\alpha}{\hbar^2}V_{\sr B}'(x)\Big)
-\Big(\partial_{\eta}-\frac{\alpha}{\hbar^2}V_{\sr F}(x)\Big)\Big) + \\
&\quad +\frac{x^{-1/2}}{\hbar^2}\bigg[V_{\sr B}'(x)\Big(V_{\sr F}(x)+\frac{\alpha}{2}\eta-\frac{\hbar^2}{2}\partial_{\xi_0}\Big)+\big(Q\hbar+\alpha\big)V_{\sr F}'(x) + \\
&\qquad\quad
+\alpha x \eta V_{\sr B}''(x)
-\frac{Q\hbar}{2x}\Big(V_{\sr F}(x)-\frac{\alpha}{2}\eta-\frac{\hbar^2}{2}\partial_{\xi_0}\Big)+\widehat{h}(x)\bigg],
\label{s_w_op_a}%(\ref{s_w_op_a})
\end{split}
\end{align}
and
\begin{align}
\begin{split}
L_{-1}&=\partial_x+\frac{\alpha}{\hbar^2}V_{\sr B}'(x)
-\frac{\alpha\eta}{\hbar^2}V_{\sr F}'(x)
-\frac{x^{-1}\eta}{2}\Big(\partial_{\eta}-\frac{\alpha}{\hbar^2}V_{\sr F}(x)\Big),
\\
L_{-2}&=
\frac{x^{-2}\eta}{2}\Big(\partial_{\eta}-\frac{\alpha}{\hbar^2}V_{\sr F}(x)\Big)
-x^{-1}\partial_x
+\frac{1}{\hbar^2}\bigg[\frac{1}{2}V_{\sr B}'(x)^2
+\frac{x^{-1}}{2}V_{\sr F}'(x)\Big(V_{\sr F}(x)-\frac{\hbar^2}{2}\partial_{\xi_0}\Big) +
\\
&\qquad\qquad
+\Big(\frac{Q\hbar}{2}+\alpha\Big)V_{\sr B}''(x)
+\frac{3x^{-1}\alpha \eta}{4}V_{\sr F}'(x)
-\frac{3\alpha \eta}{4}V_{\sr F}''(x)
+x^{-1}\widehat{f}(x)\bigg]+\frac{1}{16x^2}.
\label{em_w_op_a}%(\ref{em_w_op_a})
\end{split}
\end{align}

For the wave-function $\widehat{\chi}_{\alpha}(x,\sqrt{x}{\eta})$ with the prefactor in (\ref{chi_hat_w}) we define super-Virasoro operators $\widehat{G}_{n+\frac12}$ and $\widehat{L}_{n}$ analogously
\begin{align}
\widehat{G}_{n+\frac12}\widehat{\chi}_{\alpha}(x,\sqrt{x}{\eta})
&= \oint_{y=x}\frac{dy}{2\pi i}(y-x)^{n+1}
\Big[S_-(y;x,{\eta})-\frac{Q}{2\hbar y^{3/2}}
\Big(\xi_0-\frac{\hbar^2}{2}\partial_{\xi_0}\Big)\Big]
\widehat{\chi}_{\alpha}(x,\sqrt{x}{\eta}),
\nonumber\\
\widehat{L}_{n}\widehat{\chi}_{\alpha}(x,\sqrt{x}{\eta})
&= \oint_{y=x}\frac{dy}{2\pi i}(y-x)^{n+1}
\Big[T_-(y;x,{\eta})+\frac{1}{16y^{2}}\Big]
\widehat{\chi}_{\alpha}(x,\sqrt{x}{\eta}).
\end{align}
For example, analogously to (\ref{s_w_op_a}) and (\ref{em_w_op_a}) we obtain
\begin{align}
\begin{split}
\widehat{G}_{-\frac12}&=x^{1/2}\eta\partial_x
-x^{-1/2}\partial_{\eta}\left(=\theta \partial_x-\partial_{\theta}\right),
\\
\widehat{G}_{-\frac32}&=\frac{x^{-3/2}}{4}\Delta_{\frac{\alpha}{\hbar}}\eta
-\frac{x^{-1/2}}{2}\eta\partial_x+\frac{x^{-3/2}}{2}\partial_{\eta} +
\\
&
+\frac{x^{-1/2}}{\hbar^2}\bigg[\Big(V_{\sr B}'(x)-\frac{Q\hbar}{2x}\Big)\Big(V_{\sr F}(x)-\frac{\hbar^2}{2}\partial_{\xi_0}\Big)
+Q\hbar V_{\sr F}'(x)+\widehat{h}(x)\bigg],
\label{s_w_op_t}%(\ref{s_w_op_t})
\end{split}
\end{align}
and
\begin{align}
\begin{split}
\widehat{L}_{-1}&=\partial_x-\frac{x^{-1}\eta}{2}\partial_{\eta}\left(=\partial_x\right),
\\
\widehat{L}_{-2}&=
-x^{-1}\partial_x+\frac{x^{-2}\eta}{2}\partial_{\eta}
+\frac{1}{\hbar^2}\bigg[\frac{1}{2}V_{\sr B}'(x)^2
+\frac{x^{-1}}{2}V_{\sr F}'(x)\Big(V_{\sr F}(x)-\frac{\hbar^2}{2}\partial_{\xi_0}\Big) +
\\
&\hspace{12em}
+\frac{Q\hbar}{2}V_{\sr B}''(x)
+x^{-1}\widehat{f}(x)\bigg]+\frac{1}{16x^2},
\label{em_w_op_t}%(\ref{em_w_op_t})
\end{split}
\end{align}
where in the second equalities of $\widehat{G}_{-\frac12}$ and $\widehat{L}_{-1}$ we changed variables as in (\ref{theta-eta}).
Operators found above agree with those identified in section \ref{ssec-wavefunction-Ramond} in the operator formalism.

Furthermore, we can transform (\ref{s_w_op_t}) and (\ref{em_w_op_t}) into operators $\widehat{\mathsf{G}}_{n+\frac12}$ and $\widehat{\mathsf{L}}_{n}$ acting on the bosonic and fermionic components of the wave-function (\ref{chi_bf_def}),
\begin{align}
\begin{split}
&
\widehat{\mathsf{G}}_{n+\frac12}\widehat{\chi}_{{\sr B},\alpha}(x)-
\sqrt{x}\eta\widehat{\mathsf{G}}_{n+\frac12}\widehat{\chi}_{{\sr F},\alpha}(x)
\equiv \widehat{G}_{n+\frac12}\widehat{\chi}_{\alpha}(x,\sqrt{x}{\eta}),
\\
&
\widehat{\mathsf{L}}_{n}\widehat{\chi}_{{\sr B},\alpha}(x)+
\sqrt{x}\eta\widehat{\mathsf{L}}_{n}\widehat{\chi}_{{\sr F},\alpha}(x)
\equiv \widehat{L}_{n}\widehat{\chi}_{\alpha}(x,\sqrt{x}{\eta}).
\end{split}
\end{align}
We find
\begin{align}
\begin{split}
&
\widehat{\mathsf{G}}_{-\frac12}\widehat{\chi}_{{\sr B},\alpha}(x)=\widehat{\chi}_{{\sr F},\alpha}(x),\qquad
\widehat{\mathsf{G}}_{-\frac12}\widehat{\chi}_{{\sr F},\alpha}(x)=\partial_x\widehat{\chi}_{{\sr B},\alpha}(x),    \\
&
\widehat{\mathsf{G}}_{-\frac32}\widehat{\chi}_{{\sr B},\alpha}(x)=
-\frac{x^{-1}}{2}\widehat{\chi}_{{\sr F},\alpha}(x)+\widehat{G}\widehat{\chi}_{{\sr B},\alpha}(x),
\\
&
\widehat{\mathsf{G}}_{-\frac32}\widehat{\chi}_{{\sr F},\alpha}(x)=
\frac{x^{-2}}{4}\Delta_{\frac{\alpha}{\hbar}}\widehat{\chi}_{{\sr B},\alpha}(x)
-\frac{x^{-1}}{2}\partial_x\widehat{\chi}_{{\sr B},\alpha}(x)
+\widehat{G}\widehat{\chi}_{{\sr F},\alpha}(x),
\label{w_comp_g_hat}%(\ref{w_comp_g_hat})
\end{split}
\end{align}
and
\begin{align}
\begin{split}
&
\widehat{\mathsf{L}}_{-1}\widehat{\chi}_{{\sr B},\alpha}(x)=\partial_x\widehat{\chi}_{{\sr B},\alpha}(x),\qquad
\widehat{\mathsf{L}}_{-1}\widehat{\chi}_{{\sr F},\alpha}(x)=\partial_x\widehat{\chi}_{{\sr F},\alpha}(x),
\\
&
\widehat{\mathsf{L}}_{-2}\widehat{\chi}_{{\sr B},\alpha}(x)=
\left(-x^{-1}\partial_x+\widehat{L}\right)\widehat{\chi}_{{\sr B},\alpha}(x),
\\
&
\widehat{\mathsf{L}}_{-2}\widehat{\chi}_{{\sr F},\alpha}(x)=
\left(-x^{-1}\partial_x+\widehat{L}\right)\widehat{\chi}_{{\sr F},\alpha}(x),
\label{w_comp_l_hat}%(\ref{w_comp_l_hat})
\end{split}
\end{align}
where
\begin{align}
\begin{split}
\widehat{G}&=
\frac{x^{-1/2}}{\hbar^2}\bigg[\Big(V_{\sr B}'(x)-\frac{Q\hbar}{2x}\Big)\Big(V_{\sr F}(x)-\frac{\hbar^2}{2}\partial_{\xi_0}\Big)+Q\hbar V_{\sr F}'(x)
+\widehat{h}(x)\bigg],
\\
\widehat{L}&=
\frac{1}{\hbar^2}\bigg[\frac{1}{2}V_{\sr B}'(x)^2
+\frac{x^{-1}}{2}V_{\sr F}'(x)\Big(V_{\sr F}(x)-\frac{\hbar^2}{2}\partial_{\xi_0}\Big)+\frac{Q\hbar}{2}V_{\sr B}''(x)+x^{-1}\widehat{f}(x)\bigg]+\frac{1}{16x^2}.
\nonumber
\end{split}
\end{align}

%******************************************************************************************
%******************************************************************************************

\subsection{Super-quantum curves at level 3/2}    \label{ssec-superquantum-32}

In order to identify super-quantum curves we consider the loop equation
\begin{align}
\begin{split}
&
\Left<\big(c_1 \sqrt{x}S_+(x;x,{\eta})+c_2 \eta xT_+(x;x,{\eta})\big)\chi_{\alpha}^{\textrm{ins}}(x,\sqrt{x}{\eta})\Right> =
\\
&=
\left(-\frac{c_1Q}{2\hbar x}\Big(\xi_0-\frac{\alpha\eta}{2}-\frac{\hbar^2}{2}\partial_{\xi_0}\Big)+\frac{c_2}{16x}\right)\chi_{\alpha}(x,\sqrt{x}{\eta}),
\label{q_c_3_2}%(\ref{q_c_3_2})
\end{split}
\end{align}
and analyze for which values of $c_1$ and $c_2$ it can be written as a differential equation. We find that it happens only for $c_1=-\alpha^2/\hbar^2$ and $c_2=2\alpha^2/\hbar^2$, and only for special values of the momentum $\alpha$
\begin{equation}
\alpha=0,\quad \beta^{\frac12}\hbar,\quad \textrm{or}\quad -\beta^{-\frac12}\hbar.
\label{momenta_3_2}%(\ref{momenta_3_2})
\end{equation}
Indeed, for this choice of parameters, the loop equation (\ref{q_c_3_2}) can be written as a differential equation
\begin{align}
\begin{split}
&
\Big(\partial_x\partial_{{\eta}}+\frac{\alpha}{\hbar^2}V_{\sr B}'(x)\partial_{{\eta}}
-\frac{\alpha}{\hbar^2}V_{\sr F}(x)\partial_x
-\frac{\alpha^2}{2\hbar^2}V_{\sr B}'(x)\partial_{\xi_0}
-\frac{\alpha Q}{2\hbar x}\big(\partial_{{\eta}}-\frac{\alpha}{2}\partial_{\xi_0}\big)+\frac{\alpha^2}{\hbar^4}\widehat{h}(x) +
\\
&+
\eta\Big(x\partial_x^2+\frac{3\alpha^2}{2\hbar^2}\partial_x+\frac{2\alpha}{\hbar^2}xV_{\sr B}'(x)\partial_x
-\frac{\alpha}{\hbar^2}V_{\sr F}'(x)\big(\partial_{\eta}-\frac{\alpha}{2}\partial_{\xi_0}\big)-\frac{2\alpha^2}{\hbar^4}\widehat{f}(x)\Big)\Big)\chi_{\alpha}(x,\sqrt{x}{\eta})=0.
\label{q_c_3_2_a}%(\ref{q_c_3_2_a})
\end{split}
\end{align}
Furthermore, using the relations in appendix \ref{app:commutation}, we find that this equation, for the momenta (\ref{momenta_3_2}), can be written in terms of operators (\ref{s_w_op_t}) and (\ref{em_w_op_t}) as follows
%\begin{align}
%\begin{split}
%&
%{\eta}\Big(\partial_x\partial_{{\eta}}+
%\frac{\alpha^2}{\hbar^4}\Big(V_{\sr B}'(x)V_{\sr F}(x)+Q\hbar V_{\sr F}'(x)
%-\frac{Q\hbar}{2x}V_{\sr F}(x)+\widehat{h}(x)
%\\
%&\hspace{6em}
%-\frac{Q\hbar^3}{2\alpha x}\partial_{\eta}
%+\frac{Q\hbar^3}{4x}\partial_{\xi_0}-\frac{\hbar^2}{2}V_{\sr B}'(x)\partial_{\xi_0}\Big)\Big)\widehat{\chi}_{\alpha}(x,{\eta})=0.
%\end{split}
%\end{align}
\begin{align}
\left(-\partial_x\Big(\frac{1}{\sqrt{x}}\partial_{\eta}\Big)-\frac{\alpha^2}{\hbar^2}\widehat{G}_{-\frac32}
-\sqrt{x}\eta\Big(\partial_x^2-\frac{2\alpha^2}{\hbar^2}\widehat{L}_{-2}\Big)\right)\widehat{\chi}_{\alpha}(x,\sqrt{x}{\eta})=0.
\end{align}
This is the super-quantum curve at level 3/2. As expected, by definitions (\ref{def_y_bf}), (\ref{yByF}), (\ref{chi_hat_def}), and relations (\ref{h_hat_classic}) and (\ref{f_hat_classic}), in the large $N$ limit (\ref{large_N_lim}) with $\beta=1$ this equation reduces to the super-spectral curve $A_{\sr F}(x,y_{\sr B}|y_{\sr F})=\sqrt{x}\eta A_{\sr B}(x,y_{\sr B}|y_{\sr F})$ in (\ref{b_f_sp_curve}).
On the other hand, in terms of (\ref{w_comp_g_hat}) and (\ref{w_comp_l_hat}), this super-quantum curve can be rewritten in the form of equations for the bosonic component $\widehat{\chi}_{{\sr B},\alpha}(x)$ defined in (\ref{chi_bf_def})
\begin{align}
\begin{split}
&
\Big(\widehat{\mathsf{L}}_{-1}\widehat{\mathsf{G}}_{-\frac12}-\frac{\alpha^2}{\hbar^2}\widehat{\mathsf{G}}_{-\frac32}\Big)\widehat{\chi}_{{\sr B},\alpha}(x)=0,  \\
&
\Big(\widehat{\mathsf{L}}_{-1}^2-\frac{2\alpha^2}{\hbar^2}\widehat{\mathsf{L}}_{-2}
+\frac{\alpha^2}{\hbar^2}\widehat{\mathsf{G}}_{-\frac32}\widehat{\mathsf{G}}_{-\frac12}
\Big)\widehat{\chi}_{{\sr B},\alpha}(x)=0.
\label{rns_q_curve_bf}
\end{split}
\end{align}
These equations indeed take form of Neveu-Schwarz singular vectors at level $3/2$, whose universal form we derived in (\ref{NS:singular:opeartors:examples}). Quantum curves at higher levels can be found analogously. However, as follows from the operator formalism discussed in section \ref{sec-Ramond}, it is clear that higher level quantum curves will also take form of Neveu-Schwarz singular vectors, expressed in terms of super-Virasoro generators found above.

%******************************************************************************************
%******************************************************************************************

\subsection{Multi-Penner Ramond-NS super-quantum curves}   \label{ssec-Ramond-NS-Penner}

To provide an explicit example, we specialize now our general considerations to the case of a supersymmetric multi-Penner model. In the case of the Ramond-NS model it is natural to consider the multi-Penner potential of the form
\be
V_{\mathsf{r}}(x,\eta) = V_{\sr B}(x)+ V_{\sr F}(x)\eta=\sum_{i=1}^M \alpha_i \log(x-x_i-\frac{1}{2}(x+x_i)\eta\eta_i){+\xi_0\eta},  \label{multiPenner-R-NS}
\ee
so that
\begin{align}
V_{\sr B}(x) =\sum_{i=1}^M \alpha_i\log(x-x_i),  \qquad
V_{\sr F}(x) = \xi_0 +\frac{1}{2}\sum_{i=1}^M \alpha_i\eta_i \frac{x+x_i}{x-x_i}.
\end{align}
The operators $\hat{h}(x)$ and $\hat{f}(x)$ introduced in (\ref{h-hat-Ramond}) and (\ref{f-hat-Ramond}), in the eigenvalue model can be represented via functions $h(x)$ and $f(x)$ given in (\ref{hy-R-NS}) and (\ref{fy-R-NS}). For the multi-Penner potential (\ref{multiPenner-R-NS}) these functions take form
\begin{align}
\begin{split}
h(x) &= \hbar \sqrt{\beta} \sum_{i=1}^M \sum_{a=1}^N \frac{\alpha_i x_i(\eta_a+ \eta_i)}{(x-x_i)(z_a-x_i)},    \\
f(x) &= \hbar \sqrt{\beta} \sum_{i=1}^M \sum_{a=1}^N \left(\frac{\alpha_i x_i}{(x-x_i)(z_a-x_i)}
-\frac{\alpha_ix_i\eta_iz_a\eta_a}{(x-x_i)(z_a-x_i)^2}
-\frac{\alpha_ix_i^2\eta_i\eta_a}{2(x-x_i)^2(z_a-x_i)} \right).
\nonumber
\end{split}
\end{align}
These functions can be represented by the action of operators expressed only in terms of parameters of the potential (i.e. $x_i$, $\eta_i$, {and $\xi_0$}), by taking advantage of the following identifications
\begin{align}
\partial_{x_i} = \frac{\sqrt{\beta}}{\hbar}\sum_{i,a} \Big{(} \frac{\alpha_i}{z_a-x_i}-\frac{\alpha_i\eta_iz_a\eta_a}{(z_a-x_i)^2}\Big{)},
\quad
\partial_{\eta_i} = -\frac{\sqrt{\beta}}{\hbar}\sum_{i,a}\frac{\alpha_i\eta_a}{2}\frac{z_a+x_i}{z_a-x_i},
\quad
\partial_{\xi_0}& =-\frac{\sqrt{\beta}}{\hbar}\sum_{a}\eta_a,
\nonumber
\end{align}
so that we find
\begin{align}
\begin{split}
\hat{h}(x)&=\hbar^2\sum_{i=1}^M \frac{1}{x-x_i}\left(x_i\eta_i\partial_{x_i}
-\partial_{\eta_i}\right)+\frac{\hbar^2}{2}V_{\sr B}'(x)\partial_{\xi_0},   \\
\hat{f}(x)&=
\hbar^2\sum_{i=1}^M \left(\frac{x_i}{x-x_i}\partial_{x_i}+\frac{x_i\eta_i}{2(x-x_i)^2}\partial_{\eta_i}\right)
+\frac{\hbar^2}{4}V_{\sr F}'(x)\partial_{\xi_0}. \label{p_operators_diff}
\end{split}
\end{align}
These formulas can be equivalently expressed in terms of variables $x$ and $\theta$, through $\partial_\eta \to \sqrt{x} \partial_{\theta}$ and $\partial_x \to \frac{\theta}{2x}\partial_{\theta}+\partial_x$.

Using (\ref{p_operators_diff}), we can now write the operators $\widehat{G}_{-3/2}$ in (\ref{s_w_op_t}) and $\widehat{L}_{-2}$ in (\ref{em_w_op_t}) as
\begin{align}
\begin{split}
\widehat{G}_{-3/2}&=\frac{\Delta_{\frac{\alpha}{\hbar}}\eta}{4x^{3/2}}
-\frac{\eta}{2x^{1/2}}\partial_x+\frac{1}{2x^{3/2}}\partial_{\eta} +\frac{1}{\hbar^2x^{1/2}}\bigg[\Big(V_{\sr B}'(x)-\frac{Q\hbar}{2x}\Big)V_{\sr F}(x) + \\
&\quad  + Q\hbar V_{\sr F}'(x)
+\hbar^2\sum_{i=1}^M \frac{1}{x-x_i}\left(x_i\eta_i\partial_{x_i}
-\partial_{\eta_i}\right) +\frac{Q\hbar^3}{4x}\partial_{\xi_0}\bigg],
\label{s_w_op_t-multiP}%(\ref{s_w_op_t-multiP})
\\
\widehat{L}_{-2}&=
-\frac{1}{x}\partial_x+\frac{\eta}{2x^2}\partial_{\eta}
+\frac{1}{\hbar^2}\bigg[\frac{1}{2}V_{\sr B}'(x)^2
+\frac{1}{2x}V_{\sr F}'(x)V_{\sr F}(x) +
\\
& \quad +\frac{Q\hbar}{2}V_{\sr B}''(x)
+\frac{1}{\hbar^2x}\sum_{i=1}^M
\left(\frac{x_i}{x-x_i}\partial_{x_i}+\frac{x_i\eta_i}{2(x-x_i)^2}\partial_{\eta_i}\right)
\bigg]+\frac{1}{16x^2}.
\end{split}
\end{align}
It is also convenient to introduce the following normalization factor
\be
C = \exp\Big{(}-\frac{1}{2\hbar^2}\sum_{i\neq j}\alpha_i \alpha_j \log(x_i - x_j -\frac{1}{2}(x_i+x_j)\eta_i\eta_j) + \frac{1}{\hbar^2}\sum_i \alpha_i\eta_i\xi_0\Big{)},
\ee
and to define a modified wave-function $\widetilde{\chi}_{\alpha}=C^{-1}\widehat{\chi}_{\alpha}$, its bosonic and fermionic components $\widetilde{\chi}_{{\sr B},\alpha}(x)=C^{-1}\widehat{\chi}_{{\sr B},\alpha}(x)$
and $\widetilde{\chi}_{{\sr F},\alpha}(x)=C^{-1}\widehat{\chi}_{{\sr F},\alpha}(x)$, as well as the corresponding modified operators $\widetilde{G}_{-r} = C^{-1}\widehat{G}_{-r}C$ and $\widetilde{L}_{-n} = C^{-1}\widehat{L}_{-n}C$. For $r=3/2$ and $n=2$ these operators take form
\begin{align}
\begin{split}
\widetilde{G}_{-3/2}&=\frac{\Delta_{\frac{\alpha}{\hbar}}\eta}{4x^{3/2}}
-\frac{\eta}{2x^{1/2}}\partial_x+\frac{1}{2x^{3/2}}\partial_{\eta} + \frac{Q\hbar}{4x^{3/2}}\Big(\partial_{\xi_0}-\frac{2}{\hbar^2}\xi_0\Big)+
\\
&\quad + \frac{1}{x^{1/2}}\sum_{i=1}^M \Big{(} \frac{\Delta_{\frac{\alpha_i}{\hbar}} \eta_i(x+x_i)}{(x-x_i)^2} + \frac{1}{x-x_i}\left(x_i\eta_i\partial_{x_i}
-\partial_{\eta_i}\right)\Big{)}, \\
\widetilde{L}_{-2}&=
-\frac{1}{x}\partial_x+\frac{\eta}{2x^2}\partial_{\eta} +\frac{1}{16x^2}
 +\sum_{i=1}^M \Big{(}\frac{\Delta_{\frac{\alpha_i}{\hbar}}}{(x-x_i)^2}
+\frac{x_i}{x(x-x_i)}\partial_{x_i}+\frac{x_i\eta_i}{2x(x-x_i)^2}\partial_{\eta_i}\Big{)},
\end{split} \label{multi_penner_operators-PC}
\end{align}
where $\Delta_{\alpha}=\frac{\alpha}{2} (\alpha - Q)$. Accordingly, for operators $\widehat{G}$ and $\widehat{L}$ in (\ref{w_comp_g_hat}) and (\ref{w_comp_l_hat}) we define $\widetilde{G}=C^{-1}\widehat{G}C$ and $\widetilde{L}=C^{-1}\widehat{L}C$, which take form
\begin{align}
\begin{split}
\widetilde{G}&=
\frac{1}{x^{1/2}}\sum_{i=1}^M \Big{(} \frac{\Delta_{\frac{\alpha_i}{\hbar}} \eta_i(x+x_i)}{(x-x_i)^2} + \frac{1}{x-x_i}\left(x_i\eta_i\partial_{x_i}
-\partial_{\eta_i}\right)\Big{)}+
\frac{Q\hbar}{4x^{3/2}}\Big(\partial_{\xi_0}-\frac{2}{\hbar^2}\xi_0\Big),
\\
\widetilde{L}&= \sum_{i=1}^M \Big{(}\frac{\Delta_{\frac{\alpha_i}{\hbar}}}{(x-x_i)^2}
+\frac{x_i}{x(x-x_i)}\partial_{x_i}+\frac{x_i\eta_i}{2x(x-x_i)^2}\partial_{\eta_i}\Big{)}+\frac{1}{16x^2}.
\end{split}
\end{align}
Finally, by \eqref{rns_q_curve_bf} we can write down Ramond-NS multi-Penner quantum curve equations at level 3/2 (for $\alpha=\beta^{\frac12}\hbar$ or $-\beta^{-\frac12}\hbar$) in terms of the components
$\widetilde{\chi}_{{\sr B},\alpha}(x)=C^{-1}\widehat{\chi}_{{\sr B},\alpha}(x)$ and $\widetilde{\chi}_{{\sr F},\alpha}(x)=C^{-1}\widehat{\chi}_{{\sr F},\alpha}(x)$,
\begin{align}
\begin{split}
&\partial_x \widetilde{\chi}_{{\sr F},\alpha} + \frac{\alpha^2}{2\hbar^2x}\widetilde{\chi}_{{\sr F},\alpha} - \frac{\alpha^2}{\hbar^2}\widetilde{G}\widetilde{\chi}_{{\sr B},\alpha} = 0, \\
&\partial_x^2\widetilde{\chi}_{{\sr B},\alpha}+\frac{3\alpha^2}{2\hbar^2x}\partial_x\widetilde{\chi}_{{\sr B},\alpha} - \frac{2\alpha^2}{\hbar^2}\widetilde{L}\widetilde{\chi}_{{\sr B},\alpha} + \frac{\alpha^2\Delta_{\frac{\alpha}{\hbar}}}{4x^2\hbar^2}\widetilde{\chi}_{{\sr F},\alpha}+ \frac{\alpha^2}{\hbar^2}\widetilde{G}\widetilde{\chi}_{{\sr F},\alpha}=0.
\end{split}
\end{align}

In the particular case of $M=1$, when certain condition on the parameters $\alpha$ and $\alpha_1$ is satisfied, one can rewrite the operators (\ref{multi_penner_operators-PC}) in the form which does not contain time derivatives. To this aim equations (\ref{s_vir_const-LH}) for $n=0$ are used. Those equations can be modified by taking account of the $x$-deformation of the potential (\ref{V-tilde-BF}), the prefactor appearing in (\ref{chi_hat_w}) and $C$, as well as excluding the right hand side of the equations (\ref{s_vir_const-LH}) for $n=0$ (which appears at both sides of those). In this manner we obtain following equations (for any value of $M$):
\be
\widetilde{l}_0 \widetilde{\chi}_{\alpha} = 0, \hspace{10mm} \widetilde{g}_0 \widetilde{\chi}_{\alpha} = 0,
\ee
where
\begin{align}
\begin{split}
 \widetilde{l}_0  =& - x \partial_x - \sum_{i=1}^M x_i \partial_{x_i} -\Delta_{\frac{\alpha}{\hbar}} - \sum_{i=1}^M \Delta_{\frac{\alpha_i}{\hbar}} + \Delta_{\frac{\alpha_\infty}{\hbar}},  \\
 \widetilde{g}_0  =& \partial_{\eta} + \sum_{i=1}^M \partial_{\eta_i} - x\eta \partial_x - \sum_{i=1}^M x_i\eta_i\partial_{x_i}
-\Delta_{\frac{\alpha}{\hbar}}\eta -\sum_{i=1}^M \Delta_{\frac{\alpha_i}{\hbar}} \eta_i - \frac{\alpha_{\infty}}{2\hbar^2}\Big(\widetilde{\alpha} - \hbar \sqrt{\beta}\sum_a \eta_a\Big),
\end{split}
\end{align}
and we used additional notation $\alpha_{\infty} = \sqrt{\beta}\hbar N + Q \hbar - \alpha - \sum_{i=1}^M \alpha_i$ and $\widetilde{\alpha}=\alpha \eta + \sum_{i=1}^M \alpha_i \eta_i$.
Imposing the constraint $\alpha_{\infty}=0$ and $M=1$, the modification of operators (\ref{multi_penner_operators-PC}), when acting on the wave function $\widetilde{\chi}_{\alpha}$, takes form:
\begin{align}
\begin{split}
x^{3/2}\bar{G}_{-3/2} &= -\frac{3x+x_1}{4(x-x_1)}\Delta_{\frac{\alpha}{\hbar}} \eta + \frac{2xx_1}{(x-x_1)^2}\Delta_{\frac{\alpha_1}{\hbar}}\eta_1 +
\frac{3x-x_1}{2(x-x_1)}(\partial_{\eta}-x\eta\partial_x) +\\
&+ \frac{Q}{2\hbar}\Big{(}\frac{\hbar^2}{2}\partial_{\xi_0}-\xi_0\Big{)}, \\
x^2 \bar{L}_{-2} &= \Big(\frac{x^2x_1 \eta_1 \eta}{2(x-x_1)^2} - \frac{(2x-x_1)x}{x-x_1}\Big)\partial_x
+ \Big(\frac{xx_1\eta_1\eta}{2(x-x_1)^2}-\frac{x}{x-x_1}\Big)\Delta_{\frac{\alpha}{\hbar}} +\\
&- \frac{xx_1\eta_1}{2(x-x_1)^2}\partial_{\eta}+\frac{1}{2}\eta \partial_{\eta} + \frac{xx_1}{(x-x_1)^2}\Delta_{\frac{\alpha_1}{\hbar}} + \frac{1}{16}.
\end{split}
\end{align}
We can expand the wave function with respect to the $\xi_0$ as $\widetilde{\chi}_{\alpha} = \widetilde{\chi}_{\alpha,+} + \widetilde{\chi}_{\alpha,-} \xi_0$. Then, these operators, when acting on components $\widetilde{\chi}_{\alpha,+}$ and $\widetilde{\chi}_{\alpha,-}$, give rise to the level $3/2$ quantum curves without time derivatives, constructed via (\ref{rns_q_curve_bf}) -- one should only be careful to commute first the operator $\widehat{G}_{-3/2}$ to the right of $\widehat{G}_{-1/2}$, so that it only acts directly on the wave function.

%******************************************************************************************
%******************************************************************************************
%******************************************************************************************
%******************************************************************************************

\section{Ramond-R super-eigenvalue model and super-quantum curves}    \label{sec-matrix-Ramond-R}

In this section we reconsider the Ramond-R sector and derive super-quantum curves using the formalism of eigenvalue models. Similarly as in section \ref{sec-Ramond-R}, for brevity we consider a particular case of supersymmetric Penner potential. As the starting point we consider the super-eigenvalue model found in (\ref{Ramond-R-eigenvalue}). We stress that the form of this eigenvalue model is not obvious to postulate a priori -- so the first advantage of the CFT approach is that this model can be identified at all. In this section we find corresponding quantum curves using eigenvalue model techniques. We confirm that they have the structure of Ramond singular vectors and show that they agree with relevant supersymmetric BPZ equations in the Ramond sector.

The Ramond-R wave-function for the one-Penner potential identified in \eqref{psi-Ramond-R-2} can be written in the form of eigenvalue integral as follows
\begin{align}
\chi_{\alpha}^{\sr R}(x,\xi)= \int\! d^N\!z\, d^N\!\theta\ \Delta_{{\sr R},x}(\vect{z},\vect{\theta})^{\beta}
\mathrm{e}^{-\frac{\sqrt{\beta}}{\hbar}\sum_{a=1}^N
V_{{\sr R},x,\xi}(z_a,\theta_a)},
\label{z_ram_mat}
\end{align}
where
\begin{align}
\Delta_{{\sr R},x}(\vect{z},\vect{\theta})=
\prod_{1\le a<b\le N}\left(z_a-z_b-
\left(\sqrt{\frac{z_a(x-z_b)}{z_b(x-z_a)}}+\sqrt{\frac{z_b(x-z_a)}{z_a(x-z_b)}}
\right)\frac{\theta_a\theta_b}{2}\right),
\end{align}
and
\begin{align}
\begin{split}
V_{{\sr R},x,\xi}(z,\theta) & =
V_{{\sr B},x}(z)+V_{{\sr F},x,\xi}(z)\theta, \\
V_{{\sr B},x}(z)&=\alpha \log (x-z)+\gamma \log (z-w),\\
V_{{\sr F},x,\xi}(z)&=\frac{\gamma \eta}{2(z-w)}
\left(\sqrt{\frac{w(x-z)}{z(x-w)}}+\sqrt{\frac{z(x-w)}{w(x-z)}}\right)
+\frac{\sqrt{x}\xi}{\sqrt{z(x-z)}}.
\end{split}
\end{align}
By reference to \eqref{S:R:correlator} we consider the following shift invariance with a fermionic constant $\delta$ of the wave-function \eqref{z_ram_mat} as
\begin{align}
z_a\ \to\ z_a+\frac{\theta_a \delta\sqrt{z_a(x-z_a)}}{y-z_a},\qquad
\theta_a\ \to\ \theta_a+\frac{\delta\sqrt{z_a(x-z_a)}}{y-z_a},
\end{align}
and obtain a loop equation
\begin{align}
0= \int\! d^N\!z\, d^N\!\theta\
\sum_{a=1}^N\left(\theta_a\partial_{z_a}-\partial_{\theta_a}\right)
\left(\frac{\sqrt{z_a(x-z_a)}}{y-z_a}
\Delta_{{\sr R},x}(\vect{z},\vect{\theta})^{\beta}
\mathrm{e}^{-\frac{\sqrt{\beta}}{\hbar}\sum_{a=1}^N
V_{{\sr R},x,\xi}(z_a,\theta_a)}\right).
\label{ram_mat_fl}
\end{align}
Using the notation
\begin{align}
\Left<\mathcal{O}\Right>_{\xi}=
\int\! d^N\!z\, d^N\!\theta\ \mathcal{O} \Delta_{{\sr R},x}(\vect{z},\vect{\theta})^{\beta}
\mathrm{e}^{-\frac{\sqrt{\beta}}{\hbar}\sum_{a=1}^N
V_{{\sr R},x,\xi}(z_a,\theta_a)}
\end{align}
for an operator $\mathcal{O}$, the loop equation \eqref{ram_mat_fl} at $y=x$ can be written as
\begin{align}
\Left<\mathcal{S}_{+}(x)\Right>_{\xi}=0,
\label{ram_mat_fl_s}
\end{align}
where
\begin{align}
\mathcal{S}_{+}(x)&=
\left(\frac{1-\beta}{2}+\frac{\alpha\sqrt{\beta}}{\hbar}\right)
\sum_{a=1}^N\frac{x\theta_a}{\sqrt{z_a}(x-z_a)^{3/2}}
+\frac{\gamma\sqrt{\beta}}{\hbar}\sum_{a=1}^N
\frac{w\theta_a}{(w-z_a)\sqrt{z_a(x-z_a)}}+
\nonumber\\
&\ \ \
+\frac{\beta}{2}\sum_{a=1}^N\frac{xf(x,\vect{z},\vect{\theta})}{x-z_a}
-\frac{(\alpha+\gamma)\sqrt{\beta}}{\hbar}f(x,\vect{z},\vect{\theta})
-\frac{\sqrt{\beta}\xi}{\hbar}\sum_{a=1}^N\frac{\sqrt{x}}{x-z_a}+
\nonumber\\
&\ \ \
+\frac{\gamma\sqrt{\beta}\eta}{2\hbar}\left(
\sqrt{\frac{w}{x-w}}\sum_{a=1}^N\frac{1}{w-z_a}
+\sqrt{\frac{x-w}{w}}\sum_{a=1}^N\frac{z_a}{(w-z_a)(x-z_a)}\right).
\end{align}
Here we introduced
\begin{align}
f(x,\vect{z},\vect{\theta})=\sum_{a=1}^N\frac{\theta_a}{\sqrt{z_a(x-z_a)}},
\end{align}
%In the above and the following computations, we can use useful identities
%\begin{align}
%\begin{split}
%&
%f(x,\vect{z},\vect{\theta})^2=0,\\
%&
%\sum_{a=1}^N\frac{\theta_a f(x,\vect{z},\vect{\theta})}{(y-z_a)\sqrt{z_a(x-z_a)}}=
%\sum_{a,b=1}^N\frac{\theta_a\theta_b}{(y-z_a)\sqrt{z_az_b(x-z_a)(x-z_b)}}.
%\end{split}
%\end{align}
and note that $f(x,\vect{z},\vect{\theta})^2=0$. The loop equation \eqref{ram_mat_fl_s} is equivalent to
\begin{align}
\Left<\mathcal{S}_{1,+}(x)\Right>_{\xi=0}=
\Left<\mathcal{S}_{2,+}(x)\Right>_{\xi=0}=0,
\label{ram_mat_fl_s2}
\end{align}
where
\begin{align}
\mathcal{S}_{1,+}(x)&=
\left(\frac{1-\beta}{2}+\frac{\alpha\sqrt{\beta}}{\hbar}\right)
\sum_{a=1}^N\frac{x\theta_a}{\sqrt{z_a}(x-z_a)^{3/2}}
+\frac{\gamma\sqrt{\beta}}{\hbar}\sum_{a=1}^N
\frac{w\theta_a}{(w-z_a)\sqrt{z_a(x-z_a)}} +
\nonumber\\
&\ \ \
+\frac{\beta}{2}\sum_{a=1}^N\frac{xf(x,\vect{z},\vect{\theta})}{x-z_a}
-\frac{(\alpha+\gamma)\sqrt{\beta}}{\hbar}f(x,\vect{z},\vect{\theta}) +
\nonumber\\
&\ \ \
+\frac{\gamma\sqrt{\beta}\eta}{2\hbar}\left(
\sqrt{\frac{w}{x-w}}\sum_{a=1}^N\frac{1}{w-z_a}
+\sqrt{\frac{x-w}{w}}\sum_{a=1}^N\frac{z_a}{(w-z_a)(x-z_a)}\right),
\end{align}
and
\begin{align}
\mathcal{S}_{2,+}(x)&=
\left(\frac{1-\beta}{2}+\frac{\alpha\sqrt{\beta}}{\hbar}\right)
\sum_{a=1}^N\frac{x\theta_a f(x,\vect{z},\vect{\theta})}{\sqrt{z_a}(x-z_a)^{3/2}}
+\frac{\gamma\sqrt{\beta}}{\hbar}\sum_{a=1}^N
\frac{w\theta_a f(x,\vect{z},\vect{\theta})}{(w-z_a)\sqrt{z_a(x-z_a)}} +
\nonumber\\
&\ \ \
-\sum_{a=1}^N\frac{1}{x-z_a}
+\frac{\gamma\sqrt{\beta}\eta}{2\hbar}\left(
\sqrt{\frac{w}{x-w}}\sum_{a=1}^N\frac{f(x,\vect{z},\vect{\theta})}{w-z_a}
+\sqrt{\frac{x-w}{w}}\sum_{a=1}^N
\frac{z_af(x,\vect{z},\vect{\theta})}{(w-z_a)(x-z_a)}\right).
\end{align}

We now find differential equations for the wave-functions
\begin{align}
\chi_{+,\alpha}^{\sr R}(x)= \chi_{\alpha}^{\sr R}(x,0),\qquad
\chi_{-,\alpha}^{\sr R}(x)= -\hbar \partial_{\xi}\chi_{\alpha}^{\sr R}(x,\xi),
\end{align}
from the analysis of the loop equations \eqref{ram_mat_fl_s2}. Consider first the following combinations of these equations
\begin{align}
\Left<c_1\mathcal{S}_{2,+}(x)+c_2\eta\mathcal{S}_{1,+}(x)\Right>_{\xi=0}=
\Left<c_3\mathcal{S}_{1,+}(x)+c_4\eta\mathcal{S}_{2,+}(x)\Right>_{\xi=0}=0.
\label{ram_mat_fl_bf}
\end{align}
Here $c_{i}$ for $i=1,\ldots,4$ are functions of $x$ and $w$ which are determined by comparing these equations with differential equations
\begin{align}
\begin{split}
&
\partial_x \chi_{+,\alpha}^{\sr R}(x)
+A_1\partial_{\eta}\chi_{-,\alpha}^{\sr R}(x)
+A_2\eta\partial_{\eta}\chi_{+,\alpha}^{\sr R}(x)
+A_3\eta\partial_{w}\chi_{-,\alpha}^{\sr R}(x)
+A_4\eta\chi_{-,\alpha}^{\sr R}(x)=0,\\
&
\partial_x \chi_{-,\alpha}^{\sr R}(x)
+B_1\partial_{\eta}\chi_{+,\alpha}^{\sr R}(x)
+B_2\eta\partial_{\eta}\chi_{-,\alpha}^{\sr R}(x)
+B_3\eta\partial_{w}\chi_{+,\alpha}^{\sr R}(x)
+B_4\chi_{-,\alpha}^{\sr R}(x)=0,
\label{ram_mat_fl_p}
\end{split}
\end{align}
where $A_i$ and $B_i$ for $i=1,\ldots,4$ are functions of $x$ and $w$.
In the computation we use the following relations
\begin{align}
\begin{split}
\partial_x \chi_{+,\alpha}^{\sr R}(x)&=
\Left<\frac{\beta}{4}\sum_{a=1}^N\frac{x\theta_a f(x,\vect{z},\vect{\theta})}
{\sqrt{z_a}(x-z_a)^{3/2}}
-\frac{\alpha\sqrt{\beta}}{\hbar}\sum_{a=1}^N\frac{1}{x-z_a}\Right. +\\
&\qquad
\Left.+\frac{\gamma\sqrt{\beta}x\eta}{4\hbar\sqrt{w(x-w)}}
\sum_{a=1}^N\frac{\theta_a}{\sqrt{z_a}(x-z_a)^{3/2}}
-\frac{\gamma\sqrt{\beta}x\eta f(x,\vect{z},\vect{\theta})}
{4\hbar\sqrt{w}(x-w)^{3/2}}\Right>_{\xi=0},\\
\partial_x \chi_{-,\alpha}^{\sr R}(x)&=
\Left<-\frac{\alpha\beta}{\hbar}\sum_{a=1}^N
\frac{\sqrt{x}f(x,\vect{z},\vect{\theta})}{x-z_a}
-\frac{\sqrt{\beta}}{2}\sum_{a=1}^N\frac{\sqrt{x}\theta_a}
{\sqrt{z_a}(x-z_a)^{3/2}}
+\frac{\sqrt{\beta}f(x,\vect{z},\vect{\theta})}{2\sqrt{x}}\Right. +\\
&\qquad
\Left.+\frac{\gamma\beta x^{3/2}\eta}{4\hbar\sqrt{w(x-w)}}
\sum_{a=1}^N\frac{\theta_a f(x,\vect{z},\vect{\theta})}{\sqrt{z_a}(x-z_a)^{3/2}}
\Right>_{\xi=0},
\end{split}
\end{align}
\begin{align}
\partial_{\eta} \chi_{+,\alpha}^{\sr R}(x)&=
\Left<\frac{\gamma\sqrt{\beta}}{2\hbar}
\left(\sum_{a=1}^N\frac{2\sqrt{w(x-w)}\theta_a}
{(w-z_a)\sqrt{z_a(x-z_a)}}
+\left(\sqrt{\frac{w}{x-w}}-\sqrt{\frac{x-w}{w}}\right)
f(x,\vect{z},\vect{\theta})\right)\Right>_{\xi=0},\nonumber\\
\partial_{\eta} \chi_{-,\alpha}^{\sr R}(x)&=
\Left<\frac{\gamma\beta}{\hbar}
\sum_{a=1}^N\frac{\sqrt{xw(x-w)}\theta_a f(x,\vect{z},\vect{\theta})}
{(w-z_a)\sqrt{z_a(x-z_a)}}\Right>_{\xi=0},
\end{align}
and
\begin{align}
\eta\partial_w \chi_{+,\alpha}^{\sr R}(x)=
\Left<-\frac{\gamma\sqrt{\beta}\eta}{\hbar}
\sum_{a=1}^N\frac{1}{w-z_a}\Right>_{\xi=0},\quad
\eta\partial_w \chi_{-,\alpha}^{\sr R}(x)=
\Left<-\frac{\gamma\beta\eta}{\hbar}
\sum_{a=1}^N\frac{\sqrt{x}f(x,\vect{z},\vect{\theta})}{w-z_a}\Right>_{\xi=0}.
\end{align}
Then we find that $c_i$ for $i=1,2,3,4$ are determined as
\begin{align}
c_1=\frac{\alpha\sqrt{\beta}}{\hbar},\quad
c_2=\frac{\alpha\gamma}{\hbar^2\sqrt{w(x-w)}},\quad
c_3=-\frac{2\alpha}{\hbar\sqrt{x}},\quad
c_4=\frac{\alpha\gamma\sqrt{\beta}}{\hbar^2}\sqrt{\frac{x}{w(x-w)}},
\nonumber
\end{align}
and only for
\begin{align}
\alpha=\frac{\beta^{\frac12}\hbar}{2}\quad \textrm{or}\quad -\frac{\beta^{-\frac12}\hbar}{2},
\label{lv1_rs_mat}
\end{align}
the loop equations \eqref{ram_mat_fl_bf} are rewritten as the differential equations \eqref{ram_mat_fl_p} with
\begin{align}
\begin{split}
&
A_1=\frac{\alpha}{\hbar}\sqrt{\frac{w}{x(x-w)}},\qquad
A_2=\frac{\alpha\gamma}{\hbar^2(x-w)},\qquad
A_3=-\frac{\alpha}{\hbar}\sqrt{\frac{w}{x(x-w)}},\\
&
A_4=-\frac{\alpha}{\hbar}\sqrt{\frac{w}{x(x-w)}}
\left(\frac{x\Delta_{\frac{\gamma}{\hbar}}}{w(x-w)}
-\frac{\alpha\gamma}{\hbar^2(x-w)}\right),
\end{split}
\end{align}
and
\begin{align}
\begin{split}
&
B_1=-\frac{2\alpha}{\hbar}\sqrt{\frac{w}{x(x-w)}},\qquad
B_2=\frac{\alpha\gamma}{\hbar^2(x-w)},\qquad
B_3=\frac{2\alpha}{\hbar}\sqrt{\frac{w}{x(x-w)}},\\
&
B_4=\frac{2\alpha \Delta_{\frac{\gamma}{\hbar}}}{\gamma x}
+\frac{\alpha\gamma w}{\hbar^2x(x-w)},
\end{split}
\end{align}
where $\Delta_{\gamma}=\frac{\gamma}{2}(\gamma-Q)$ and $Q=\beta^{-1/2}-\beta^{1/2}$.

We can rewrite the above differential equations as differential equations for
\begin{align}
\begin{split}
\widetilde{\chi}_{+,\alpha}^{\sr R}(x)&=
(x-w)^{\frac{\alpha\gamma}{\hbar^2}}\chi_{+,\alpha}^{\sr R}(x),
\\
\widetilde{\chi}_{-,\alpha}^{\sr R}(x)&=
\frac{1}{s\sqrt{2}}\Big((x-w)^{\frac{\alpha\gamma}{\hbar^2}}\chi_{-,\alpha}^{\sr R}(x)
-\frac{\gamma\eta}{\hbar}\frac{\sqrt{x}}{\sqrt{w(x-w)}}
\widetilde{\chi}_{+,\alpha}^{\sr R}(x)\Big),
\label{wave_r_def}
\end{split}
\end{align}
where $s$ is a constant. Then we find differential equations
\begin{align}
\begin{split}
\partial_x \widetilde{\chi}_{+,\alpha}^{\sr R}(x)&=
\frac{s \alpha}{\hbar}\sqrt{\frac{2w}{x(x-w)}}\left(
\eta\partial_w-\partial_{\eta}+
\frac{\Delta_{\frac{\gamma}{\hbar}}x\eta}{w(x-w)}
\right)\widetilde{\chi}_{-,\alpha}^{\sr R}(x),\\
\left(\partial_x -\frac{\alpha Q}{\hbar x}
\right)\widetilde{\chi}_{-,\alpha}^{\sr R}(x)&=
-\frac{\alpha}{s \hbar}\sqrt{\frac{2w}{x(x-w)}}\left(
\eta\partial_w-\partial_{\eta}+\frac{\Delta_{\frac{\gamma}{\hbar}}x\eta}{w(x-w)}
\right)\widetilde{\chi}_{+,\alpha}^{\sr R}(x),
\label{ramond_q_c_mat}
\end{split}
\end{align}
which hold for special values of $\alpha$ in \eqref{lv1_rs_mat}. These are one-Penner Ramond-R quantum curves at level 1 we wished to find. As expected, by multiplying the factor $x^{-1/8}$ by $\widetilde{\chi}_{\pm,\alpha}^{\sr R}(x)$, and taking $s=-\mathrm{e}^{\frac{i\pi}{4}}$ as in \eqref{ramond_theta_def}, they reproduce differential equations (\ref{NVD:Ramond}) found using CFT techniques, and in fact take form of Ramond versions of BPZ equations \cite{Zamolodchikov:1988nm}.

In \eqref{ram_mat_fl} we considered the fermionic type loop equation which is generated by the superconformal current. We can also consider the bosonic type loop equation
\begin{align}
0= \int\! d^N\!z\, d^N\!\theta\
\sum_{a=1}^N \left(\partial_{z_a} - \frac12\partial_{\theta_a}\frac{\theta_a}{y-z_a}\right)
\left(\frac{1}{y-z_a}
\Delta_{{\sr R},x}(\vect{z},\vect{\theta})^{\beta}
\mathrm{e}^{-\frac{\sqrt{\beta}}{\hbar}\sum_{a=1}^N
V_{{\sr R},x,\xi}(z_a,\theta_a)}\right),
\label{ram_mat_bos}
\end{align}
which is obtained by an infinitesimal shift as
\begin{equation}
z_a\ \to\ z_a+\frac{\epsilon}{y-z_a},\qquad
\theta_a\ \to\ \theta_a+\frac{\theta_a \epsilon}{2(y-z_a)^2},
\label{f_shift_r}
\end{equation}
and this shift is generated by the energy-momentum tensor. The loop equation \eqref{ram_mat_bos} can be written as
\begin{align}
\Left<\mathcal{T}_{+}(y)\Right>_{\xi}=0,
\label{ram_mat_bo_s}
\end{align}
where
\begin{align}
\mathcal{T}_{+}(y)&=
\frac{1-\beta}{2}\sum_{a=1}^N \frac{1}{(y-z_a)^2}
+\frac{\beta}{2}\sum_{a,b=1}^N \frac{1}{(y-z_a)(y-z_b)} +
\nonumber\\
&\ \ \
-\frac{\beta}{4}\sum_{a,b=1}^N \frac{x^2 \theta_a\theta_b}
{z_a^{3/2}(x-z_a)^{3/2}\sqrt{z_b(x-z_b)}(y-z_a)} +
\nonumber\\
&\ \ \
-\frac{\beta}{8}\sum_{a,b=1}^N \frac{\theta_a\theta_b (z_a-z_b)}{(y-z_a)^2(y-z_b)^2}
\left(\sqrt{\frac{z_a(x-z_b)}{z_b(x-z_a)}}+\sqrt{\frac{z_b(x-z_a)}{z_a(x-z_b)}}\right) +
\nonumber\\
&\ \ \
+\frac{\sqrt{\beta}}{\hbar}\sum_{a=1}^N
\Bigg[
\frac{\alpha}{(x-z_a)(y-z_a)}+\frac{\gamma}{(w-z_a)(y-z_a)}
-\frac{\gamma x^2\eta \theta_a}{4\sqrt{w(x-w)}z_a^{3/2}(x-z_a)^{3/2}(y-z_a)} +
\nonumber\\
&\ \ \ \ \ \
+\frac{\gamma\eta\theta_a}{2}
\left(\frac{1}{(w-z_a)^2(y-z_a)}+\frac{1}{2(w-z_a)(y-z_a)^2}\right)
\left(\sqrt{\frac{w(x-z_a)}{z_a(x-w)}}+\sqrt{\frac{z_a(x-w)}{w(x-z_a)}}\right) +
\nonumber\\
&\ \ \ \ \ \
+\frac{(x-2z_a)\sqrt{x}\xi \theta_a}{2z_a^{3/2}(x-z_a)^{3/2}(y-z_a)}
-\frac{\sqrt{x}\xi\theta_a}{2\sqrt{z_a(x-z_a)}(y-z_a)^2}\Bigg].
\end{align}
By expanding the loop equation \eqref{ram_mat_bo_s} around $y=\infty$ we get Virasoro constraints for the wave-function \eqref{z_ram_mat}. Especially, by taking the expansion coefficient of $y^{-2}$ (subleading order) we get
\begin{align}
0&=
\Left<
\frac{(1-\beta)N+\beta N^2}{2}
-\frac{\sqrt{\beta}(\alpha+\gamma)N}{\hbar}
+\frac{\sqrt{\beta}\alpha x}{\hbar}\sum_{a=1}^N\frac{1}{x-z_a}
-\frac{\beta x^2}{4}\sum_{a=1}^N
\frac{\theta_a f(x,\vect{z},\vect{\theta})}{\sqrt{z_a}(x-z_a)^{3/2}}
\Right. +
\nonumber\\
&\ \ \
-\frac{\sqrt{\beta}\gamma x^2\eta}{4\hbar \sqrt{w(x-w)}}
\sum_{a=1}^N \frac{\theta_a}{\sqrt{z_a}(x-z_a)^{3/2}}
-\frac{\sqrt{\beta}\xi \sqrt{x}}{2\hbar}
\sum_{a=1}^N \frac{\sqrt{z_a}\theta_a}{(x-z_a)^{3/2}} +
\nonumber\\
&\ \ \
+\frac{\sqrt{\beta}\gamma w}{\hbar}\sum_{a=1}^N \frac{1}{w-z_a}
+\frac{\sqrt{\beta}\gamma w\eta}{2\hbar}\sum_{a=1}^N
\frac{\theta_a}{(w-z_a)^2}
\left(\sqrt{\frac{w(x-z_a)}{z_a(x-w)}}+\sqrt{\frac{z_a(x-w)}{w(x-z_a)}}\right) +
\nonumber\\
&\ \ \
\Left.
-\frac{\sqrt{\beta}\gamma \eta}{4\hbar}\sum_{a=1}^N
\frac{\theta_a}{w-z_a}
\left(\sqrt{\frac{w(x-z_a)}{z_a(x-w)}}+\sqrt{\frac{z_a(x-w)}{w(x-z_a)}}\right)
\Right>_{\xi}.
\end{align}
We find that this equation can be written as a differential equation for the wave-function \eqref{z_ram_mat}:
\begin{align}
\left(x\partial_x+w\partial_w+\frac{1}{2}\eta\partial_{\eta}\right)
\chi_{\alpha}^{\sr R}(x,\xi)=
\left(\Delta_{\frac{\alpha_0}{\hbar}}-\Delta_{\frac{\alpha}{\hbar}}
-\Delta_{\frac{\gamma}{\hbar}}-\frac{\alpha\gamma}{\hbar^2}\right)
\chi_{\alpha}^{\sr R}(x,\xi),
\end{align}
where $\alpha_0$ is given in \eqref{PsiR:definition}:
\begin{align}
\frac{\alpha_0}{\hbar}=
-\frac{\alpha+\gamma}{\hbar}+N\sqrt{\beta}+\beta^{-\frac12}-\beta^{\frac12}.
\end{align}
We then obtain the following differential equation for the wave-function \eqref{wave_r_def}:
\begin{align}
\left(x\partial_x+w\partial_w+\frac{1}{2}\eta\partial_{\eta}\right)
\widetilde{\chi}_{\pm,\alpha}^{\sr R}(x)=
\left(\Delta_{\frac{\alpha_0}{\hbar}}-\Delta_{\frac{\alpha}{\hbar}}
-\Delta_{\frac{\gamma}{\hbar}}\right)
\widetilde{\chi}_{\pm,\alpha}^{\sr R}(x).
\label{ramond_l0_vir}
\end{align}
By including the factor $x^{-1/8}$ we see that this equation agrees with
the equation \eqref{L0:Ward:identity:Ramond} obtained by the scaling covariance.

As discussed in section \ref{subsec_rr_sq}, let us apply the constraint equation
\eqref{ramond_l0_vir} to the equations in \eqref{ramond_q_c_mat}. Then by defining
\begin{equation}
\widetilde{\chi}^{\sr R}_{\pm,\alpha}(x)
=
\widetilde{f}^{\sr R}_{\pm,\alpha}(x) \mp \eta \widetilde{g}^{\sr R}_{\pm,\alpha}(x),
\hskip 1cm
\widetilde{g}^{\sr R}_{\pm,\alpha}(x) = \mp \partial_\eta \widetilde{\chi}^{\sr R}_{\pm,\alpha}(x),
\end{equation}
as in \eqref{fg_ramond}, we obtain two pairs of coupled ordinary differential equations \eqref{BPZ:R:1} and \eqref{BPZ:R:2}:
\begin{align}
\begin{split}
\partial_x \widetilde{f}^{\sr R}_{+,\alpha}(x)
& =
-\frac{s \alpha}{\hbar}\sqrt{\frac{2w}{x(x-w)}}
\widetilde{g}^{\sr R}_{-,\alpha}(x),
\\
\left(\partial_x - \frac{\alpha Q}{\hbar x}\right)
\widetilde{g}^{\sr R}_{-,\alpha}(x)
& =
-\frac{\alpha}{s \hbar}\sqrt{\frac{2w}{x(x-w)}}
\left(\frac{\Delta_{\frac{\alpha_0}{\hbar}}
-\Delta_{\frac{\alpha}{\hbar}}  -x\partial_x}{w}+\frac{\Delta_{\frac{\gamma}{\hbar}}}{x-w}\right)
\widetilde{f}^{\sr R}_{+,\alpha}(x),
\\
\left(\partial_x - \frac{\alpha Q}{\hbar x}\right)
\widetilde{f}^{\sr R}_{-,\alpha}(x)
& =
-\frac{\alpha}{s \hbar}\sqrt{\frac{2w}{x(x-w)}}
\widetilde{g}^{\sr R}_{+,\alpha}(x),
\\
\partial_x \widetilde{g}^{\sr R}_{+,\alpha}(x)
& =
-\frac{s \alpha}{\hbar}\sqrt{\frac{2w}{x(x-w)}}
\left(\frac{\Delta_{\frac{\alpha_0}{\hbar}}
-\Delta_{\frac{\alpha}{\hbar}}  -x\partial_x}{w}+\frac{\Delta_{\frac{\gamma}{\hbar}}}{x-w}\right)
\widetilde{f}^{\sr R}_{-,\alpha}(x),
\label{r_q_curve_reduced}
\end{split}
\end{align}
for $\frac{\beta^{1/2}\hbar}{2}$ or $-\frac{\beta^{-1/2}\hbar}{2}$.
Then we find that the wave-functions $\widetilde{f}^{\sr R}_{\pm,\alpha}(x)$ obey the following hypergeometric differential equations:
\begin{align}
\begin{split}
&
\left[
\partial_x^2+\left(\frac{2\alpha^2}{\hbar^2 x}
+\frac{\alpha Q}{\hbar (x-w)}\right)\partial_x
-\frac{2\alpha^2}{\hbar^2}\left(
\frac{\Delta_{\frac{\gamma}{\hbar}}}{(x-w)^2}+
\frac{\Delta_{\frac{\alpha_0}{\hbar}}
-\Delta_{\frac{\gamma}{\hbar}}-\Delta_{\frac{\alpha}{\hbar}}}{x(x-w)}
\right)\right]
\widetilde{f}^{\sr R}_{+,\alpha}(x)=0,
\\
&
\left[
\partial_x^2+\left(\frac{2\alpha^2}{\hbar^2 x}
+\frac{\alpha Q}{\hbar (x-w)}\right)\partial_x
+\frac{\alpha Q}{\hbar x^2}
-\frac{2\alpha^2}{\hbar^2}\left(
\frac{\Delta_{\frac{\gamma}{\hbar}}}{(x-w)^2}+
\frac{\Delta_{\frac{\alpha_0}{\hbar}}
-\Delta_{\frac{\gamma}{\hbar}}-\Delta_{\frac{\alpha}{\hbar}}}{x(x-w)}
\right)\right]
\widetilde{f}^{\sr R}_{-,\alpha}(x)=0.
\end{split}
\end{align}
Assuming that the WKB expansion of $\widetilde{f}^{\sr R}_{\pm,\alpha}(x)$ for $\frac{\beta^{1/2}\hbar}{2}$ or $-\frac{\beta^{-1/2}\hbar}{2}$
around $\hbar=0$ takes form
\begin{align}
\widetilde{f}^{\sr R}_{\pm,\alpha}(x)
\sim
C\,\exp\left(\frac{1}{2\hbar}\int^x y_{\pm}(x')dx' + O(\hbar^0) \right),
\end{align}
where $C$ is an $x$-independent factor, we obtain a ``spectral curve''
\be
\Sigma=\left\{(x,y_{\pm})\in {\IC}^2\, |\, A(x,y_{\pm})=0 \right\}
\ee
for the one-Penner Ramond-R wave-function \eqref{z_ram_mat} at $\eta=0$, where
\begin{align}
A(x,y)=y^2
-\left(\frac{\gamma^2}{(x-w)^2}
+\frac{\alpha_0^2-\gamma^2}{x(x-w)}\right).
\end{align}
We see that this spectral curve is the same as the spectral curve for the hermitian eigenvalue model \eqref{ZV-intro} with the one-Penner potential $V(x)=\gamma \log(x-w)$.

%******************************************************************************************
%******************************************************************************************
%******************************************************************************************
%******************************************************************************************

\newpage

\appendix

\section{Proofs and computations}  \label{sec-app}

%******************************************************************************************
%******************************************************************************************

\subsection{Computations in the Ramond-NS sector: the supercurrent $S(y)$}      \label{ssec-S-Ramond}

In this appendix we compute, in the operator formalism, the expectation value (\ref{vev-S-Ramond}) of the supercurrent $S(y)$ in the Ramond-NS sector. First, for the supercurrent $S_+(y)$ defined in (\ref{TS-Ramond}), by (\ref{positive:GL:on:vacuum}), (\ref{TS_comm_screen_R_sp}) and (\ref{TS_comm_screen_R}) we get
\begin{align}
&
\left\langle V^+_{N\sqrt{\beta}-\frac{\alpha}{\hbar},\hbox{{\boldmath $t$},{\boldmath $\xi$}}}\right|\sqrt{y}S_{+}(y)
\Phi^{\frac{\alpha}{\hbar}}(x,\theta)\prod\limits_{a=1}^N\left.\left.\Phi^{-\sqrt{\beta}}(z_a,\theta_a)\right|0,+\right\rangle =
\nonumber\\
& =  \Big(\frac{\sqrt{x}}{y-x}\left(\theta\partial_x-\partial_\theta\right)  + \frac{\Delta_{\frac{\alpha}{\hbar}}\theta(y+x)}{\sqrt{x}(y-x)^2}\Big) \left\langle V^+_{N\sqrt{\beta}-\frac{\alpha}{\hbar},\hbox{{\boldmath $t$},{\boldmath $\xi$}}}\right|
\Phi^{\frac{\alpha}{\hbar}}(x,\theta)\prod\limits_{a=1}^N\left.\left.\Phi^{-\sqrt{\beta}}(z_a,\theta_a)\right|0,+\right\rangle +
\nonumber\\
& \quad + \sum\limits_{a=1}^N\Big(\theta_a\partial_{z_a}- \partial_{\theta_a}\Big)
\Big(\frac{\sqrt{z_a}}{y-z_a}
\left\langle V^+_{N\sqrt{\beta}-\frac{\alpha}{\hbar},\hbox{{\boldmath $t$},{\boldmath $\xi$}}}\right|
\Phi^{\frac{\alpha}{\hbar}}(x,\theta)\prod\limits_{a=1}^N\left.\left.\Phi^{-\sqrt{\beta}}(z_a,\theta_a)\right|0,+\right\rangle
\Big) +
\nonumber\\
& \quad + \frac{1}{y}\left\langle V^+_{N\sqrt{\beta}-\frac{\alpha}{\hbar},\hbox{{\boldmath $t$},{\boldmath $\xi$}}}\right|
\Phi^{\frac{\alpha}{\hbar}}(x,\theta)\prod\limits_{a=1}^N\left.\left.\Phi^{-\sqrt{\beta}}(z_a,\theta_a)G_0\right|0,+\right\rangle.
\label{s_p_exp_ram}
\end{align}

In the next step we compute the expectation value of $S_-(y)$. Using the notation (\ref{Phi:m-Ramond}) we introduce
\be
\begin{split}
\ket{x,\theta,\vect{z},\vect{\theta}}
%& =
%{\rm e}^{-\frac{\alpha}{\hbar}\frac{\theta}{\sqrt{x}}\psi_0}
%\Phi^{\frac{\alpha}{\hbar}}_<(x,\theta)
%\Big(\prod\limits_{a=1}^N{\rm e}^{\sqrt{\beta}\frac{\theta_a}{ \sqrt{z_a}} \psi_0}\Big)
%\prod\limits_{a=1}^N \Phi^{-\sqrt{\beta}}_<(z_a,\theta_a)\ket{0,+} = \\
%&
= \Phi^{\frac{\alpha}{\hbar}}_{\le}(x,\theta) \prod\limits_{a=1}^N \Phi^{-\sqrt{\beta}}_{\le}(z_a,\theta_a)\ket{0,+}.
\end{split}
\ee
Then it follows that
\begin{align}
&\Phi^{\frac{\alpha}{\hbar}}(x,\theta)\prod\limits_{a=1}^N\Phi^{-\sqrt{\beta}}(z_a,\theta_a)\ket{0,\pm}  =
\nonumber\\
& = \prod\limits_{a=1}^N\Big(x-z_a-\theta\theta_a\sqrt{\frac{z_a}{x}}\Big)^{-\frac{\alpha\sqrt{\beta}}{\hbar}}
\prod\limits_{1\le a < b \le N}  \Big(z_a-z_b-\theta_a\theta_b\sqrt{\frac{z_b}{z_a}}\Big)^\beta \ket{x,\theta,\vect{z},\vect{\theta}}.
\end{align}
For $m < 0$ we have
\be
\left\{G_m,\psi_<(x)\right\} = \sum\limits_{n=1}^\infty x^{n-\frac12}{\sf a}_{m-n}, \qquad
\left[G_m,\phi_<(x)\right] = \sum\limits_{n=0}^\infty x^n\psi_{m-n},
\ee
and consequently
\begin{align*}
\sqrt{y}\left\{S_-(y),\psi_<(x)\right\}
& = \sum\limits_{m=1}^\infty y^{m-1}\left\{G_{-m},\psi_<(z)\right\}
= \sum\limits_{m,n=0}^\infty \hskip -2pt y^{m} x^{n+\frac12}{\sf a}_{-m-n-2}
= \sqrt{x}\,\frac{\partial\phi_<(y) - \partial\phi_<(x)}{y-x},  \\
\sqrt{y}\left[S_-(y),\phi_<(x)\right]
& = \sum\limits_{m,n=0}^\infty \hskip -2pt y^{m} x^{n}\psi_{-m-n-1} = \frac{\sqrt{y}\,\psi_<(y) - \sqrt{x}\,\psi_<(x)}{y-x},
\end{align*}
so that
\begin{equation}
\sqrt{y}\left[S_-(y),\Phi_<^{\frac{\alpha}{\hbar}}(x,\theta)\right]
=
\frac{\alpha}{\hbar}
\left(\frac{\sqrt{y}\,\psi_<(y) - \sqrt{x}\,\psi_<(x)}{y-x} +\frac{\partial\phi_<(y) - \partial\phi_<(x)}{y-x}\theta\sqrt{x}\right)
\Phi_<^{\frac{\alpha}{\hbar}}(x,\theta).
\end{equation}
Further
\begin{align}
\sqrt{y}\left\{S_-(y),\psi_0\right\} = \sum\limits_{m=0}^\infty y^{m}{\sf a}_{-m-1} \; = \; \partial\phi_<(y),
\nonumber
\end{align}
so that
\begin{align}
& \sqrt{y}\left[S_-(y),\Phi^{\frac{\alpha}{\hbar}}_{\le}(x,\theta)\right] =
\nonumber\\
& \quad =  \frac{\alpha}{\hbar}\left\{\frac{\sqrt{y}\,\psi_<(y) - \sqrt{x}\,\psi_<(x)}{y-x} +
\left(\frac{\partial\phi_<(y) - \partial\phi_<(x)}{y-x} + \frac{\partial\phi_<(y)}{x}\right)\theta\sqrt{x}\right\}
\Phi_{\le}^{\frac{\alpha}{\hbar}}(x,\theta) = \nonumber \\
& \quad =  \frac{\alpha}{\hbar} \left(\frac{\sqrt{y}\,\psi_<(y) - \sqrt{x}\,\psi_<(x)}{y-x} + \frac{y\partial\phi_<(y) - x\partial\phi_<(x)}{y-x} \frac{\theta}{\sqrt{x}}\right) \Phi_{\le}^{\frac{\alpha}{\hbar}}(x,\theta).
\end{align}
Finally
\be
\sqrt{y}S_-(y)\ket{0,+} =
\big(\psi_0\partial\phi_<(y) + \sqrt{y}\psi_<(y)\partial\phi_<(y) + Q\sqrt{y}\partial\psi_<(y)\big)\ket{0,+},
\ee
and
\begin{align}
\frac{1}{y}G_0\ket{0,+} = -\frac{1}{2y}Q\psi_0\ket{0,+}.
\nonumber
\end{align}
Combining the above ingredients we get
\begin{align*}
&\sqrt{y}S_-(y)\ket{x,\theta,\vect{z},\vect{\theta}} + \frac{1}{y}\Phi^{\frac{\alpha}{\hbar}}_{\le}(x,\theta)
\prod\limits_{a=1}^N \Phi^{-\sqrt{\beta}}_{\le}(z_a,\theta_a)G_0\ket{0,+} = \\
& =  \frac{\alpha}{\hbar} \left\{\frac{\sqrt{y}\,\psi_<(y) - \sqrt{x}\,\psi_<(x)}{y-x} +
\frac{y\partial\phi_<(y) - x\partial\phi_<(x)}{y-x} \frac{\theta}{\sqrt{x}}\right\} \ket{x,\theta,\vect{z},\vect{\theta}} + \\
& \qquad - \sqrt{\beta}\sum\limits_{a=1}^N  \left\{\frac{\sqrt{y}\,\psi_<(y) - \sqrt{z_a}\,\psi_<(z_a)}{y-z_a} +
\frac{y\partial\phi_<(y) - z_a\partial\phi_<(z_a)}{y-z_a}\frac{\theta_a}{\sqrt{z_a}}\right\}
\ket{x,\theta,\vect{z},\vect{\theta}}  + \\
& \qquad + \sqrt{y}\Big(\psi_<(y)\partial\phi_<(y) + Q\partial\psi_<(y)\Big)\ket{x,\theta,\vect{z},\vect{\theta}} +  \\
& \qquad + \left(\partial\phi_<(y)-Q/2y\right)\Phi^{\frac{\alpha}{\hbar}}_{\le}(x,\theta) \prod\limits_{a=1}^N \Phi^{-\sqrt{\beta}}_{\le}(z_a,\theta_a)\psi_0\ket{0,+}.
\end{align*}
On the other hand, note that
\begin{align*}
\left\langle V^+_{N\sqrt{\beta}-\frac{\alpha}{\hbar},\hbox{{\boldmath $t$},{\boldmath $\xi$}}}\right|\sqrt{y}\,\psi_<(y)
& =
\left\langle V^+_{N\sqrt{\beta}-\frac{\alpha}{\hbar},\hbox{{\boldmath $t$},{\boldmath $\xi$}}}\right|\frac{V_{\sr F}(y) -\xi_0}{\hbar},  \\
\left\langle V^+_{N\sqrt{\beta}-\frac{\alpha}{\hbar},\hbox{{\boldmath $t$},{\boldmath $\xi$}}}\right|\partial\phi_<(y)
& =
\left\langle V^+_{N\sqrt{\beta}-\frac{\alpha}{\hbar},\hbox{{\boldmath $t$},{\boldmath $\xi$}}}\right|\frac{V'_{\sr B}(y)}{\hbar},   \\
\left\langle V^+_{N\sqrt{\beta}-\frac{\alpha}{\hbar},\hbox{{\boldmath $t$},{\boldmath $\xi$}}}\right|\sqrt{y}\partial\psi_<(y)
& =
\left\langle V^+_{N\sqrt{\beta}-\frac{\alpha}{\hbar},\hbox{{\boldmath $t$},{\boldmath $\xi$}}}\right|\frac{\sqrt{y}}{\hbar}\frac{\partial}{\partial y}\left(\frac{V_{\sr F}(y)-\xi_0}{\sqrt{y}}\right).
\end{align*}
This implies that
\begin{align*}
&
\Big\langle V^+_{N\sqrt{\beta}-\frac{\alpha}{\hbar},\hbox{{\boldmath $t$},{\boldmath $\xi$}}}\Big|
\sqrt{y}S_-(y)\Big|x,\theta,\vect{z},\vect{\theta}\Big\rangle
+
\frac{1}{y}\Big\langle V^+_{N\sqrt{\beta}-\frac{\alpha}{\hbar},\hbox{{\boldmath $t$},{\boldmath $\xi$}}}\Big|\Phi^{\frac{\alpha}{\hbar}}_{\le}(x,\theta)
\prod\limits_{a=1}^N \Phi^{-\sqrt{\beta}}_{\le}(z_a,\theta_a)G_0\ket{0,+} = \\
& = \frac{\alpha}{\hbar^2}
\left\{\frac{V_{\sr F}(y) - V_{\sr F}(x)}{y-x} +
\frac{yV'_{\sr B}(y) - xV'_{\sr B}(x)}{y-x}\frac{\theta}{\sqrt{x}}\right\}
\Big\langle V^+_{N\sqrt{\beta}-\frac{\alpha}{\hbar},\hbox{{\boldmath $t$},{\boldmath $\xi$}}}\Big|x,\theta,\vect{z},\vect{\theta}\Big\rangle  +  \\
& \quad - \frac{\sqrt{\beta}}{\hbar}\sum\limits_{a=1}^N
\left\{\frac{V_{\sr F}(y) - V_{\sr F}(z_a)}{y-z_a} +
\frac{yV'_{\sr B}(y) - z_aV'_{\sr B}(z_a)}{y-z_a}\frac{\theta_a}{\sqrt{z_a}}\right\}
\Big\langle V^+_{N\sqrt{\beta}-\frac{\alpha}{\hbar},\hbox{{\boldmath $t$},{\boldmath $\xi$}}}\Big|x,\theta,\vect{z},\vect{\theta}\Big\rangle  +  \\
& \quad + \left(\frac{(V_{\sr F}(y)-\xi_0)V'_{\sr B}(y)}{\hbar^2} + \frac{Q \left(V'_{\sr F}(y)- V_{\sr F}(y)/2y + \xi_0/2y\right)}{\hbar}\right)
\Big\langle V^+_{N\sqrt{\beta}-\frac{\alpha}{\hbar},\hbox{{\boldmath $t$},{\boldmath $\xi$}}}\Big|x,\theta,\vect{z},\vect{\theta}\Big\rangle +  \\
& \quad + \left(\frac{V'_{\sr B}(y)}{\hbar} - \frac{Q}{2y}\right)
\Big\langle V^+_{N\sqrt{\beta}-\frac{\alpha}{\hbar},\hbox{{\boldmath $t$},{\boldmath $\xi$}}}\Big|\Phi^{\frac{\alpha}{\hbar}}_{\le}(x,\theta)
\prod\limits_{a=1}^N \Phi^{-\sqrt{\beta}}_{\le}(z_a,\theta_a)\psi_0\ket{0,+}.
\end{align*}
We also note that
\begin{align*}
& \bra{0,+}{\rm e}^{\frac{2}{\hbar}\xi_0\psi_0}{\rm e}^{\psi_0\left(\frac{\alpha}{\hbar}\frac{\theta}{\sqrt{x}} - \sqrt{\beta}\sum_{a=1}^{N} \frac{\theta_a}{\sqrt{z_a}}\right)}\psi_0\ket{0,+} = \\
& = \frac{\xi_0}{\hbar} - \frac12\left(\frac{\alpha}{\hbar}\frac{\theta}{\sqrt{x}} - \sqrt{\beta}\sum_{a=1}^{N} \frac{\theta_a}{\sqrt{z_a}}\right)
= \left(\frac{\xi_0}{\hbar} - \frac12\hbar\partial_{\xi_0}\right){\rm e}^{\frac{\xi_0}{\hbar}\left(\frac{\alpha}{\hbar}\frac{\theta}{\sqrt{x}} - \sqrt{\beta}\sum_{a=1}^{N} \frac{\theta_a}{\sqrt{z_a}}\right)}.
\end{align*}
Since
\begin{align*}
\sum\limits_{a=1}^Nz_a^n\, {\rm e}^{-\frac{\sqrt{\beta}}{\hbar}\sum_{a=1}^N\! V_{\sr R}(z_a,\theta_a)}
& = -\frac{\hbar}{\sqrt{\beta}}\,\partial_{t_n}  {\rm e}^{-\frac{\sqrt{\beta}}{\hbar}\sum_{a=1}^N\! V_{\sr R}(z_a,\theta_a)},    \\
\sum\limits_{a=1}^Nz_a^{n-\frac12}\theta_a\, {\rm e}^{-\frac{\sqrt{\beta}}{\hbar}\sum_{a=1}^N\! V_{\sr R}(z_a,\theta)}
& = -\frac{\hbar}{\sqrt{\beta}}\,\partial_{\xi_n}  {\rm e}^{-\frac{\sqrt{\beta}}{\hbar}\sum_{a=1}^N\! V_{\sr R}(z_a,\theta_a)},
\end{align*}
and
\begin{align*}
\frac{V_{\sr F}(y) - V_{\sr F}(z_a)}{y-z_a}
& =
\sum\limits_{m=1}^\infty\xi_m\frac{y^m - z_a^m}{y-z_a}
\; = \;
\sum\limits_{m=1}^\infty\sum\limits_{n=0}^{m-1}\xi_m y^n z_a^{m-n-1},     \\
\frac{yV'_{\sr B}(y) - z_aV'_{\sr B}(z_a)}{y-z_a}\frac{\theta_a}{\sqrt{z_a}}
& =
\sum\limits_{m=1}^\infty m t_m\frac{y^m-z_a^m}{y-z_a}\frac{\theta_a}{\sqrt{z_a}}
\; = \;
\sum\limits_{m=1}^\infty\sum\limits_{n=0}^{m-1} m t_m y^n z_a^{m-n-\frac32}\theta_a,
\end{align*}
we get
\begin{align*}
&
-\frac{\sqrt{\beta}}{\hbar}\sum\limits_{a=1}^N
\left\{\frac{V_{\sr F}(y) - V_{\sr F}(z_a)}{y-z_a} +
\frac{yV'_{\sr B}(y) - z_aV'_{\sr B}(z_a)}{y-z_a}\frac{\theta_a}{\sqrt{z_a}}\right\}
{\rm e}^{-\frac{\sqrt{\beta}}{\hbar}\sum_{a=1}^N\! V_{\sr R}(z_a,\theta_a)} = \\
& =
\sum\limits_{m=1}^\infty\sum\limits_{n=0}^{m-1} y^n
\left(\xi_m \partial_{t_{m-n-1}} + m t_m \partial_{\xi_{m-n-1}}\right)
{\rm e}^{-\frac{\sqrt{\beta}}{\hbar}\sum_{a=1}^N\! V_{\sr R}(z_a,\theta_a)}  =  \\
& = \sum\limits_{n=0}^\infty y^n \sum\limits_{m=n+1}^\infty
\left(\xi_m \partial_{t_{m-n-1}} + m t_m \partial_{\xi_{m-n-1}}\right)
{\rm e}^{-\frac{\sqrt{\beta}}{\hbar}\sum_{a=1}^N\! V_{\sr R}(z_a,\theta_a)} = \\
& \equiv \hbar^{-2}\,\widehat h(y)\,
{\rm e}^{-\frac{\sqrt{\beta}}{\hbar}\sum_{a=1}^N\! V_{\sr R}(z_a,\theta_a)},
\end{align*}
which defines the operator  (\ref{h-hat-Ramond})
\be
\widehat h(y) = \hbar^2  \sum_{n=0}^\infty y^n \sum_{m=n+1}^\infty
\big(\xi_m \partial_{t_{m-n-1}} + m t_m \partial_{\xi_{m-n-1}}\big).    \nonumber
\ee
Similarly
\be
\frac{\alpha}{\hbar^2}
\left\{\frac{V_{\sr F}(y) - V_{\sr F}(x)}{y-x} +
\frac{yV'_{\sr B}(y) - xV'_{\sr B}(x)}{y-x}\frac{\theta}{\sqrt{x}}\right\}
{\rm e}^{\frac{\alpha}{\hbar^2} V_{\sr R}(x,\theta)}
= \hbar^{-2}\,\widehat h(y)\,{\rm e}^{\frac{\alpha}{\hbar^2} V_{\sr R}(x,\theta)}.
\ee
Ultimately we find
\begin{align*}
&
\Big\langle V^+_{N\sqrt{\beta}-\frac{\alpha}{\hbar},\hbox{{\boldmath $t$},{\boldmath $\xi$}}}\Big|
\sqrt{y}S_-(y)\Big|x,\theta,\vect{z},\vect{\theta}\Big\rangle
+ \frac{1}{y}\Big\langle V^+_{N\sqrt{\beta}-\frac{\alpha}{\hbar},\hbox{{\boldmath $t$},{\boldmath $\xi$}}}\Big|\Phi^{\frac{\alpha}{\hbar}}_{\le}(x,\theta)
\prod\limits_{a=1}^N \Phi^{-\sqrt{\beta}}_{\le}(z_a,\theta_a)G_0\ket{0,+}  =
\\
& =   \frac{1}{\hbar^2}\left(V_{\sr F}(y)V'_{\sr B}(y) + Q\hbar V'_{\sr F}(y)+ \widehat h(y)\right)
\Big\langle V^+_{N\sqrt{\beta}-\frac{\alpha}{\hbar},\hbox{{\boldmath $t$},{\boldmath $\xi$}}}\Big|x,\theta,\vect{z},\vect{\theta}\Big\rangle   +
\\
& \quad -  \frac{1}{\hbar^2}\left( \frac{1}{2}\left(V'_{\sr B}(y) - \frac{Q\hbar}{2y}\right)\hbar^2\partial_{\xi_0}
+ \frac{Q\hbar V_{\sr F}(y)}{2 y}  \right)
\Big\langle V^+_{N\sqrt{\beta}-\frac{\alpha}{\hbar},\hbox{{\boldmath $t$},{\boldmath $\xi$}}}\Big|x,\theta,\vect{z},\vect{\theta}\Big\rangle.
\end{align*}
Combined with (\ref{s_p_exp_ram}) this finally gives the equation (\ref{vev-S-Ramond}).
%\be
%\begin{split}
%& \Left< \sqrt{y}S(y)\Phi^{\frac{\alpha}{\hbar}}(x,\theta)\Right> =
%\Big(\frac{\sqrt{x}}{y-x}\left(\theta\partial_x-\partial_\theta\right)  + \frac{\Delta_{\frac{\alpha}{\hbar}}\theta(y+x)}{\sqrt{x}(y-x)^2}\Big)
%\widehat\chi_\alpha(x,\theta) +  \\
%& \qquad \qquad +  \frac{1}{\hbar^2} \Big(\Big(V'_{\sr B}(y)- \frac{Q\hbar}{2 y}\Big)\Big(V_{\sr F}(y) - \frac12\hbar^2\partial_{\xi_0}\Big) + Q\hbar V'_{\sr F}(y)+ \widehat h(y)  \Big)\widehat\chi_\alpha(x,\theta).      \nonumber
%\end{split}
%\ee

%******************************************************************************************
%******************************************************************************************

\subsection{Computations in the Ramond-NS sector: the energy-momentum tensor $T(y)$}      \label{ssec-T-Ramond}

In this appendix, in the operator formalism we compute the expectation value (\ref{vev-T-Ramond}) of the energy-momentum tensor in the Ramond-NS sector. To start with, for the energy-momentum tensor $T_+(y)$ defined in (\ref{TS-Ramond}),
by (\ref{positive:GL:on:vacuum}), (\ref{zero:GL:on:vacuum}), (\ref{TS_comm_screen_R_sp}) and (\ref{TS_comm_screen_R}) we get
\begin{align}
&
\left\langle V^+_{N\sqrt{\beta}-\frac{\alpha}{\hbar},\hbox{{\boldmath $t$},{\boldmath $\xi$}}}\right|yT_{+}(y)
\Phi^{\frac{\alpha}{\hbar}}(x,\theta)\prod\limits_{a=1}^N\left.\left.\Phi^{-\sqrt{\beta}}(z_a,\theta_a)\right|0,+\right\rangle =
\nonumber\\
& = \Big(\frac{x}{y-x}\partial_x + \frac{y\left(\Delta_{\frac{\alpha}{\hbar}} + {\textstyle\frac12}\theta \partial_\theta\right)}{(y-x)^2}\Big) \left\langle V^+_{N\sqrt{\beta}-\frac{\alpha}{\hbar},\hbox{{\boldmath $t$},{\boldmath $\xi$}}}\right|
\Phi^{\frac{\alpha}{\hbar}}(x,\theta)\prod\limits_{a=1}^N\left.\left.\Phi^{-\sqrt{\beta}}(z_a,\theta_a)\right|0,+\right\rangle +
\nonumber\\
&\quad +  \sum\limits_{a=1}^N
\Big(\partial_{z_a}\frac{z_a}{y-z_a} - \partial_{\theta_a}\frac{\frac12\theta_a y}{(y-z_a)^2}\Big)
\left\langle V^+_{N\sqrt{\beta}-\frac{\alpha}{\hbar},\hbox{{\boldmath $t$},{\boldmath $\xi$}}}\right|
\Phi^{\frac{\alpha}{\hbar}}(x,\theta)\prod\limits_{a=1}^N\left.\left.\Phi^{-\sqrt{\beta}}(z_a,\theta_a)\right|0,+\right\rangle
+ \nonumber \\
&\quad +  \frac{1}{16y}\left\langle V^+_{N\sqrt{\beta}-\frac{\alpha}{\hbar},\hbox{{\boldmath $t$},{\boldmath $\xi$}}}\right|
\Phi^{\frac{\alpha}{\hbar}}(x,\theta)\prod\limits_{a=1}^N\left.\left.\Phi^{-\sqrt{\beta}}(z_a,\theta_a)\right|0,+\right\rangle.
\label{t_p_exp_ram}
\end{align}

Using the same calculational techniques as in appendix \ref{ssec-S-Ramond} we get
\[
[yT_-(x),\phi_<(x)] =
\sum\limits_{m=1}^{\infty} y^{m-1}[L_{-m},\phi_<(x)]
= \sum\limits_{m=1}^\infty\sum\limits_{n=0}^\infty {\sf a}_{-m-n} y^{m-1}x^n
=  \frac{y\partial\phi_<(y) - x \partial\phi_<(x)}{y-x},
\]
and
\begin{align*}
&
[yT_-(x),\psi_{\le}(x)] = \sum\limits_{m=1}^{\infty} y^{m-1}[L_{-m},\psi_{\le}(x)]
= \sum\limits_{m=1}^\infty\sum\limits_{n=0}^\infty\left(n+{\textstyle\frac12}m\right)\psi_{-m-n}y^{m-1}x^{n-\frac12} = \\
& \quad = \frac{\sqrt{x}}{2}\frac{\sqrt{y}\psi_{\le}(y) - \sqrt{x}\psi_{\le}(x) - (y-x)\partial\big(\sqrt{x}\psi_{\le}(x)\big)}{(y-x)^2} + \frac{1}{2\sqrt{x}} \frac{y\partial\big(\sqrt{y}\psi_{\le}(y))- x\partial\big(\sqrt{x}\psi_{\le}(x)\big)}{y-x},
\end{align*}
where we denoted $\psi_{\le}(x) = \frac{\psi_0}{\sqrt{x}} + \psi_<(x)$. Consequently
\begin{align}
\nonumber
\left[y T_-(y),\Phi^{\frac{\alpha}{\hbar}}_\le (z,\theta)\right]
= &  \frac{\alpha}{\hbar} \Big( \frac{y\partial\phi_<(y) - x\partial\phi_<(x)}{y-x}
+ \frac{y\partial_y\left(\sqrt{y}\psi_{\le}(y)\right) - x\partial_x\left(\sqrt{x}\psi_{\le}(x)\right)}{y-x}\frac{\theta}{2\sqrt{x}}  + \\
\nonumber
&
+ \frac{\sqrt{y}\psi_{\le}(y) - \sqrt{x}\psi_{\le}(x) - (y-x)\partial_x\left(\sqrt{x}\psi_{\le}(x)\right)}{(y-x)^2}\frac{\sqrt{x}\theta}{2}
\Big)
\Phi^{\frac{\alpha}{\hbar}}_\le(z,\theta).
\end{align}
Further
\begin{align}
yT_-(y)\ket{0,+}
= & \frac{y}{2} \Big( \left(\partial\phi_<(y)\right)^2 + Qy\partial^2\phi_<(y)
+\partial\psi_<(y)\psi_<(y) \Big)\ket{0,+} +
\nonumber\\
& +
\frac12 \Big( \sqrt{y}\partial\psi_<(y) + \frac{1}{2\sqrt{y}}\psi_<(y) \Big)\psi_0 \ket{0,+}.
\end{align}
This gives
\begin{align*}
&
\Big\langle V^+_{N\sqrt{\beta}-\frac{\alpha}{\hbar},\hbox{{\boldmath $t$},{\boldmath $\xi$}}}\Big|
yT_-(y)\Big|x,\theta,\vect{z},\vect{\theta}\Big\rangle = \\
& =
\left\{
\frac{\alpha}{\hbar^2}
\left(
\frac{yV'_{\sr B}(y) - xV'_{\sr B}(x)}{y-x}
+
\frac{yV'_{\sr F}(y) - xV'_{\sr F}(x)}{2(y-x)}\frac{\theta}{\sqrt{x}}
+
\frac{V^{(2)}_{\sr F}(y,x)\sqrt{x}\theta}{2(y-x)^2}
\right)
\right. +
\\
&
\quad -\frac{\sqrt{\beta}}{\hbar}
\sum\limits_{a=1}^N
\left(
\frac{yV'_{\sr B}(y) - z_aV'_{\sr B}(z_a)}{y-z_a}
+
\frac{yV'_{\sr F}(y) - z_aV'_{\sr F}(z_a)}{2(y-z_a)}\frac{\theta_a}{\sqrt{z_a}}
+
\frac{V^{(2)}_{\sr F}(y,z_a)\sqrt{z_a}\theta}{2(y-z_a)^2}
\right) +
\\
&
\quad \left.+
\frac1{2\hbar^2} \left(y\left(V'_{\sr B}(y)\right)^2 + Qy V''_{\sr B}(y) +
V'_{\sr F}(y)\left(V_{\sr F}(y)-\frac{\hbar^2}{2}\partial_{\xi_0}\right)
\right)
\right\}
\Big\langle V^+_{N\sqrt{\beta}-\frac{\alpha}{\hbar},\hbox{{\boldmath $t$},{\boldmath $\xi$}}}\Big|x,\theta,\vect{z},\vect{\theta}\Big\rangle =
\\
& =
\frac{1}{\hbar^2}
\left\{
{\widehat f}(y)
+
\frac12
\left(y\left(V'_{\sr B}(y)\right)^2 + Qy V''_{\sr B}(y) +
V'_{\sr F}(y)\left(V_{\sr F}(y)-\frac{\hbar^2}{2}\partial_{\xi_0}\right)
\right)
\right\}
\Big\langle V^+_{N\sqrt{\beta}-\frac{\alpha}{\hbar},\hbox{{\boldmath $t$},{\boldmath $\xi$}}}\Big|x,\theta,\vect{z},\vect{\theta}\Big\rangle,
\end{align*}
where ${\widehat f}(y) $ was defined in (\ref{f-hat-Ramond})
\[
{\widehat f}(y)
=
\hbar^2\sum\limits_{n=0}^\infty y^n \sum\limits_{k=n+1}^\infty \left(kt_k\partial_{t_{k-n-1}} + \left(k-\frac{n+1}{2}\right)\xi_k\partial_{\xi_{k-n-1}}\right).
\]
Ultimately, combined with (\ref{t_p_exp_ram}) we find the equation (\ref{vev-T-Ramond}).
%\be
%\begin{split}
%& \Left< yT(y)\Phi^{\frac{\alpha}{\hbar}}(x,\theta)\Right> =
%\Big(\frac{x}{y-x}\partial_x + \frac{y\left(\Delta_{\frac{\alpha}{\hbar}} + {\textstyle\frac12}\theta \partial_\theta\right)}{(y-x)^2} + \frac{1}{16y}\Big)  \widehat\chi_\alpha(x,\theta) +  \\
%&   \qquad \qquad + \frac{1}{\hbar^2} \Big(  {\widehat f}(y) + \frac12
%\Big(y\left(V'_{\sr B}(y)\right)^2 + Qy V''_{\sr B}(y) +
%V'_{\sr F}(y)\Big(V_{\sr F}(y)-\frac{\hbar^2}{2}\partial_{\xi_0}\Big) \Big) \Big)
%\widehat\chi_\alpha(x,\theta).  \nonumber
%\end{split}
%\ee

%******************************************************************************************
%******************************************************************************************

\subsection{Computations in the Ramond-NS super-eigenvalue model} \label{app:commutation}

In this appendix we present results relevant for the computations in the eigenvalue model in the Ramond-NS sector.
First, we find that commutation relations for operators $\widehat{h}(x)$ in (\ref{h_hat_op}) and $\widehat{f}(x)$ in (\ref{f_hat_op}) take form
\begin{align}
\begin{split}
\left[\widehat{h}(x), \partial_x^nV_{\sr B}(x)\right] &=\frac{1}{n+1}\hbar^2\partial_x^{n+1}V_{\sr F}(x),
\\
\left\{\widehat{h}(x), \partial_x^nV_{\sr F}(x)\right\} &=
\left[\widehat{f}(x), \partial_x^nV_{\sr B}(x)\right]=
\frac{1}{n+1}\hbar^2\partial_x^{n+1}\big(xV_{\sr B}'(x)\big),
\\
\left[\widehat{f}(x), \partial_x^nV_{\sr F}(x)\right] &=\frac{1}{2(n+1)(n+2)}\hbar^2\partial_x^{n+1}\big((2n+3)xV_{\sr F}'(x)-V_{\sr F}(x)\big).
\end{split}
\end{align}

Second, consider $\chi_{\alpha}^{\textrm{ins}}(x,\sqrt{x}\eta)$ defined in (\ref{chi_ins_def})
\begin{equation}
\chi_{\alpha}^{\textrm{ins}}(x,\sqrt{x}\eta)=\Big(1+\frac{\alpha\sqrt{\beta}\eta}{\hbar}\sum_{a=1}^N\Big(\frac{1}{2}\eta_a+\frac{z_a\eta_a}{x-z_a}\Big)\Big)\prod_{a=1}^N(x-z_a)^{-\frac{\sqrt{\beta}}{\hbar}\alpha}.
\nonumber
\end{equation}
Its derivatives with respect to $\eta$ and $x$ take form
\begin{align}
\partial_{\eta}\chi_{\alpha}^{\textrm{ins}}(x,\sqrt{x}\eta)&=
\frac{\alpha\sqrt{\beta}}{2\hbar}\sum_{a=1}^N\frac{(x+z_a)\eta_a}{x-z_a}\chi_{\alpha}^{\textrm{ins}}(x,\sqrt{x}\eta),
\nonumber\\
\partial_x\chi_{\alpha}^{\textrm{ins}}(x,\sqrt{x}\eta)&=
-\frac{\alpha\sqrt{\beta}}{\hbar}\sum_{a=1}^N
\Big(\frac{1}{x-z_a}-\frac{z_a\eta_a\eta}{(x-z_a)^2}\Big)\chi_{\alpha}^{\textrm{ins}}(x,\sqrt{x}\eta),
\nonumber\\
\partial_x\partial_{\eta}\chi_{\alpha}^{\textrm{ins}}(x,\sqrt{x}\eta)&=
-\frac{\alpha\sqrt{\beta}}{\hbar}\sum_{a=1}^N\frac{z_a\eta_a\chi_{\alpha}^{\textrm{ins}}(x,\sqrt{x}\eta)}{(x-z_a)^2}
-\frac{\alpha^2\beta}{2\hbar^2}\sum_{a,b=1}^N\frac{(x+z_a)\eta_a\chi_{\alpha}^{\textrm{ins}}(x,\sqrt{x}\eta)}{(x-z_a)(x-z_b)} +
\nonumber\\
&\qquad
+\frac{\alpha^2\beta\eta}{2\hbar^2}\sum_{a,b=1}^N\frac{(x+z_a)z_b\eta_a\eta_b\chi_{\alpha}^{\textrm{ins}}(x,\sqrt{x}\eta)}{(x-z_a)(x-z_b)^2}.
\end{align}

%*********************************************************************
%*********************************************************************
%*********************************************************************
%*********************************************************************

\acknowledgments{
This work is supported by the ERC Starting Grant no. 335739 \emph{``Quantum fields and knot homologies''} funded by the European Research Council under the European Union's Seventh Framework Programme, and the Ministry of Science and Higher Education in Poland.} The work of PC is also supported by the NCN Preludium grant no. 2016/23/N/ST1/01250 \emph{``Quantum curves and Schr\"odinger equations in matrix models''}. The work of MM is also supported by the Max-Planck-Institut f\"ur Mathematik in Bonn.
%We thank for discussions and comments on the manuscript.

\newpage

\bibliographystyle{JHEP}
\bibliography{abmodel}

\end{document}